\documentclass[useAMS,usenatbib]{mn2e}

\let\jnlstyle=\rm
\def\refjnl#1{{\jnlstyle#1}}

\def\apj{\refjnl{ApJ}}                 
\def\apjl{\refjnl{ApJ}}                
\def\apjs{\refjnl{ApJS}}               
\def\aap{\refjnl{A\&A}}                
\def\mnras{\refjnl{MNRAS}}             
\def\prd{\refjnl{Phys.~Rev.~D}}        
\def\jcap{\refjnl{JCAP}}          

\input{psfig.sty}

\usepackage{graphicx}
\usepackage{epsfig}
\usepackage{amssymb}
\usepackage{amsmath}
\usepackage{amsfonts}
\usepackage{txfonts}
\usepackage{multirow}
\usepackage{afterpage}
\usepackage{xspace}
\usepackage{hyperref}
\usepackage{ulem} 

\newcommand{\mycomment}[1]{}
\newcommand{\dNLA}{extended\text{-NLA}\xspace}
\newcommand{\dTT}{extended\text{-TT}\xspace}
\newcommand{\gtwopcf}{$\gamma$\text{-2PCF}\xspace}
\newcommand{\eg}{\textit{e.g. }\xspace}

\newcommand{\lcdm}{$\Lambda$CDM\xspace}

\title[Lensing beyond 2pt: accounting for IA]{
Non-linear infusion of intrinsic alignment and source clustering: impact on non-Gaussian cosmic shear statistics }

\author[J. Harnois-D\'{e}raps et al.]{J. Harnois-D\'{e}raps$^{1}$
\thanks{E-mail: joachim.harnois-deraps@ncl.ac.uk}
N. \v{S}ar\v{c}evi\'c,$^{1,2}$
L. Medina Varela$^{3}$,
J. Armijo$^{4,5}$
C. T. Davies$^{6}$,
\newauthor 
N. van Alfen$^{7}$
J. Blazek$^{7}$
L. Castiblanco$^{1,8}$,
A. Halder$^{6,9}$,
K. Heitmann$^{10}$,
P. Larsen$^{10}$,\newauthor
L. Linke$^{11}$,
J. Liu$^{4,5}$,
C. MacMahon-Gell\'er$^1$,
L. Porth$^{12}$,
S. Rangel$^{10}$,
C. Uhlemann$^{1,8}$\newauthor
and the LSST Dark Energy Science Collaboration
\\
$^{1}$School of Mathematics, Statistics and Physics, Newcastle University, Herschel Building, NE1 7RU, Newcastle-upon-Tyne, UK\\
$^{2}$Duke University, Durham, NC 27708, USA\\
$^{3}$Department of Physics, The University of Texas at Dallas, TX 75080, USA\\
$^{4}$Kavli IPMU (WPI), UTIAS, The University of Tokyo, Chiba 277-8583, Japan \\
$^{5}$Center for Data-Driven Discovery (CD3), UTIAS, The University of Tokyo, Chiba 277-8583, Japan\\
$^{6}$University Observatory, Faculty of Physics, Ludwig-Maximilians-Universität, Scheinerstr. 1, 81679, Munich, Germany\\
$^{7}$Department of Physics, Northeastern University, Boston, MA 02115, USA\\
$^{8}$Fakultät für Physik, Universität Bielefeld, Postfach 100131, 33501 Bielefeld, Germany\\
$^{9}$Institute of Astronomy and Kavli Institute for Cosmology, University of Cambridge, Madingley Road, Cambridge CB3 0HA, United Kingdom\\
$^{10}$Argonne National Laboratory, Lemont, IL 60439, USA\\
$^{11}$Universit\"at Innsbruck, Institut f\"ur Astro- und Teilchenphysik, Technikerstr. 25/8, 6020 Innsbruck, Austria\\
$^{12}$Argelander-Institut für Astronomie,
Auf dem Hügel, 71
53121, Bonn,
Germany\\
}

\date{Accepted XXX. Received YYY; in original form ZZZ}

\pubyear{2024}

\begin{document}
\label{firstpage}
\pagerange{\pageref{firstpage}--21}
\maketitle


\begin{abstract}
Intrinsic alignments (IA) of galaxies is one of the key secondary signals to cosmic shear measurements, and must be modeled to interpret weak lensing data and infer the correct cosmology. There are large uncertainties in the physical description of IA, and analytical calculations are often out of reach for weak lensing statistics beyond two-point functions. We present here a set of six flexible IA models infused directly into weak lensing simulations, constructed from the mass shells, the projected tidal fields and, optionally, dark matter halo catalogues. We start with the non-linear linear alignment (NLA) and progressively sophisticate the galaxy bias and the tidal coupling models, including the commonly-used extended NLA  (also known as the e-NLA or $\delta$-NLA) and the  tidal torque (TT) models.  We validate our methods with MCMC analyses from two-point shear statistics,  then compute the impact on non-Gaussian cosmic shear probes from these catalogues as well as from reconstructed convergence maps. 
We find that the $\delta$-NLA model has by far the largest impact on most probes, at times more than twice the strength of the NLA. We also observe large differences between the IA models in under-dense regions, which makes minima, void profiles and lensing PDF the best probes for model rejection. Furthermore, our bias models allow us to separately study the source-clustering term for each of these probes, finding good agreement with the existing literature, and extending the results to these new probes. The third-order aperture mass statistics ($M_{\rm ap}^3$) and the integrated three-point functions are particularly sensitive to this when including low-redshift data, often exceeding a 20\% impact on the data vector. Our IA models are straightforward to implement and rescale from a single simulated IA-infused galaxy catalogue, allowing for fast model exploration.

\end{abstract}

\begin{keywords}
Gravitational lensing: weak -- Methods: numerical -- Cosmology: dark matter, dark energy \& large-scale structure of Universe 
\end{keywords}



\section{Introduction}
\label{sec:intro}

Recent cosmic shear measurements from the Kilo Degree Survey\footnote{KiDS:kids.strw.leidenuniv.nl}, the Dark Energy Survey\footnote{DES:www.darkenergysurvey.org}, and the Hyper Suprime Camera Survey\footnote{HSC:www.naoj.org/Projects/HSC} have established weak gravitational lensing as a competitive probe of cosmology \citep[see \eg][]{KiDS1000_Asgari, KiDS1000_vdB, KiDS1000_Li, DESY3_Secco, DESY3_Amon, HSCY3_Cl, HSCY3_2pcf, KiDSLegacy}, achieving percent-level measurements of the structure growth parameter $S_8\equiv\sigma_8\sqrt{\Omega_{\rm m}/0.3}$.
The parameters $\Omega_{\rm m}$ and $\sigma_8$, which respectively describe the abundance of non-relativistic matter and the amplitude of the linear fluctuations in the matter density fluctuations on scales of $8 \: h^{-1}$ Mpc, are highly degenerate when probed by lensing alone, and additional data (\eg galaxy clustering data) are required to break the degeneracy in photometric galaxy surveys \citep{KiDS1000_Heymans, DESY3_3x2, HSCY3_3x2}.
Despite these successful achievements, the current precision of cosmic shear cosmology is mainly limited by two dominant effects: (i) uncertainty in the intrinsic alignment (IA) of galaxies, a secondary signal that tends to contaminate some of the shape correlations produced by lensing \citep[see. \eg][for reviews on IA]{Troxel_IA_review_2015, Kirk_IA_review_2015, Joachimi_IA_review_2015, Kiessling_IA_review_2015,Lamman_IA_guide}, and (ii) uncertainty due to baryonic feedback, which significantly redistributes the matter distribution at the halo scale and is absent from predictions based on collisionless dark matter models  \citep{Semboloni11, HWVH15, HorizonAGN}. Whereas both are important and interconnected, the current paper focuses on the former effect.   
If unaccounted for, the IA can bias the inferred cosmological parameters by 4-5$\sigma$ \citep{Kirk2012, Krause2016}.
Additionally, using an inaccurate IA model can substantially impair the inference process, as demonstrated by \citet{DESY3_Secco} in their study on DES-Y3 analysis and by \citet{Paopiamsap2024} in the context of an LSST-like cosmic shear analysis.

Different physical models have been developed to describe the origin and impact of the IA, including the non-linear linear alignment model \citep[NLA hereafter, ][]{NLA}, the density-weighted NLA (aka $\delta$-NLA or extended-NLA), and the tidal torque model \citep[][TT hereafter]{TT_V0, TATT}, which respectively assume a linear and quadratic coupling between galaxy ellipticities and the local tidal field. The Tidal Alignment and Tidal Torque model (TATT) aims to describe the contributions from a mixed sample of galaxies and for this assumes a linear combination of $\delta$-NLA and TT.
Alternatively, IA can also be described by the halo model, which models the signal as a function of halo properties such as their mass, concentration, and shapes \citep{Fortuna2020}, or by the effective field theory of large-scale structure \citep{IA_EFT}.

Several observations have sought to place constraints on the parameters of these different IA models \citep[\eg][]{BlueIA, Singh_IA_LOWZ, DESY1_IA_Samuroff, Johnston_IA, Linke_IA_3pt, Georgiou2025}.
These observations seem to converge towards a strong colour-dependence: red bright galaxies (often elliptical) are generally strongly aligned, with hints of a radial dependence, while blue galaxies (typically spiral) are consistent with the no-alignment scenario.
The selection effects of a given survey play a crucial role in these measurements and the subsequent inference, making it difficult to generalise these measurements to a different galaxy sample, therefore leaving behind a large uncertainty on the IA parameters.

While most theoretical methods have been developed to provide prescriptions for modelling IA in two-point statistics, some have been applied to infuse galaxy alignments directly into cosmological simulations, as in \citet{Joachimi_IA2013, Fluri2019, Tidalator2D, MICE_IA, Lanzieri2023, vanAlfen2023}. 
Access to such IA-infused numerical simulations is crucial for several applications, including 
validating theoretical models deep in the non-linear regime or utilising non-linear galaxy bias models \citep{NonlinearBiasNicola}, 
 predicting the impact of IA on non-Gaussian lensing probes ({\it i.e.} beyond-2pt  or higher-order statistics), for which no models exist \citep{Zuercher2020a, Tidalator2D, Lee2025_TNG-IA},
testing IA mitigation techniques such as self-calibration \citep[][]{SelfCalibrationYao1, SelfCalibrationYao2, SelCalibrationPedersen,Bera2025} or exploring the connection between large dark matter haloes and IA \citep{MICE_IA,IA_ML,Pandya2025}.
It is necessary to consider each of these factors in order to correctly interpret cosmological data from photometric redshift surveys.

This paper addresses several of the above-mentioned limitations, as we present a novel pipeline with which we infuse the NLA, the \dNLA, and the TT model on the same underlying large $N$-body simulation, producing in each case distinct cosmic shear catalogues. 
In addition, we introduce the \dTT model (which takes into account the fact that galaxies trace dark matter even in the TT model), then proceed to couple the cosmic tidal fields with galaxies taken from halo occupation distributions (HOD), thereby probing the impact of realistic non-linear galaxy bias on the IA signal.  Based on the SkySim5000 simulation -- an upgrade from cosmoDC2 \citep{cosmoDC2},  these new models are both physically motivated and challenging to describe theoretically as they require perturbation expansions beyond the second order. Existing beyond-2pt analyses of weak lensing data only include the NLA model \citep{DESY3_Zuercher, KiDS1000_JHD}, which is bound to be insufficient with the precision increase provided with the new generation of cosmic shear surveys such as those from the Vera Rubin Observatory \citep{LSST-Design} or {\it Euclid} \citep{RedBook}. 

Additionally, our different catalogues allow us to study the impact of source clustering on different weak lensing probes, a topic which has been explored in the literature in {\it e.g.} \citet{SC_Bernardeau}, \citet{
source_lens_clustering} and more recently in \citet{DESY3_Gatti_source_clust}, for the case of linear biased tracers. We reproduce some of these results and push the investigation further, including non-linear biased tracers and additional weak lensing statistics. 

This paper is structured as follows: after reviewing the theory and measurement of cosmic shear data in Sec. \ref{sec:theory}, we describe our IA models in Sec. \ref{sec:IA_th}.  
Their full numerical implementation is described in Sec. \ref{sec:sims} and validated against theoretical predictions at the level of  two-point shear correlation functions in Sec. \ref{sec:validation}.
Since our infusion method acts on shear galaxy catalogues and on convergence maps, we are in an ideal position to quantify the impact of multiple IA models on different non-Gaussian statistics, which we report in Sec. \ref{sec:HOWLS}, before concluding in Sec. \ref{sec:conclusion}. 
 
\section{Cosmic shear statistics}
\label{sec:theory}

Due to non-linear structure formation, two-point statistics do not provide a complete description of the density distribution of matter in the Universe. Therefore, cosmic shear data can be analysed using various summary statistics, each with its own advantages and disadvantages, making many alternative methods highly complementary, as recently shown in \citet{HOWLS_paper1}.
We start here  with a description of the shear two-point functions, then present other complementary cosmic shear statistics in the second part of this section.

\subsection{\texorpdfstring{$\gamma$-2PCF}{gamma-2PCF}}
\label{subsec:wl-th}

\begin{figure}
\includegraphics[width=\columnwidth]{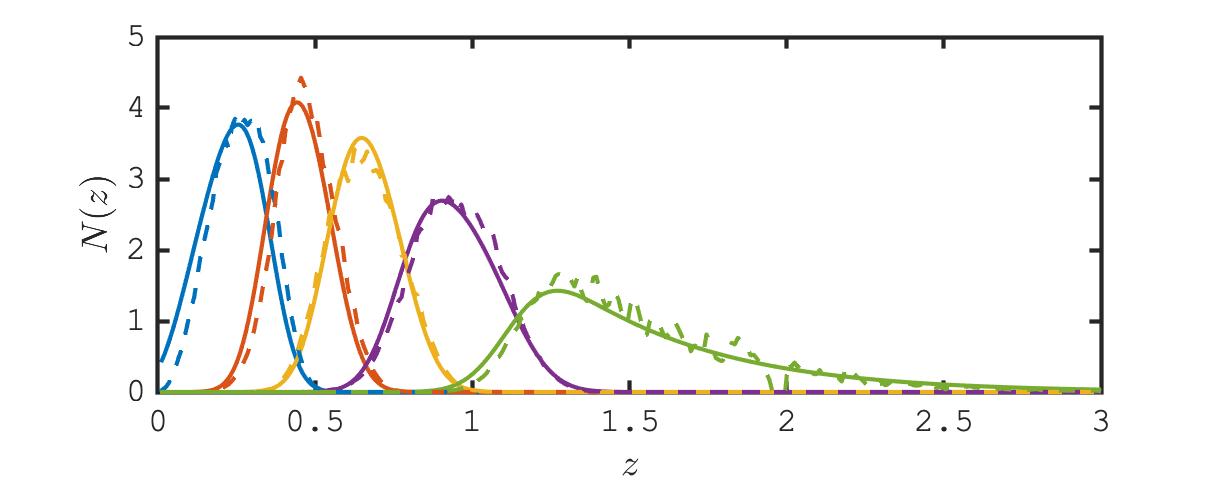}
\caption{Tomographic redshift distributions in our simulations, either taken from the Year-1 specifications for LSST (solid, Eq. \ref{eq:nz}) or from a matched selection applied to an HOD galaxy catalogue (dashed, see Sec. \ref{subsec:HOD}).}
\label{fig:Nz}
\end{figure}

Shear two-point correlation functions (\gtwopcf hereafter) can be predicted from semi-analytical theory with percent-level precision and are therefore an ideal quantity to validate weak lensing simulations.
In the Limber approximation\footnote{See \citet{Kilbinger17} for a comparison between the Limber approximation and the exact calculations.}, the tomographic lensing power spectrum $C^{ij}_{\ell}$, obtained for combinations of redshift bins $i$ and $j$, is calculated from an integral over the three-dimensional matter power spectrum $P_{\delta}(k, z)$ as:
  \begin{eqnarray}
C_{\ell}^{ij} = \int_0^{\chi_{\rm H}}  \frac{q^i(\chi) \,q^j(\chi)}{\chi^2} \, P_{\delta}\, \bigg(k = \frac{\ell+1/2}{\chi},z(\chi)\bigg) \ {\rm d}\chi,
\label{eq:C_ell}
\end{eqnarray}
where $\ell$ are angular multipoles, $k$ are angle-averaged Fourier modes, $\chi_{\rm H}$ is the comoving distance to the horizon, and the lensing kernels $q^{i}$ and $q^j$ are given by:
\begin{eqnarray}
q^{i}(\chi) = \frac{3}{2}\Omega_{\rm m} \, \bigg(\frac{H_0}{c} \bigg)^2 \frac{\chi}{a(\chi)} \int_{\chi}^{\chi_{\rm H}} n^i(\chi')\frac{\chi' - \chi}{\chi'}{\rm d}\chi' \, .
\label{eq:q_lensing}
\end{eqnarray}
In the previous expression, $c$ is the speed of light, $H_0$ the Hubble parameter, $n^i(z) =\frac{{\rm d}z}{{\rm d} \chi}n^i(\chi)$ refers to the redshift distribution in tomographic bins `$i$', while $a(\chi)$ is the scale factor at comoving distance $\chi$ from the observer, and we assume a flat universe.
The matter power spectrum is computed from {\sc Halofit} \citep{Takahashi2012} in this work, however other public tools provide accurate predictions, including \eg  {\sc HMcode} \citep{HMCode2020}, the {\sc EuclidEmulator} \citep{EuclidEmulator}, the {\sc Bacco} emulator \citep{BACCOEmulator}, or the {\sc MiraTitan} emulator \citep{miratitan_final}. 
Predictions for the \gtwopcf are finally computed from Eq. (\ref{eq:C_ell}) as:
\begin{eqnarray}
\xi_{+/-}^{ij}(\vartheta) = \frac{1}{2\pi} \int_0^{\infty} C_{\ell}^{ij} \, J_{0/4}(\ell \vartheta) \, \ell \, {\rm d}\ell,
\label{eq:xipm_th}
\end{eqnarray}
where $J_{0/4}(x)$ are Bessel functions of the first kind. 
In this paper, these calculations are carried out by the public\footnote{{\sc CosmoSIS}:https://cosmosis.readthedocs.io/en/latest/ } {\sc CosmoSIS} cosmological inference package \citep{cosmoSIS}.
The redshift distribution $n (z)$ is taken from the LSST Year-1 forecast \citep{LSST-SRD}:
\begin{eqnarray}
n (z) =  z^2 \: {\rm exp} \left[-\left(\frac{z}{z_0}\right)^\alpha \right]\, ,
\label{eq:nz}
\end{eqnarray}
with pivot redshift $z_0=0.13$ and a power-law index $\alpha = 0.78$.
The distribution is normalised to provide a number density of $n_{\rm eff} = 3.0 $  galaxies per arcmin$^{2}$.
This is lower than the expected number density for the first data release ($n_{\rm eff}\sim10$ galaxies arcmin$^{-2}$), but is large enough\footnote{The full predicted galaxy density for Rubin approaches 30 gal arcmin$^{-2}$, but including shape noise would counterbalance this gain; our conclusions are therefore realistic. }  to validate our methods since we turn off shape noise for most of our measurements; this choice is primarily driven to make the calculations more tractable (see Sec. \ref{sec:sims}). 
Source clustering will be impacted by this, however, since it affects both the mean signal and its noise, as shown in \citet{DESY3_Gatti_source_clust}, and the latter term will be under-estimated in our noise-free scenario.
This global $n(z)$ is further split into five equi-populated tomographic bins, and smoothed with a Gaussian filter of width $\sigma = 0.05 \: (1 + z)$ to mimic the photometric selection process, shown by the solid lines in Fig. \ref{fig:Nz}\footnote{The LSST SRD-Y1 forecasted redshift distribution can be generated here: github.com/LSSTDESC/CCLX/blob/master/srd\_redshift\_distributions.py}. 

The weak lensing signal is measured from the ellipticities $\epsilon_{1/2}$ of simulated or observed galaxies, which, in the absence of systematics and IA, are unbiased estimators of the cosmic shear components $\gamma_{1/2}$.
In particular, the \gtwopcf are estimated as:
\begin{eqnarray}
\widehat{\xi_{\pm}^{ij}}(\vartheta) = \frac{\sum_{a,b} w_a w_b \left[\epsilon_{a, \rm +}^i\epsilon_{b, \rm +}^j     \pm \epsilon_{a, \times}^i\epsilon_{b, \times}^j    \right]\Delta_{ab}(\vartheta)}{\sum_{a,b} w_a w_b} \, ,
\label{eq:2PCF_estimator}
\end{eqnarray}
where the sum is over all pairs of galaxies $a, b$ separated by an angular distance $\vartheta$ on the sky, respectively belonging to the tomographic bins $i$ and $j$.
Here, $w_{a (b)}$ represent the weights, describing the precision of the shape measurement, which we set to one for all galaxies in this work, while the tangential/cross components of ellipticities are denoted as $\epsilon_{+/\times}$, respectively. The binning operator $\Delta_{ab}(\vartheta)$ is equal to unity if the angular separation between the two galaxies falls within the $\vartheta$-bin, and zero otherwise.
In this work, we construct lensing catalogues from numerical simulations (described in Sec. \ref{sec:sims}), from which we measure $ \widehat{\xi_{\pm}^{ij}}(\vartheta)$ with {\sc Treecorr} \citep{TreeCorr}.
We set therein the {\sc bin\_slop} accuracy parameter to 0.05, then compute the correlations in 20 logarithmically-spaced angular bins with outer edges ranging from $0.5$ to $475.5$ arcmin.

\subsection{Non-Gaussian lensing statistics}
\label{subsec:beyond-2pt}

Non-Gaussian statistics have the potential to unlock cosmological information stored in the complex phases of the density field, outperforming the \gtwopcf that can only access information encoded in the complex amplitudes \citep{2002MNRAS.337..488C}.
Due to the complexity of the non-linear structure formation, no optimal estimator has been identified to-date as being able to capture `all' information, but numerous studies show a clear gain in constraining power, especially when used in combination with two-point functions \citep{Fu2014, Gruen2017, HD21, DESY3_Zuercher, HSCY1_peaks_sims, KiDS1000_Burger, KiDS1000_Map3, HOWLS_paper1}.
While some of these act on galaxy catalogues directly, many non-Gaussian statistics are measured on convergence maps ($\kappa$) or aperture mass maps ($M_{\rm ap}$), requiring the post-processing of galaxy catalogues with a mass-reconstruction algorithm.
To accommodate a variety of cases, we therefore generate all three types, galaxy shear catalogues $M_{\rm ap}$ and $\kappa$, the latter being produced from the mock galaxies' ellipticities with the standard Kaiser \& Squires (KS) inversion technique \citep{KaiserSquires}, which relates the shear and convergence to the lensing potential. This is achieved by assigning the two ellipticity components of every galaxy to spherical {\sc Healpix}\footnote{{\sc Healpix}: http://healpix.sf.net} maps \citep{healpix},  then converting the shear maps to convergence maps by solving the KS inversion in spherical harmonic space \citep[][see their equation 10]{Gatti20}:
\begin{eqnarray}
\gamma_{\ell m} = - \sqrt{\frac{(\ell+2)(\ell - 1)}{\ell(\ell+1)}}\left( \kappa_{E,\ell m} + {\rm i} \kappa_{B,\ell m} \right) \, .
\label{eq:KS}
\end{eqnarray}
In the above, $\kappa_{E}$ and $\kappa_{B}$ are the $E/B$ mode decomposition of the convergence maps, the latter being generally only second order and hence set to zero in this work\footnote{Small non-zero $B$-modes near the map boundary can occur when carrying out the {\sc map2alm} transform over non-periodic or partial sky coverage, which we neglect in this work.}. 
The inversion process is carried out with the polarised {\sc map2alm} functions in-built in {\sc Healpy}, and we further include smoothing to suppress numerical noise caused by empty pixels, accomplished by convolving the $\kappa_E$ map with a Gaussian beam with width $\sigma$=2.0 arcmin. We choose a pixel scale of 0.85arcmin given by $N_{\rm side}=4096$, i.e. slightly less than $\sigma$, to keep most of the small scale information. 
It is worth noting that this smoothing scale is relatively small, posing challenges for the theoretical modelling of some probes within this regime.
For this reason, many analyses opt for modelling directly from simulations \citep[see][]{HD21, DESY3_Zuercher, HSCY1_peaks_sims}, bypassing some of the theoretical challenges.

Aperture mass maps are constructed either from the  ellipticity maps, but computing instead 
\begin{eqnarray}
M_{\rm ap}(\boldsymbol{\theta}) = \int {\rm d}\boldsymbol{\theta}' \gamma_{\rm t} (\boldsymbol{\theta}',\boldsymbol{\theta}) Q(|\boldsymbol{\theta}'|, \theta_{\rm ap})\,,
\label{eq:Map_gamma_t}
\end{eqnarray}
or from the $\kappa$-maps, with 
\begin{eqnarray}
M_{\rm ap}(\boldsymbol{\theta}) = \int {\rm d}\boldsymbol{\theta}' \kappa (\boldsymbol{\theta}) U(|\boldsymbol{\theta}-\boldsymbol{\theta}'|, \theta_{\rm ap})\,,
\label{eq:Map_kappa}
\end{eqnarray}
where  $Q(\theta,\theta_{\rm ap})$ and $U(\theta,\theta_{\rm ap})$ are the aperture filter functions pair \citep{Schneider1998}. The aperture angle $\theta_{\rm ap}$, also called the `smoothing scale' is a free parameter that we vary in our different statistics.

Finally, while tomographic \gtwopcf data include auto-correlation and cross-redshift correlations, certain non-Gaussian statistics help us to access further information from analysing pairs, triplets, quadruplets, or quintuplets of redshift bins \citep{Martinet20}.
This work focuses only on auto-tomographic bins and bin pairs, yet our methods and results can be straightforwardly extended to include these higher-order redshift combinations.





Here we consider the following higher-order statistics:
\begin{enumerate}
\item {\it Integrated $\gamma$-3PCFs}: The integrated cosmic shear three-point correlation functions  $\zeta_{\pm}$ \citep[][$\gamma$-3PCFs hereafter]{Halder2021, Halder2023} are natural extensions of the $\gamma$-2PCFs;  they can be directly estimated from shear catalogue data by measuring the $\gamma$-2PCFs $\xi_{\pm}$ locally inside spatial patches on the sky and correlating these with the shear aperture mass $M_{\rm ap}$ signal \citep{Schneider1998} within the same patches. In equation form, this is given by: $\zeta_{\pm}(\vartheta) = \langle M_{\rm ap} \ \xi_{\pm}(\vartheta)\rangle$ where the average is taken over many patches on the sky. This admits a clear physical interpretation --- \textit{modulation of the small-scale $\gamma$-2PCFs by large-scale mean fluctuations of the shear field} --- that in turn probes an \textit{integrated} form of the full 3PCF (in the squeezed-limit of the lensing bispectrum in Fourier space). As $\zeta_{\pm}$ are sensitive to the squeezed lensing bispectrum, they can be modelled down to non-linear scales using the response function approach to perturbation theory \citep{Halder2022}, and hence forms a practical and physically interpretable catalogue-based higher-order cosmic shear statistic. In our lensing simulations, we 
use patches with 90 arcminutes radii, inside which we measure the local $\xi_{\pm}(\vartheta)$ in separations of 5-170 arcminutes, split in 15 log-spaced bins; the aperture mass $M_{\rm ap}$ is computed using the compensated $U$ filter function of \citet{Crittenden2002} with smoothing scale size the same as the patch radius, i.e. $\theta_{\rm ap}=90'$. These choices are physically motivated and well suited to differentiate between IA models, as seen later. Note that 
the statistics can be computed in auto- and cross-redshift bin: $\zeta_{\pm}^{ijk}(\vartheta) = \langle M_{\rm ap}^i \ \xi_{\pm}^{jk}(\vartheta)\rangle$.
%
%
    \item {\it Third-order aperture mass}:  
    This statistic, $M_{\rm ap}^3$, describes the third moment of the convergence field smoothed by a compensated filter function $U$ (see Eq. \ref{eq:Map_kappa}), which we choose to be the widely used exponential filter from \citet{Crittenden2002}. The construction of aperture mass maps requires repeated usage of smoothing, which is performed on the curved sky with help of the {\sc smoothing} and {\sc beam2bl} functions of the {\sc healpy} package \citep{Zonca2019}. We estimate the multiscale and tomographic third-order aperture mass statistics by performing a spatial average over the product of three aperture mass maps with smoothing radii $\theta_{\rm ap} = (\theta_i,\theta_j,\theta_k)$ for the tomographic redshift bin combination $(l,m,n)$. Due to the symmetry properties of $M_{\rm ap}^3$ we only consider the configurations for which $\theta_i \leq \theta_j \leq \theta_k$ and $l\leq m\leq n$. See Sect. 5.3.1 of \citet{Heydenreichetal2023}
    for a more detailed description of this method. 
    \item \textit{Peaks and Minima}: In presence of position dependent shape noise, local maxima and minima in the {\sc Healpix} convergence maps are often counted in bins of signal-to-noise ratios, $\nu\equiv\kappa/\sigma$, computed from the global noise levels in the survey.   In the noise-free case, or whenever the shape noise is constant as in this paper, these extrema can be counted in bins of $\kappa$ directly;  measurements presented in the following sections are for the noise-free case, however we discuss the impact of noise in Appendix \ref{app:figs}. Peak count statistics are amongst the most widely used non-Gaussian lensing statistics, as it is simple to measure yet highly sensitive to non-Gaussian structures.
    Some theoretical models have been developed to describe the largest peaks \citep[see][]{Shan18, HSCY1_Peaks_th}, as these are connected to largest over-densities that emerge from the peak-background split model. However, we report here the results on a much wider range of signal-to-noise ratios, which can only be accurately predicted from numerical simulations themselves.
    \item \textit{Lensing PDF}: The lensing one-point function, also referred to as the $\kappa$-PDF, is computed directly from the histogram of the pixel values of a lensing map, and can be modelled by large deviation theory, as shown in \citet{LensingPDF_Cosmo} and recently extended to tomography in  \citet{LensingPDF}. This analytical modelling, further improved by nulling techniques and including the effects of survey masking \citep{LensingPDF_Nulling, Barthelemy2024}, is accurate provided that the map is smoothed on scales of $\sim10$ arcmin to erase the highly non-linear structures, however we prefer here to use a Gaussian beam of 5 arcmin to access smaller scales. In a data analysis, this would require us to replace the analytical modelling with a simulation-based inference model, as in \citet{Giblin_PDF}  and \citet{Thiele_HSC_PDF}. 
    \item \textit{Weak lensing void profiles}: Under-dense regions of the density fields, and hence the convergence field, are of high physical interest as they offer complementary information to that extracted from overdensities \citep{Davies2022}. Similar to the minima introduced above \citep[iii, see][]{Davies2021}, the lensing voids, and their lensing profiles are sensitive to the distribution of matter away from clusters and filaments, and therefore probe the intrinsic alignment signal in a region of low tides. Our implementation follows that of \citet{Davies2018}, where lensing voids are identified from the spatial distribution of lensing peaks (iii) through a Delaunay Triangulation, which returns circumcircles that are the largest circles that are empty of tracers (the peaks here). The resulting circumcircles correspond to the lensing void catalogue, which can be summarised with summary statistics such as the void size ($R_v$) distribution, and the void lensing profile ($\kappa(r/R_v)$). The profiles are recorded in 20 linear bins ranging from  $0 < r/R_V < 2.0$, covering the interior, the ridge and the outside of the voids.

\end{enumerate}

These non-Gaussian statistics explore various physical scales and non-linear phenomena, leading to differences in their sensitivity to cosmology and systematic errors.
For instance, a statistical method effective in recovering cosmological information could be significantly influenced by secondary effects like IA or baryonic feedback, reducing the reliability of the extracted information.
Thus, incorporating IA at the field level, with enough flexibility to account for the current uncertainty in our knowledge of the IA physics, is essential for addressing these concerns (see Sec. \ref{sec:HOWLS}). 

\section{Intrinsic alignment models}
 \label{sec:IA_th}

Galaxy shapes are influenced by the gravitational forces produced from the vast structures in which they reside, leading to a coupling between their intrinsic orientations and the local cosmic tidal field.
This intrinsic alignment is distinct from the weak lensing signal and introduces a secondary correlation that contaminates the cosmic shear measurements.
The underlying physical principles that govern the IA remain unclear, and current models that attempt to describe these alignments have parameters that are not definitively determined by existing data, as discussed in various review articles \citep[see][]{Joachimi_IA_review_2015, Kirk_IA_review_2015, Troxel_IA_review_2015, Kiessling_IA_review_2015, Lamman_IA_guide}.
 
In this study, we explore two models of coupling, each implemented across three scenarios of galaxy bias, namely no bias, linear, and non-linear bias.
This approach yields six distinct models for intrinsic alignments, which are detailed in the subsequent subsections. 
Our models establish a connection between the intrinsic shapes of galaxies and the local density fluctuations as well as the projected tidal fields. 
This defines an intrinsic ellipticity tensor, $\gamma_{ij}^{\rm IA}$, from which the intrinsic ellipticities are extracted\footnote{Note that the intrinsic ellipticity tensor is sometimes written as $\gamma^I_{ij}$, as in \citet[][see their Eq. 8]{TATT}.}:
\begin{eqnarray}
\epsilon_{1}^{\rm IA} = \gamma_{11}^{\rm IA} - \gamma_{22}^{\rm IA} \, , 
\end{eqnarray}
\begin{eqnarray}
\epsilon_{2}^{\rm IA} =  \gamma_{12}^{\rm IA} \, .
\label{eq:eps_IA}
\end{eqnarray}

\subsection{Non-linear linear alignment (aka NLA) Model}

The NLA model of \citet{Hirata2004,NLA} is the most widely used intrinsic alignment model in the cosmic shear literature thus far.
In this model, IA are caused by a linear coupling between galaxy shapes and the non-linear large-scale tidal field at the galaxy positions, which are assumed to be uncorrelated with the underlying matter field (i.e. the galaxy bias is set to zero).
The intrinsic ellipticities $\epsilon_{1,2}^{\rm NLA}$ are related to tidal field $s_{ij}$ by:
\begin{eqnarray}
\begin{split}
\epsilon_1^{\rm NLA} &= C_1 (s_{11} - s_{22}) , \\    \epsilon_2^{\rm NLA} &= C_1 s_{12}\;,
\label{eq:tidal_th}
\end{split}
\end{eqnarray}
with
\begin{eqnarray}
    C_1 = - \frac{A_{\rm IA}\bar{C_1}\bar{\rho}(z)}{D(z)} \, ,
\end{eqnarray}
where $s_{ij} = \partial_{ij}\phi$ are the Cartesian components of the tidal tensor of the gravitational potential,  $D(z)$ is the linear growth factor, $\bar{\rho}(z)$ is the mean matter density at redshift $z$, and $\bar{C_1}=5\times 10^{-14} M_{\odot}^{-1} h^{-2} {\rm Mpc}^3$ is a constant calibrated in \citet{Brown2002}. 
The strength of tidal coupling is controlled by $A_{\rm IA}$, an effective parameter in the NLA model that is well measured by current cosmic shear studies, although the reported values vary widely due to the complex dependence on the survey-specific  galaxy sample \citep{KiDS1000_Asgari, DESY3_Amon,DESY3_Secco}.

While this model can incorporate the dependence on redshift and luminosity \citep[as in the latter two references, see also][]{Krause2016}, such adjustments are not applied in our analysis.
It is important to clarify that the ``non-linear" aspect of the model's name can be misleading; it actually pertains to the use of the non-linear matter power spectrum $P(k)$ in its computations. The relationship between the intrinsic shapes of galaxies and the tidal field remains linear. Equation (\ref{eq:tidal_th}) is employed to determine the intrinsic ellipticities of galaxies using the three tidal field components $s_{ij}$ constructed in Section \ref{subsec:IA_infusion}.


The observed ellipticities are, to linear order, the sum of the intrinsic shape ($I$) and the cosmic shear $G$. Then, in the context of two-point functions, these intrinsic shapes contribute to an intrinsic-intrinsic ($II$) term and an intrinsic-shear coupling ($GI$) term \citep{Hirata2004}, both secondary signals to the true cosmic shear ($GG$) term, with the $GI$ typically dominating the IA sector in cross-tomographic measurements. 
The $II$ and $GI$ terms can be both computed from the matter power spectrum as: 
 \begin{eqnarray}
P_{II}(k,z) =  \left(\frac{A_{\rm IA}\bar{C_1}\bar{\rho}(z)}{D(z)}\right)^2a^4(z) P_{\delta}(k,z)
\label{eq:Pk_II_th}
\end{eqnarray}
and
\begin{eqnarray}
P_{GI}(k,z) = - \frac{A_{\rm IA}\bar{C_1}\bar{\rho}(z)}{D(z)}a^2(z) P_{\delta}(k,z) \, ,
\label{eq:Pk_GI_th}
\end{eqnarray}
which can then be passed to the Limber and Bessel integration (Eqs. \ref{eq:C_ell} and \ref{eq:xipm_th}) to compute the secondary signals $\xi_{\pm}^{II}(\vartheta)$ and $\xi_{\pm}^{GI}(\vartheta)$, with $n^i(z)$ replacing the line-of-sight projection kernel $q^i(z)$ for every $I$ instance.

\subsection{Extended-NLA (aka $\delta$-NLA) Model}
\label{subsec:IA_th_extNLA}




The NLA model, as discussed in the previous section, is a common tool for analysing cosmic shear data but is an effective model with significant limitations (average over full populations, no galaxy bias, etc.), and hence might not accurately capture the intricacies of the underlying physical intrinsic alignment signal. Recognising the potential importance of more complex interactions, extensions to the NLA model that incorporate higher-order corrections have been proposed by \citet{Blazek2015}, \citet{Schmidt2015} and \citet{TATT}. 
A key enhancement is the addition of an over-density weighting term that accounts for the fact that galaxies, from which we sample the tidal interactions, are not randomly distributed on the sky but rather follow the underlying matter density distribution. The theoretical framework for including this term employs one-loop perturbation theory, as outlined by \citep{TATT}, under the assumption that galaxies linearly trace the over-density field `$\delta$,' up to a linear bias factor $b$.\footnote{Non-linear galaxy biasing with IA has also been studied, e.g.\ in \citet{Blazek2015} and \citet{IA_EFT}.}
In essence, this approach modifies the NLA model predictions by applying a $\delta$-weight to the tidal field, which under the linear bias assumption, corresponds to evaluating the tidal field at the locations of galaxies, thereby refining the model's accuracy in representing the physical reality.
The extended-NLA, referred to as the $\delta$-NLA model in this work, departs from the NLA model at small scales, as noted by \citet{TATT}, where the stronger alignments are more efficient at contaminating the cosmic shear signal. 
It also appears to align more closely with the outcomes of certain hydro-dynamical simulations, as indicated by \citet{Hilbert_IA2017}.

Implementing this model in numerical simulations could theoretically be straightforwardly achieved by enhancing the calculated NLA ellipticities with the aforementioned over-density weight:
\begin{eqnarray}
\epsilon_{1/2}^{\delta-\rm NLA} = \epsilon_{1/2}^{\rm NLA}\times(1 + \delta \: b_{\rm TA}) \, , 
\label{eq:tidal_th_deltaNLA}
\end{eqnarray}
with the term $b_{\rm TA}$ representing the bias relationship between galaxies and the local matter field.\footnote{Although the bias itself is straightforward to measure, $b_{\rm TA}$ is often left as a free parameter to capture related alignment mechanisms \citep{TATT}.}
In practical applications,  when utilizing the 2D projected density field, this method tends to produce unreliable results due to the large number of galaxies located in areas with negative or low $\delta$ values when placed randomly.
A more effective approach, which we adopt in this work, is to create mock catalogues by directly applying linear biasing to determine galaxy positions, which eliminates the need for subsequent weighting. For minor adjustments on the bias of the sample, Eq. (\ref{eq:tidal_th_deltaNLA}) can be applied to alter the effective value of $b_{\rm TA}$ without creating a new mock catalogue. This allows for the recalibration of the intrinsic alignment contribution from an original bias value, $b_{\rm TA}^{\rm orig}$, to a new desired bias, $b_{\rm TA}^{\rm new}$, as:
\begin{eqnarray}
{\boldsymbol \epsilon}^{\rm new} = \frac{1 + \delta \: b_{\rm TA}^{\rm new}}{1 + \delta \: b_{\rm TA}^{\rm orig}}{\boldsymbol \epsilon}^{\rm orig} \, .
\label{eq:bta_rescale}
\end{eqnarray}
Equation (\ref{eq:bta_rescale}) has limitations in its accuracy, as it is again based on projected $\delta$ which can be negative and therefore needs recalibration, but we will show later that it can nevertheless be applied to modify existing $\delta$-NLA mock data and yields good approximate solutions. 


Finally, we note that  the accuracy in modelling our infusion method is limited by the order in perturbation to which theoretical calculations are performed. The two-point function correlating ellipticities (defined in Eq. \ref{eq:tidal_th_deltaNLA}) between redshift bins $i,j$ is given by:
 \begin{eqnarray}
\langle \epsilon_{i}^{\delta-\rm NLA} \epsilon_{j}^{\delta-\rm NLA}\rangle &=& \langle \epsilon_{i}^{\rm NLA}\times(1 + b_{\rm TA}\delta_i)\  \epsilon_{j}^{\rm NLA}\times(1 + b_{\rm TA}\delta_j)\rangle\\
&=& \langle \epsilon_{i}^{\rm NLA}  \epsilon_{j}^{\rm NLA}\rangle \nonumber \\
&+& b_{\rm TA} \times \left(\langle \epsilon_{i}^{\rm NLA}  \epsilon_{j}^{\rm NLA}\delta_j  \rangle + \langle \epsilon_{i}^{\rm NLA}  \epsilon_{j}^{\rm NLA}\delta_i  \rangle \right)\nonumber\\
&+& b_{\rm TA}^2 \langle \epsilon_{i}^{\rm NLA}  \epsilon_{j}^{\rm NLA}\delta_i  \delta_j \rangle \, ,
\label{eq:2pcf_deltaNLA}
\end{eqnarray}
where {\rm both $\epsilon^{\rm NLA}$ and $\delta$ can be expressed as perturbative expansions of the linear density field}\footnote{These four correlation terms can be written to arbitrary orders.
Calculations at one-loop order include terms up to $\mathcal{O}(\delta^4_{\rm lin})$, however higher-order corrections are important at small scales. Although it is not strictly consistent, we follow the typical practice of evaluating the first term $\langle \epsilon_{i}^{\rm NLA}  \epsilon_{j}^{\rm NLA}\rangle$ using the fully non-linear power spectrum, while other terms are evaluated at one-loop.
}. 
For this reason, we expect that the $\delta$-NLA model of \citet{TATT} will deviate from the corresponding infusion result in our simulations due to non-linearities on smaller scales. In particular, due to how our infusion is done with linear biasing and smoothing of the density fields, the non-linearities in the tidal field are expected to be less-well described by the one-loop calculations, explaining the large deviations seen for the $II$ term; the $GI$ term is more robust as it includes only terms up to third order in the field (e.g $\langle \epsilon_{i}^{\rm NLA}\delta_i \gamma_j \rangle$, ...), suppressing the importance of these differences. 
Note that this issue is not present in the NLA-only method, since the fully non-linear power spectrum is used to describe the tidal field correlations.


\subsection{Tidal Torquing (aka TT) Model}
\label{subsec:IA_th_TT}

\citet{TATT} demonstrate that one-loop perturbative calculations introduce an additional term, accounting for how galaxies develop intrinsic alignments through the interaction between their angular momentum and the tidal field.
This interaction can alternatively be described as a quadratic coupling between the tidal field and the shapes of the galaxies. 
Within this tidal torquing theory (TT), the intrinsic ellipticities of galaxies are determined as \citep{TATT}:

\begin{eqnarray}
\gamma_{ij}^{\rm IA, TT} = C_2 \left[ \sum_{k=1,2,3} s_{ik} s_{kj} -\frac{1}{3} \delta_{ij} s^2 \right] \, ,
\label{eq:tidal_th_TT}
\end{eqnarray}
where 
\begin{eqnarray}
C_2 = \frac{5 A_2 \bar{C_1} \Omega_{\rm m} \rho_{\rm crit}}{D^2(z)} \,  = \left[ \frac{-5 A_2}{A_{\rm IA} D(z)} \right] C_1 \, ,
\end{eqnarray}
where $s^2 = \sum s_{ij} s^{ij}$, and $A_2$ is a free parameter, set to 1.0 in the fiducial TT model.\footnote{The additional factor of 5 is added for approximate consistency in normalisation, as discussed in \cite{TATT}.}
Assuming that alignments along the line of sight ({\it i.e.} components involving $k$=3) are largely suppressed in cosmic shear measurements due to the broad lensing kernels, we demonstrate in Appendix \ref{app:2d_TT} that the terms within the square brackets of Eq. (\ref{eq:tidal_th_TT}) reduce to:

\begin{eqnarray}
\begin{split}
\epsilon_1^{\rm TT} &= C_2  \left[ s_{11}^2 - s_{22}^2\right] \, ,\\
 \epsilon_2^{\rm TT} &=  C_2 s_{12}\left[s_{11}+s_{22}  \right]\, .
\end{split}
\label{eq:eps_TT}
\end{eqnarray}
In this model, galaxies are also assumed to be randomly distributed on the sky. 
Incorporating the $\delta$-weighting necessitates third-order perturbation theory. Note that we treat in this paper the TT model as an independent contribution to the overall IA model, however the full TATT model is a linear combination, and includes $\epsilon_i^{\rm TATT} = \epsilon_i^{\rm \delta-NLA} + \epsilon_i^{\rm TT}$, which we include in our TATT theoretical calculations in Sec. \ref{sec:validation}.  For additional terms which can contribute to IA at one-loop order, see 
\cite{Schmitz18,IA_EFT}.



\subsection{Extended-TT (aka $\delta$-TT) Model}
\label{subsec:IA_th_extTT}


Calculating the next level of contribution to the TT model requires engaging in two-loop perturbation theory, a process that remains undone due to its considerable complexity.
However, in simulations, applying the $\delta$-weighting term to the TT model is relatively straightforward.
This step merely requires using Eq. (\ref{eq:eps_TT}) with galaxies that trace the matter field with a non-zero bias factor. 
Similar to the \dNLA model, the ellipticities within the \dTT framework can be connected back to the original TT model as follows:
\begin{eqnarray}
\epsilon_{1/2}^{\delta-\rm TT} = \epsilon_{1/2}^{\rm TT}\times(1 + \delta \: b_{\rm TA}) \, .
\label{eq:tidal_th_deltaTT}
\end{eqnarray}
In this study, we opt to utilise galaxy positions that are subject to linear biasing, mirroring our earlier approach.
Given the lack of established theoretical frameworks for the \dTT model, it currently exists solely in numerical simulations.

Note that this extended-TT model is inconsistent with the order of expansion in the galaxy bias terms, the same way the $\delta$-NLA model is already inconsistent in the PT framework as seen before. 

\subsection{Halo Occupation Distribution-TATT Model}
\label{subsec:HOD-TATT}

The previous four models associate either the linear or quadratic couplings to the cosmic tidal forces  (Eqs. \ref{eq:tidal_th} and \ref{eq:tidal_th_TT}, respectively) with galaxies positioned either at random or linearly tracing the total matter distribution; these models can further be combined to make hybrid models such as the TATT.  
While these are interesting and useful approximations, the connection between galaxies and dark matter is far more complex, and a more accurate picture consists of galaxies populating dark matter haloes, typically with a relaxed, older galaxy close to the centre, and a number of other satellite galaxies orbiting the former.
This halo occupation distribution (HOD hereafter) formalism has been used to describe many galaxy samples \citep[\eg][]{2013MNRAS.430..767C,2015MNRAS.454.3938S, 2016MNRAS.459.3251V, ABACUS_HOD, DESI_HOD} and is therefore routinely used to in-paint galaxies in gravity-only simulations \citep{MICE_IA,Balrog, SLICS,  ABACUS_HOD}.
The remaining two models in this paper exploit such HOD galaxy samples, extracted from the same underlying $N$-body simulations. In this case the underlying lensing signal is the same to first order, but the galaxy bias is non-linear (labelled $b_{\rm nl}$), with levels on non-linearity that vary with HOD parameters (see Sec. \ref{subsec:HOD}).
We couple these galaxies with both Eqs. (\ref{eq:tidal_th}) and (\ref{eq:tidal_th_TT}), resulting in two final models which we name HOD-NLA and  HOD-TT, and the combination of both being referred to as the HOD-TATT model.



\section{Simulations}
\label{sec:sims}

We construct simulations that are designed to approximate upcoming cosmic shear data, with some simplifying assumptions. Our final products are galaxy catalogues with shear and IA values, as well as convergence maps constructed from these catalogues, covering up to a full octant on the sky with a redshift distribution shown in Fig. \ref{fig:Nz}. We neglect the complex masking and observational effects such as PSF and depth variations, and assume a galaxy density of 3.0 gal arcmin$^{-2}$. As mentioned before, this is lower than expected from Rubin but is enough to validate our IA infusion methods, especially since we report our main results from noise-free galaxy shapes; selected results with shape noise are presented in Appendix \ref{app:figs}. At full density, the infusion is still tractable since it scales linearly with $N_{\rm gal}$; statistical computation such as {\sc TreeCorr} calculations, however,  will take longer as the scaling is typically less ideal. 

\subsection{Creation of cosmic shear galaxy catalogues} 
\label{subsec:WL_cats}

The weak lensing simulations developed for this work are based on the {\it Outer Rim} $N$-body simulation \citep{OuterRim}, which evolved 10,240$^3$ particles in a $(4.225 \mathrm{Gpc})^3$ cosmological volume, assuming a flat \lcdm cosmology with $\Omega_{\rm m}=0.2648$, $\Omega_{\rm b}=0.0448$, $h=0.71$, $\sigma_8 = 0.801$, $n_{\rm s}= 0.963$, $w_0=-1.00$.
A total of $101$ particle snapshots were originally saved out to redshift $z=10$, of which we use only those with $z\le3.0$. Particles from each snapshot are assigned to curved mass shells approximately 114 Mpc thick, producing a sequence of 57 {\sc Healpix} maps $\delta_i(\theta, \phi)$ with {\sc nside} = 8192 and $i=1...57$, filling up a light-cone over an octant up to $z = 3$ with an angular resolution of 0.4 arcmin.  Given the relatively low number density used in this work, we downgraded these maps to {\sc nside} = 4096 for most of our calculations, but saved the high-resolution maps for future work on denser samples. Some of these maps have been used in the cosmoDC2 simulation \citep{cosmoDC2}, however the survey area was limited to 440 deg$^2$, compared to the 5157 deg$^2$ available in SkySim5000.

Ray-tracing is performed using the Born approximation, summing over the mass shells using a $\chi$-integral similar to Eq. (\ref{eq:C_ell}): 
\begin{eqnarray}
\kappa_i(\theta, \phi) = \int_0^{\chi_{\rm H}}  q^i(\chi) \, \delta(\theta,\phi, \chi)\,\ {\rm d}\chi \, .
\end{eqnarray}
Source planes are placed at the high-redshift edge of every mass plane, each resulting in a convergence map $\kappa_i(\theta,\phi)$ that is subsequently transformed into shear maps $\gamma_{1/2, i}(\theta,\phi)$ using Eq. (\ref{eq:KS}).

\subsection{Galaxy bias models}

We position galaxies in the light cone following three distinct algorithms, each impacting the strength of the IA signal:
\begin{enumerate}
\item {\it Random:} galaxies are distributed randomly on the octant,  i.e. not tracing the underlying matter field, thereby reproducing one of the fundamental assumptions in the NLA and TT models.
\item {\it Linear bias:} galaxy positions are sampled from the mass sheets smoothed\footnote{We also tried sampling the field with $0.1$ and $0.5$ $h^{-1}\mathrm{Mpc}$ but this resulted in noisier results.} with a 1.0 $h^{-1}\mathrm{Mpc}$ (comoving) beam, assuming a linear bias of $b_{\rm TA}$, thereby implementing one of the key assumptions of the \dNLA and \dTT models. 
We assume $b_{\rm TA}=1.0$ as our fiducial case but also consider $b_{\rm TA}=2.0$ to test the model flexibility. In practice, we Poisson-sample the number of galaxies in a given pixel with mean $\bar{n}(1+b_{TA}\delta)$, where $\bar{n}$ is extracted from the overlap between the tomographic source $N^i(z)$ and the edges of the mass shell under consideration. 
\item{\it Non-linear bias:} SkySim5000 galaxy positions are obtained by populating dark matter haloes with the HOD prescription described in \citet{cosmoDC2} and used recently in \citet{TXPipe} for a $3\times2$pt analysis of cosmoDC2\footnote{We use version: {\tt skysim5000\_v1.2}.}. Specifically, we select objects over a patch of 732.20 deg$^2$ given by 0 $<$RA$<$ 20 deg and -36.61 $<$DEC$<$0 deg, we apply a magnitude cut of mag$_r<24.8$, and further downsample randomly to retain 20\% of the galaxies, approaching the global galaxy redshift distribution and number density used with the other simulation catalogues (Eq. \ref{eq:nz}). This selected sample is further split into five tomographic bins by calculating the probability a given galaxy in the HOD sample lies in each bin, based on its true redshift. This is simply done by interpolating each of the tomographic $N(z)$ shown in Fig. 1 at the galaxies' true redshift, then using these as weights to assign a tomographic bin in a random draw. After a galaxy is assigned, it is removed from the sample to avoid double counting. This method allows us to capture some of the effect of photo-$z$ uncertainty, as it gives all galaxies a small but non-zero probability of being included in tomographic bins outside of their true redshifts\footnote{The full photometric uncertainty from the Rubin data will be far more complex, and will capture possible deviations in the mean and in the width of the $N(z)$, as well as outliers.}. This results in the $N(z)$ presented in Fig. \ref{fig:Nz}, with dashed lines closely matching the target $n(z)$. Compared to the analytical SRD-Y1, the mean redshifts of the five tomographic bins are shifted by [-0.027, -0.018,  0.003, -0.010,  0.038], respectively, which we take into account when making predictions for these HOD lensing mocks. This HOD scheme assumes a non-perturbative galaxy bias and captures stochasticity as well, encoded in the distribution of the halos and the galaxy population algorithm. For simplicity we refer to these mocks as having a non-linear bias.
\end{enumerate}
Note that both the {\it linear bias} and the {\it non-linear bias} mocks include source clustering, which can affect the cosmological signal even in absence of intrinsic alignment. We come back to this in Sec. \ref{sec:HOWLS}, where we turn on and off these different effects. All the catalogues described here are turned into convergence maps using the Kaiser-Squires method described in Sec. \ref{subsec:beyond-2pt} 

\subsection{Extraction of projected tidal fields}
\label{subsec:IA_sims}

\begin{figure}
\includegraphics[width=\columnwidth]{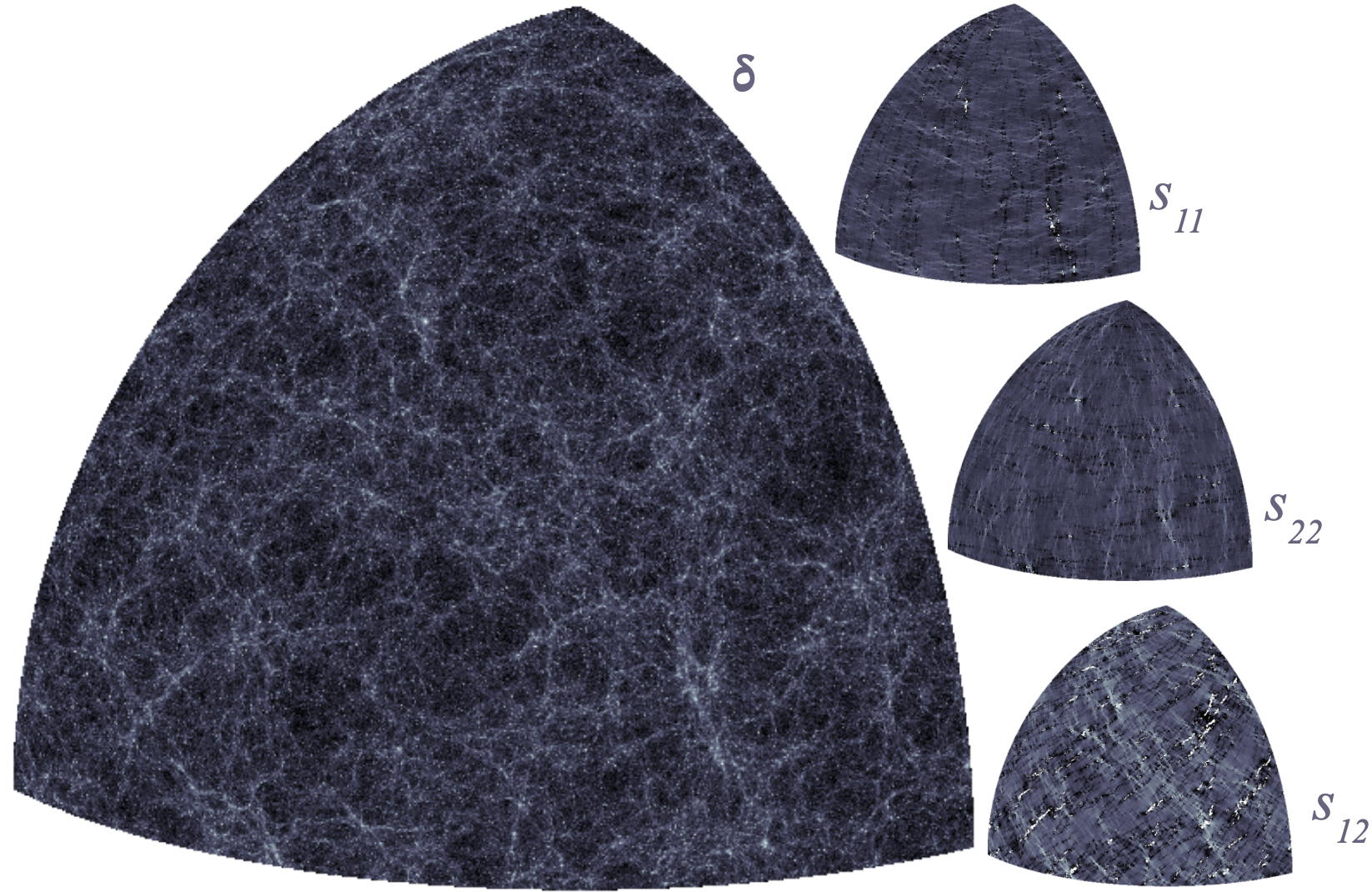}
\caption{Density field (left) and the associated projected tidal field tensors $s_{11}$, $s_{22}$ and $s_{12}$ (right), for $z=0.18$. These cosmic tides are computed from Eq. (\ref{eq:sij_2D_sph}), and used in Eq. (\ref{eq:tidal_th})  and Eq. (\ref{eq:eps_TT}) to infuse IA with a linear or quadratic coupling, respectively.}
\label{fig:maps}
\end{figure}

Our infusion method relies on couplings between intrinsic galaxy shapes and the local tidal field, hence the first step in our method consists of extracting the tidal field maps $s_{ij}(\theta,\phi)$ from the mass maps $\delta(\theta,\phi)$ that source them\footnote{Note that we have removed the redshift indexing of the maps here to simplify notation, {\it i.e.} $\delta_i(\theta,\phi) \rightarrow \delta(\theta,\phi)$.}.
In three dimensions, the trace-free tidal tensor $s_{ij}(\boldsymbol x)$ can be obtained from the matter over-density field $\delta(\boldsymbol x)$ as \citep{Catelan_IA_Tidal}:
\begin{eqnarray}
 \widetilde{s}_{ij} (\boldsymbol k)  = \left[\frac{k_i k_j}{k^2} - \frac{\delta_{ij}}{3}\right]  \widetilde {\delta}(\boldsymbol k) \mathcal{G}(\sigma_{\rm G}) \, ,
 \label{eq:sij}
\end{eqnarray}
where $\mathcal{G}(\sigma_{\rm G})$ is a three-dimensional Gaussian function described by a single (free) parameter $\sigma_{\rm G}$ that  controls the physical scales which are allowed to affect the IA term in our model. 
Tilde symbols denote Fourier-transformed quantities, the indices $(i,j)$ label the components of the Cartesian wave-vector $\boldsymbol{k}^T = (k_1,k_2,k_3)$, and $k^2 = k_1^2 + k_2^2 + k_3^2$. 
As shown in \citet{Tidalator} in the flat-sky approximation, projected tidal fields  computed from projected mass sheets provide an excellent agreement with the theoretical NLA model, which in contrast computes the full tidal fields from the three-dimensional matter density and project along the radial dimension at the end.
We promote here this transformation to curved-sky maps, exploiting the polarisation {\sc alm2map\_spin} operations built in {\sc Healpy}:  we define $\delta({\boldsymbol \theta})$ as our $E$-mode signal, assign zero $B$-mode, then compute the $U({\boldsymbol \theta})$ and $Q({\boldsymbol \theta})$ Stokes parameter maps.
Then, noting that $Q({\boldsymbol \theta})=s_{11}({\boldsymbol \theta})-s_{22}({\boldsymbol \theta})$, $U({\boldsymbol \theta})=s_{12}({\boldsymbol \theta})$, and $\delta({\boldsymbol \theta})=s_{11}({\boldsymbol \theta})+s_{22}({\boldsymbol \theta})$, we compute the curved-sky projected tidal field tensors ${s}_{ij}({\boldsymbol \theta})$ from each mass shell as:
\begin{eqnarray}\label{eq:sij_2D_sph}
\begin{split}
    s_{11}({\boldsymbol \theta})  &=  \frac{1}{\Delta \chi_{\rm shell}}\left[\frac{ \delta + Q}{2}  - \frac{\delta}{3}\right] \, , \\ s_{22}({\boldsymbol \theta})  &=   \frac{1}{\Delta \chi_{\rm shell}}\left[\frac{\delta - Q}{2}  - \frac{\delta}{3}\right] \, ,\\ s_{12}({\boldsymbol \theta}) &=  \frac{U}{\Delta \chi_{\rm shell}}  \, ,
\end{split}
\end{eqnarray}
where the $U({\boldsymbol \theta})$ and $Q({\boldsymbol \theta})$ maps are smoothed by the Gaussian beam with width $\sigma_{\rm G}$, and the normalisation by $\Delta \chi_{\rm shell}$ is required to account for the comoving thickness of the shells. We suppress large artificial tidal fields at the boundary of our simulated octant by replicating 8$\times$ the $\delta$ maps and carrying out these harmonic calculations on full sky densities; we re-apply the octant mask on the tidal field maps after the last operation.
Note that the value of $\sigma_{\rm G}$ is a free parameter both in the infusion technique described in this paper and in the NLA and TATT models.
We therefore explore two cases, $\sigma_{\rm G}=0.1$ and $0.5$ $h^{-1}$Mpc, however this may be further optimised in the future. If we wanted to infuse instead the Linear Alignment model \citep{Catelan_IA_Tidal}, we could substitute the simulated $\delta$ maps by linearize versions, obtained for example with linear perturbation theory.
Finally, the full-sky density field is downgraded from $N_{\rm side}=8192$ to $N_{\rm side}=4096$ since the smoothing removes information on the smallest angular scales.


Projected tidal field maps $s_{11}({\boldsymbol \theta})$, $s_{22}({\boldsymbol \theta})$, and $s_{12}({\boldsymbol \theta})$ are constructed using this procedure for each mass sheet; Fig. \ref{fig:maps} shows the three tidal fields and the underlying density maps for the $z=0.18$ shell.
We can clearly see the connection between all maps around over-dense regions.


\subsection{Infusion of intrinsic alignments}
\label{subsec:IA_infusion}

Having now produced shear catalogues and tidal field maps, we can use Eq. (\ref{eq:tidal_th}) to linearly couple the alignment of galaxies with the local tidal field, or Eq. (\ref{eq:tidal_th_TT}) to use a quadratic coupling instead. These allow us to compute the intrinsic ellipticities ${\boldsymbol \epsilon}^{\rm int}$ for the six IA models described in Sec. \ref{sec:IA_th}, which we combine with the cosmic shear signal ${\boldsymbol g}$ to compute observed ellipticities: 
\begin{eqnarray}
{\boldsymbol \epsilon}^{\rm obs} = \frac{{\boldsymbol \epsilon}^{\rm int} + {\boldsymbol g}}{1 + {\boldsymbol \epsilon}^{\rm int}{\boldsymbol g^*}} 
\label{eq:eps_obs}
\end{eqnarray}
with
\begin{eqnarray}
{\boldsymbol \epsilon}^{\rm int}  = \frac{{\boldsymbol \epsilon}^{\rm IA} + {\boldsymbol \epsilon}^{\rm ran}}{1 + {\boldsymbol \epsilon}^{\rm IA}{\boldsymbol \epsilon^{\rm ran, *}}}.
\label{eq:eps_int}
\end{eqnarray}
In the above expressions, the denominators ensure that the combined ellipticities never exceed unity.
The complex spin-2 reduced shear
\begin{eqnarray}
{\boldsymbol g} \equiv \frac{\gamma_1 + {\rm i} \gamma_2}{1 + \kappa}
\end{eqnarray}
is computed from the shear  $(\gamma_{1/2}$) and convergence ($\kappa$) maps, interpolated at the galaxy positions and redshifts. 
The random orientation term ${\boldsymbol \epsilon}^{\rm ran}$ is drawn from two Gaussians (one per component\footnote{We further constrain the random ellipticity to satisfy $|{\boldsymbol \epsilon}^{\rm ran}| \le 1.0$.}) with their standard deviations matching the LSST-Y1 forecast, $\sigma_{\epsilon} = 0.27$, although in most calculations we work with noise-free shapes to better resolve the IA signal.  
Finally, once all galaxies have been placed in the light-cone, we interpolate the shear and IA quantities at their exact location.
Note that our current IA models make no differentiation between galaxy colors or type, and instead treats the full sample as a single population that has a single, effective, alignment signal \citep[see][for an example with a red/blue split]{DESY1_IA_Samuroff}.

\section{Validation with shear correlations $\xi_{\pm}$}
\label{sec:validation}

This section presents a comparison between theoretical predictions and measurements in the simulations for each of our six models. We include here the statistical error obtained from a covariance matrix computed analytically as described in \citet{KiDS1000_Joachimi}, which includes contributions from the Gaussian, non-Gaussian and super-sample covariance terms, assuming survey properties that match our simulation in terms of area, shape noise, galaxy density, tomographic $n(z)$ and cosmology. The diagonal elements are used to assign the error bars in our figures, and the full matrix is used in the MCMC analyses presented in Sec. \ref{sec:inference}.

\begin{figure*}
\includegraphics[width=\columnwidth]{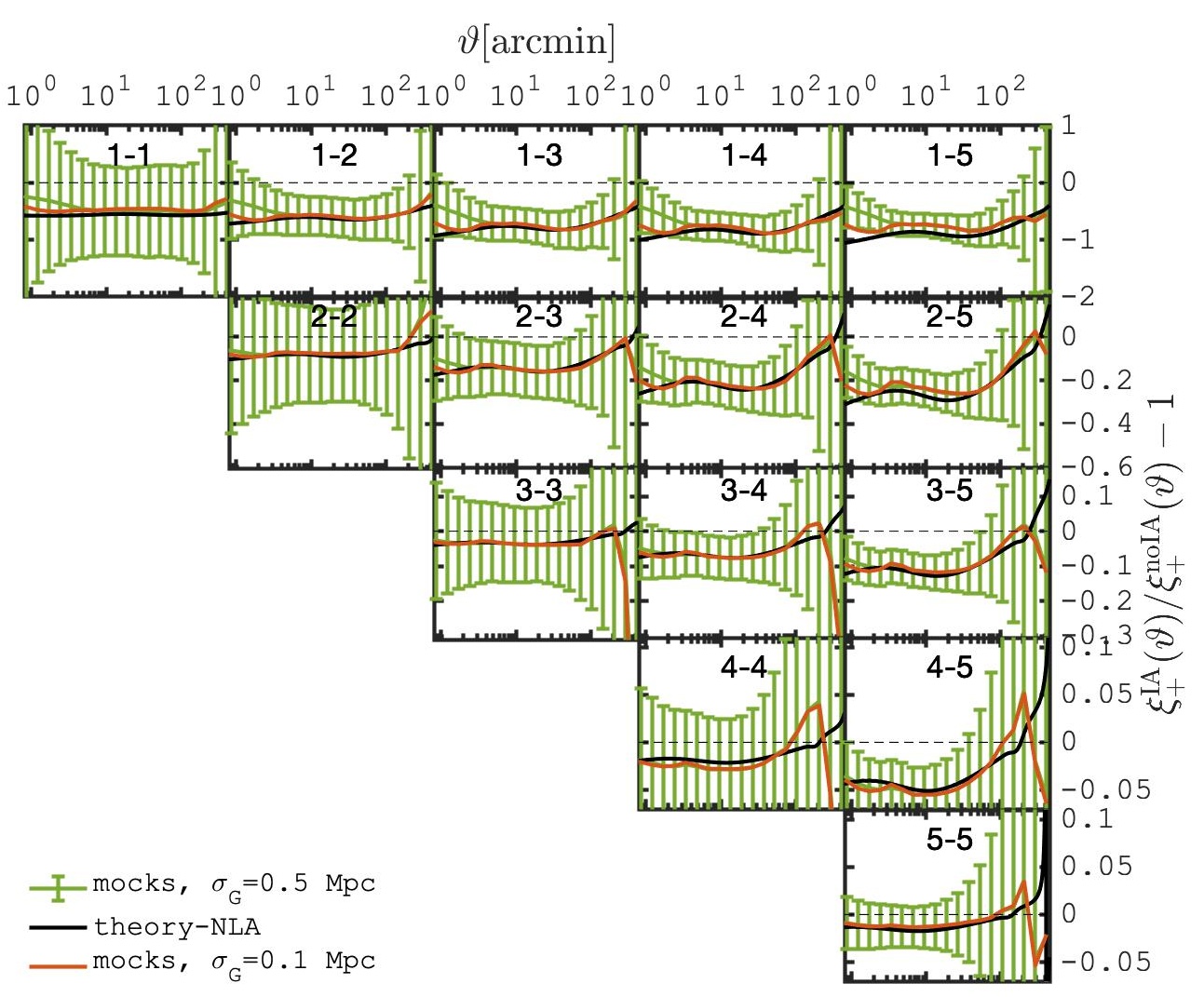}
\includegraphics[width=\columnwidth]{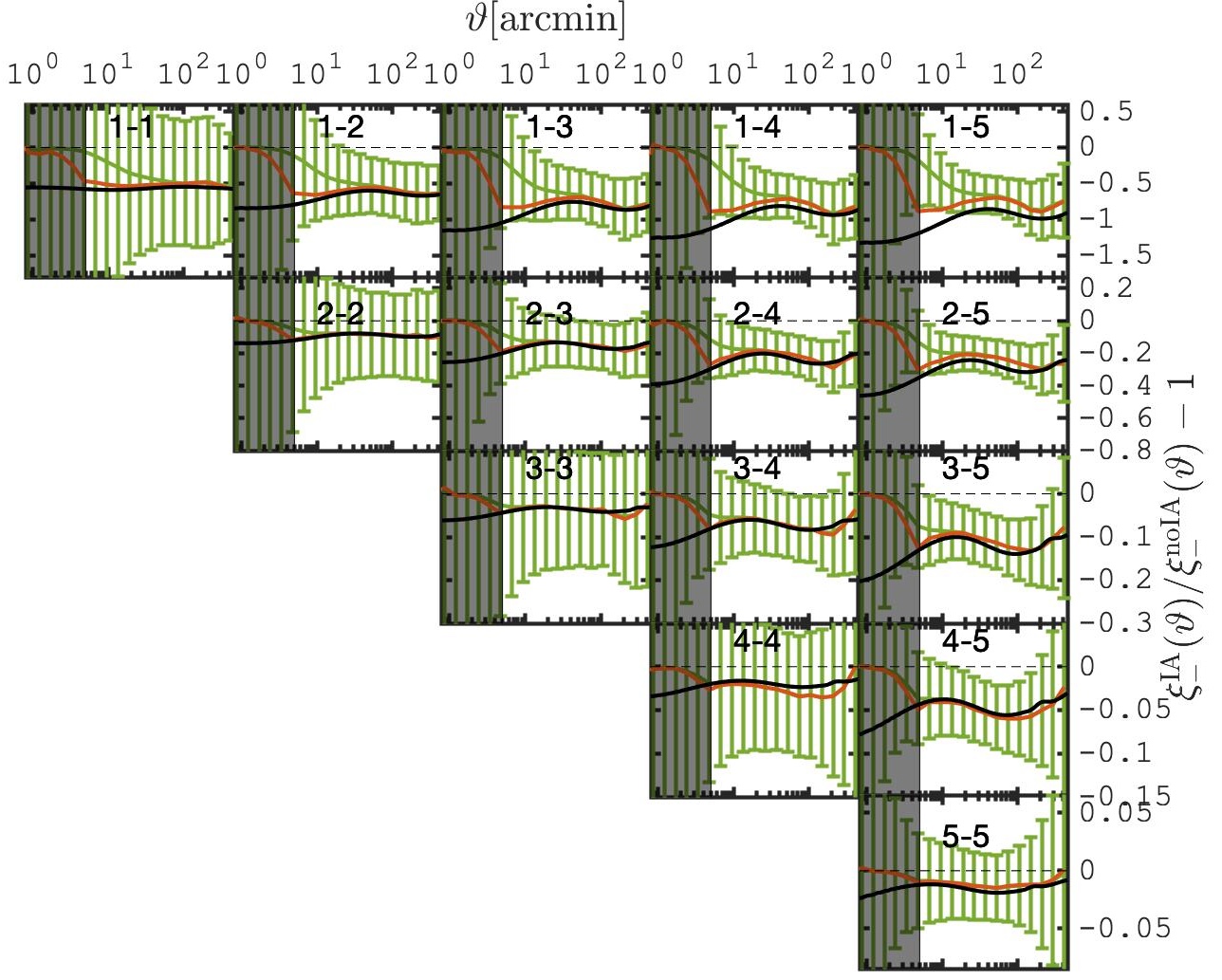}
\caption{Residual between the shear correlation functions with and without IA for $\xi_+$ (left) and $\xi_-$ (right), assuming the NLA model with $A_{\rm IA}=1.0$ both in the simulation and theory. Measurements shown in orange and green correspond to smoothing scales of 0.1 and 0.5 $h^{-1}$Mpc in the tidal field. There is no shape noise in the simulations,  but it is included in the analytical covariance matrix, from which the error bars are obtained. The gray bands represent scale cuts applied to the data vectors in our likelihood analyses. }
\label{fig:xi_NLA}
\end{figure*}

\subsection{Data vector}
\label{subsec:data_vector}
\subsubsection*{NLA model}

We start by presenting a comparison between the relative impact of IA in the NLA model as measured in the simulations (mocks) and as modelled by {\sc cosmoSIS} (theory). Specifically, we show in Fig. \ref{fig:xi_NLA} the ratio between the $\gamma$-2PCF with and without IA, for all combinations of tomographic bins as indicated in the panels, and for two smoothing scales. The agreement between the mocks and theory is remarkable over all scales except for the smallest angular separations in $\xi_-$, where the deviations are weaker in the simulations due to limits in resolution.   Inspired by the analysis choices made for the  KiDS-1000 analysis of \citet{KiDS1000_Asgari}, the gray bands indicate angular scales that are the most difficult to model and should be avoided, which, as shown by this figure, are those with $\vartheta<5$ arcmin. The tidal fields produced with a smaller smoothing scales (0.1 $h^{-1}$Mpc) show a better agreement with the model at the smallest angular scales compared to the fiducial smoothing case (0.5 $h^{-1}$Mpc).

\subsubsection*{Extended-NLA model}
\begin{figure*}
\includegraphics[width=\columnwidth]{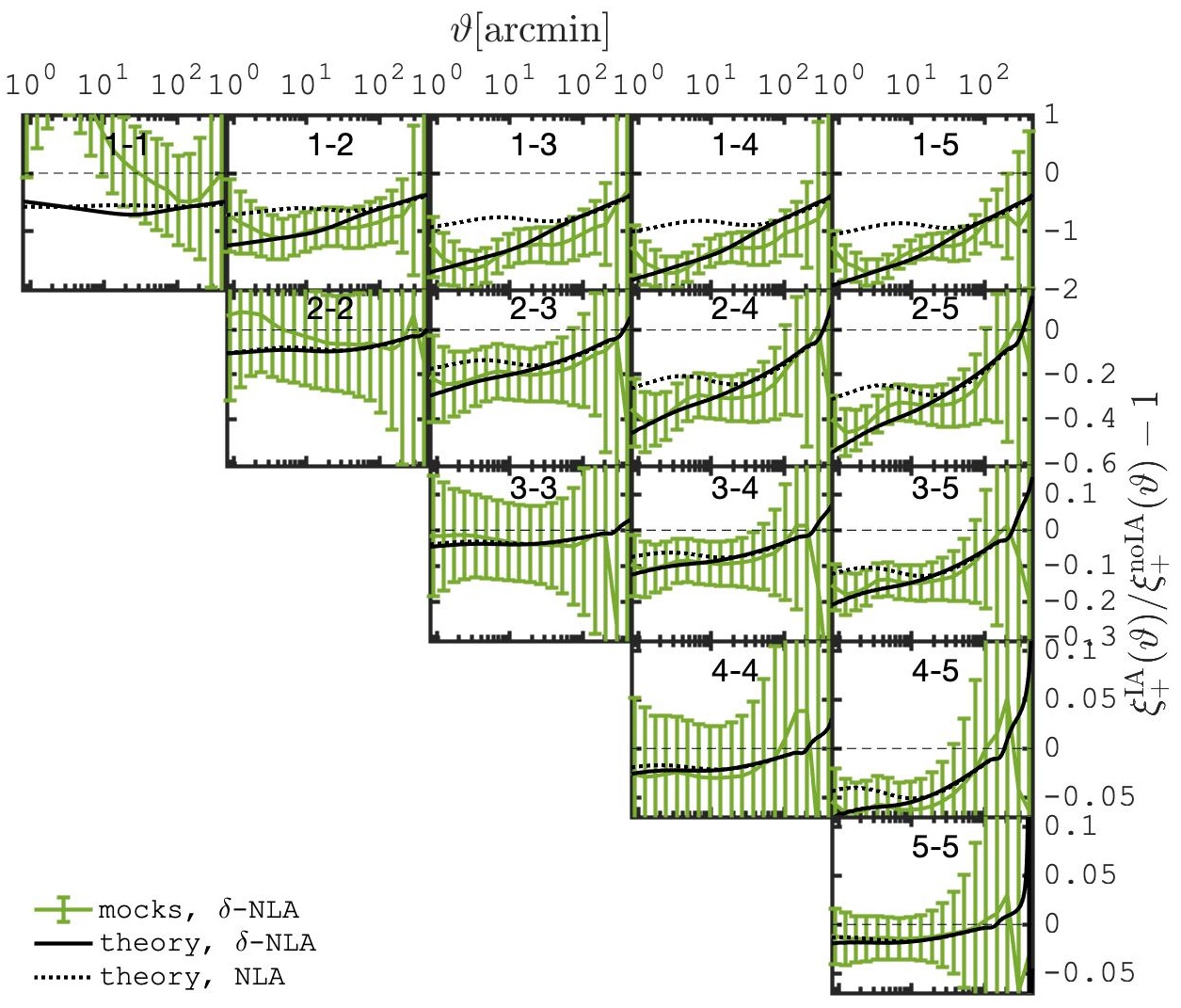}
\includegraphics[width=\columnwidth]{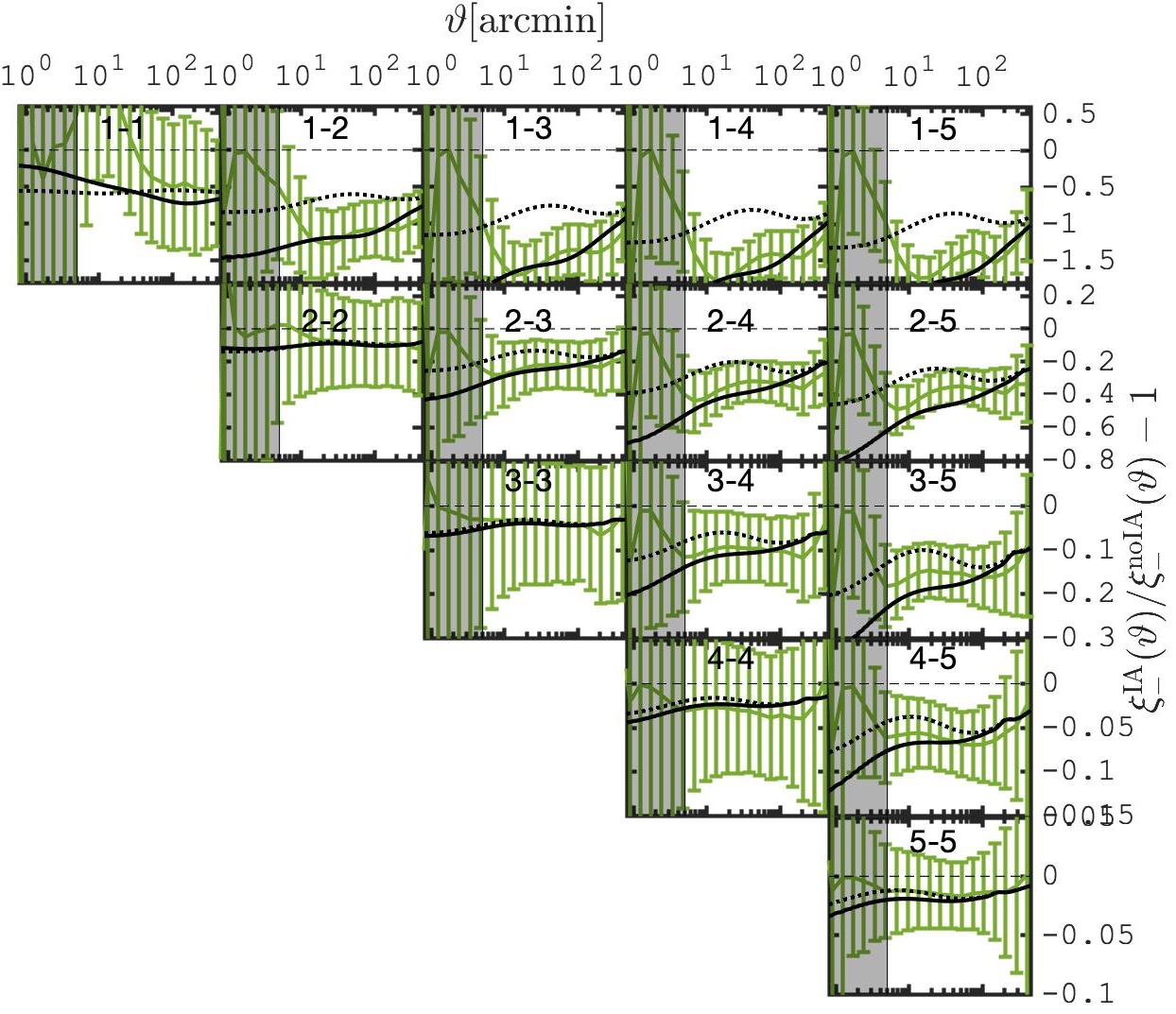}
\caption{Same as Fig. \ref{fig:xi_NLA}, but for the $\delta$-NLA model with $A_{\rm IA}=1.0$ and $b_{\rm TA}=1.0$, and only for smoothing of 0.5 $h^{-1}$Mpc. The dotted  black lines show the NLA predictions to better highlight the differences. Source coupling is included in the $\delta$-NLA simulations but not in the `noIA' case, consistent with the $\delta$-NLA theory curves.}
\label{fig:xi_deltaNLA}
\end{figure*}

\begin{figure}
\includegraphics[width=\columnwidth]{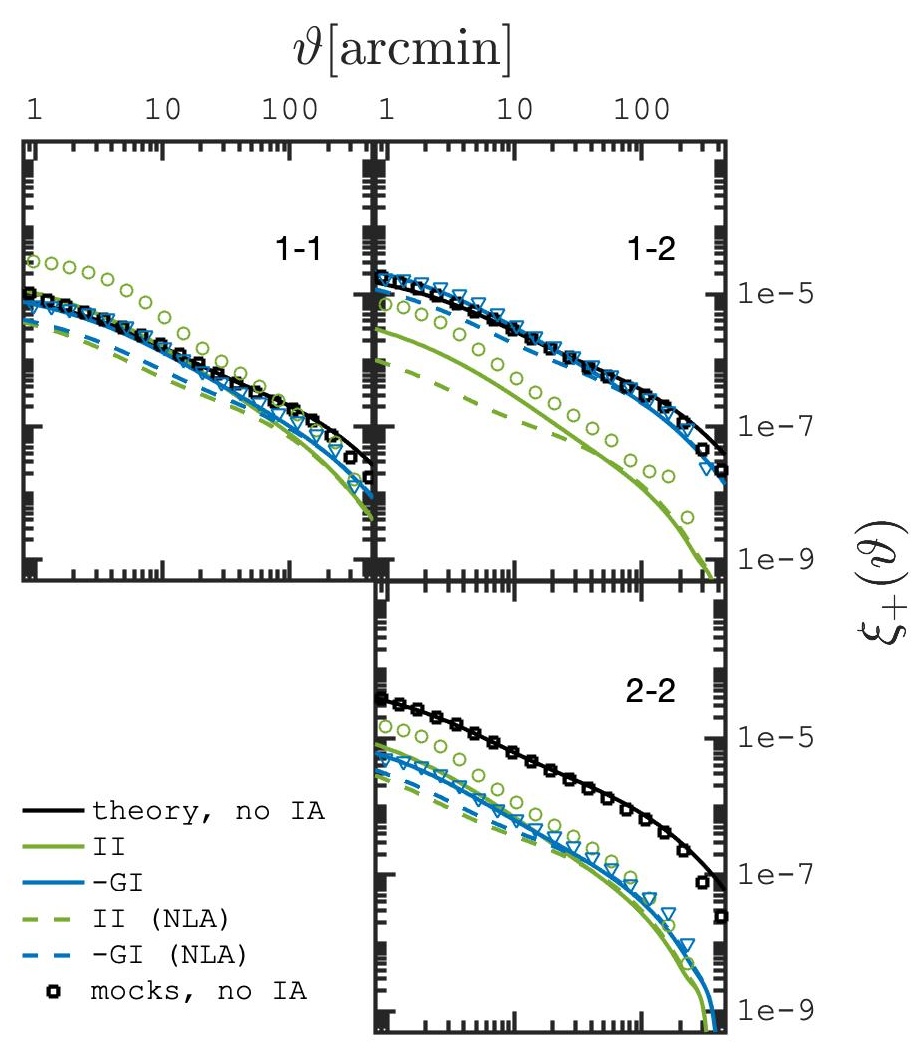}
\caption{Shear 2PCF $\xi_+$ in the extended-NLA model for the two lowest redshift bins, showing the good agreement between simulations (symbols) and theory (lines) for the $GG$ (black) and $GI$ (blue) terms, while large deviations are observed in the $II$ term (green). Similar results are obtained for the HOD-NLA model.} 
\label{fig:xi_deltaNLA_II}
\end{figure}

\begin{figure}
\includegraphics[width=\columnwidth]{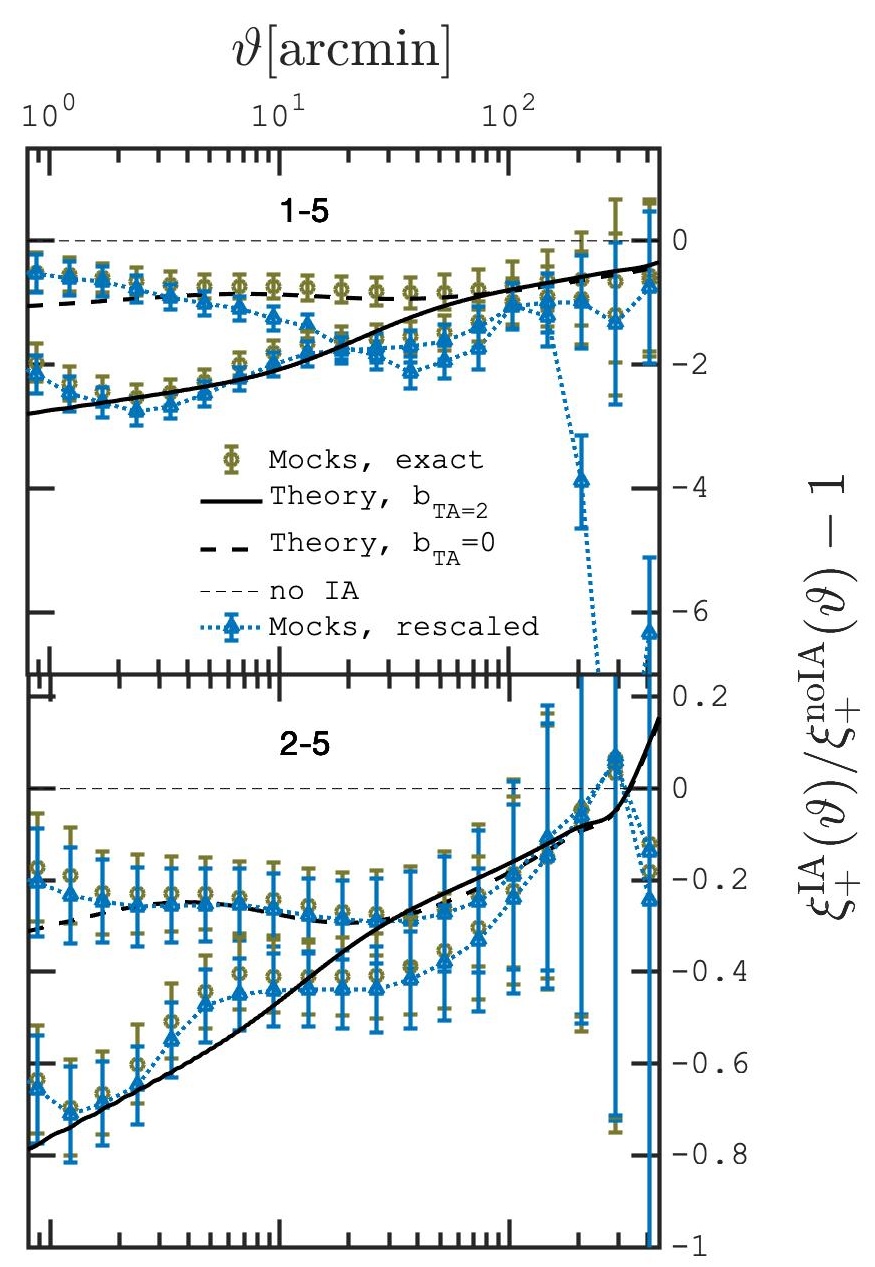}
\caption{Ratio between the shear correlation functions with and without IA in tomographic bin combination 1-5 and 2-5.
The brown symbols show measurements from simulations constructed with $b_{\rm TA}$ = 0.0 and 2.0 respectively (the exact method), while the blue symbols are obtained by rescaling $b_{\rm TA}$ = 1.0 simulations using Eq. (\ref{eq:bta_rescale}); the dashed and solid black lines show the $\delta$-NLA predictions for the two bias values. }
\label{fig:xi_deltaNLA_rescaled}
\end{figure}

The equivalent comparison for the extended-NLA model is presented in Fig. \ref{fig:xi_deltaNLA}, where we also find an excellent match with the theoretical predictions except for redshift bins 1-1 and 2-2, where the $II$ term is important and much larger in the simulations than in the theory.  This is expected since this calculation includes terms that are up to fourth order in the density field, as discussed in Sec. \ref{subsec:IA_th_extNLA}, which in the theoretical predictions are only expected to hold up to $k\sim 0.2$Mpc. Indeed, the agreement improves rapidly for larger angular separation. This excess is shown in Fig. \ref{fig:xi_deltaNLA_II}, where the different contributions are separated, clearly highlighting the disagreement in the $II$ term. 
This disagreement is further enhanced when the tidal field smoothing scale $\sigma_{\rm G}$ is lowered, which we therefore avoid in this work. 

The results are for $b_{\rm TA}$ = 1.0, and we verified that the infusion model works equally well for $b_{\rm TA}$ = 2.0. Using the former, we test the $b_{\rm TA}$-rescaling method introduced in Eq. (\ref{eq:bta_rescale}) to generate $b_{\rm TA}=2.0$ and $b_{\rm TA}=0.0$ mocks, and compare the outcome with mocks constructed directly with these bias values. Results are shown in Fig. \ref{fig:xi_deltaNLA_rescaled} for two of the tomographic bins for which IA has the strongest effect. The match is excellent here except for the lower redshift rescaling to $b_{\rm TA}=0.0$, which is not accurate for $\vartheta>5$ arcmin. We therefore recommend producing new mocks instead of using this rescaling method whenever possible. 

Note that since source clustering (SC) is a higher-order effect in cosmic shear, the measured $\xi_{\pm}$ from mocks with $b_{\rm TA}$=0.0, 1.0 and 2.0 show negligible differences in $\xi_+$, and becomes detectable only at the smallest scales (i.e. $\vartheta<5$ arcmin) in $\xi_-$, consistent with \citet{source_lens_clustering} who find an effect of a few percent for $\ell>3000$, and with \citet{DESY3_Gatti_source_clust}, who find minor impact for second moments of aperture mass maps. However, the impact of SC on higher-order lensing statistics and on the overall IA contamination is significant and can be double the secondary signal in certain circumstances, due to the up-sampling of regions with strong tidal forces. As noted in \citet{source_lens_clustering} and \citet{Linke_SC}, the impact of SC can also vary with the choice of estimator.

\begin{table}
   \centering
    \caption{IA models infused in this work, along with the respective values of the model parameters. The two HOD models, at the bottom of the table, follow a non-linear galaxy biasing, which we label here as $b_{\rm nl}$ and is partly fitted by the $b_{\rm TA}$ parameter in the MCMC, see Sec. \ref{sec:inference} for details. }
   \begin{tabular}{@{} lc @{}} 
      \hline
      \hline

      model   		& ($A_{\rm IA}, \: b_{\rm TA}, \: C_2)$ \\
      \hline
      NLA     		& {(1, 0, 0) }  \\
      \multirow{2}{*}{$\delta$-NLA }  	&   {(1, 1, 0)}   \\
      							&  {(1, 2, 0)}   \\
      TT 			&  {(0, 0, 1)}   \\
      $\delta$-TT	&  {(0, 1 ,1)}   \\
				HOD-NLA & {(1, $b_{\rm nl}$, 0)} \\
				HOD-TT & {(0, $b_{\rm nl}$, 1)} \\

      \hline
      \hline
   \end{tabular}
   \label{table:IAmodels}
\end{table}


\subsubsection*{TT model}
\label{subsec:TT}

\begin{figure*}

\includegraphics[width=\columnwidth]{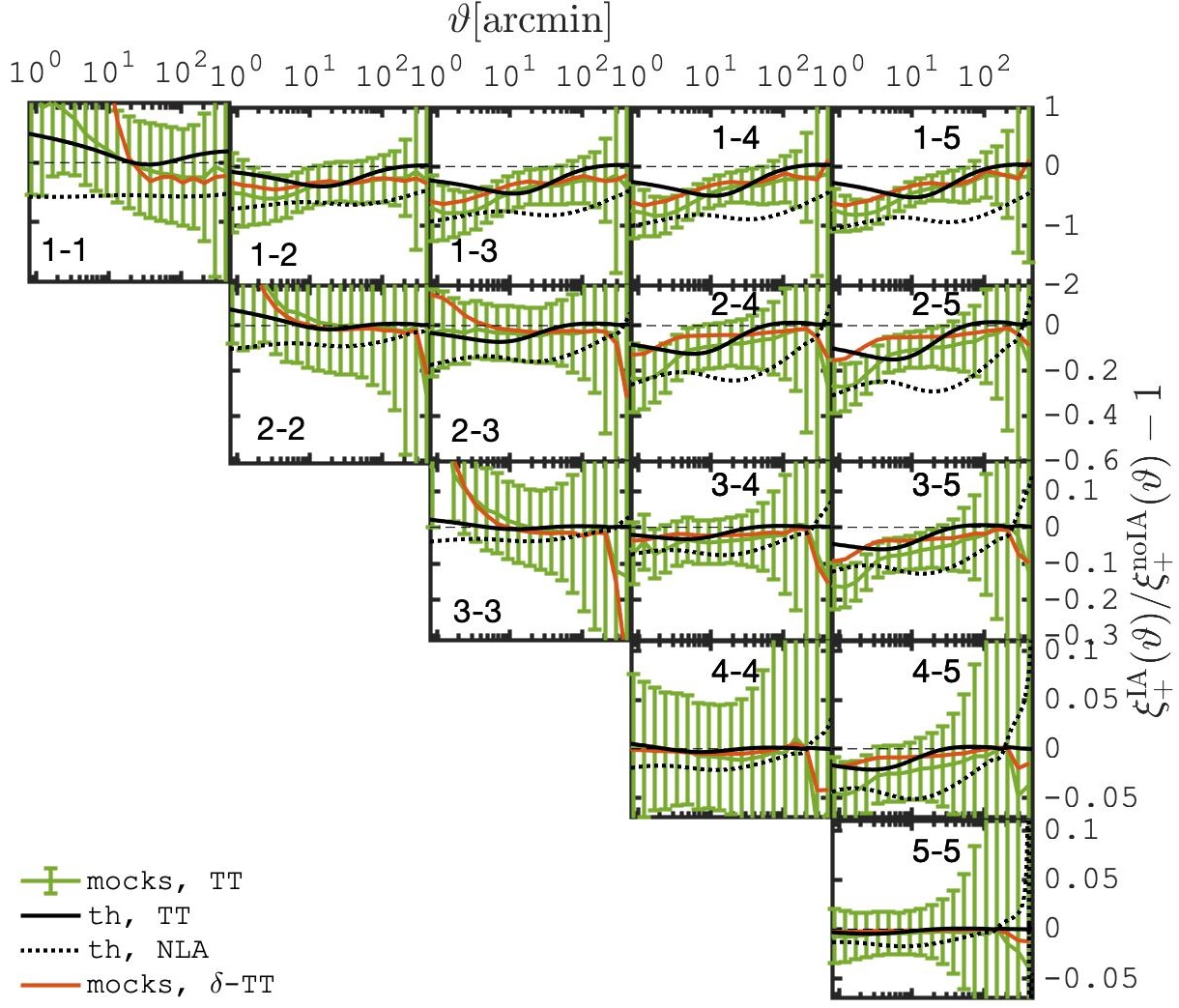}
\includegraphics[width=\columnwidth]{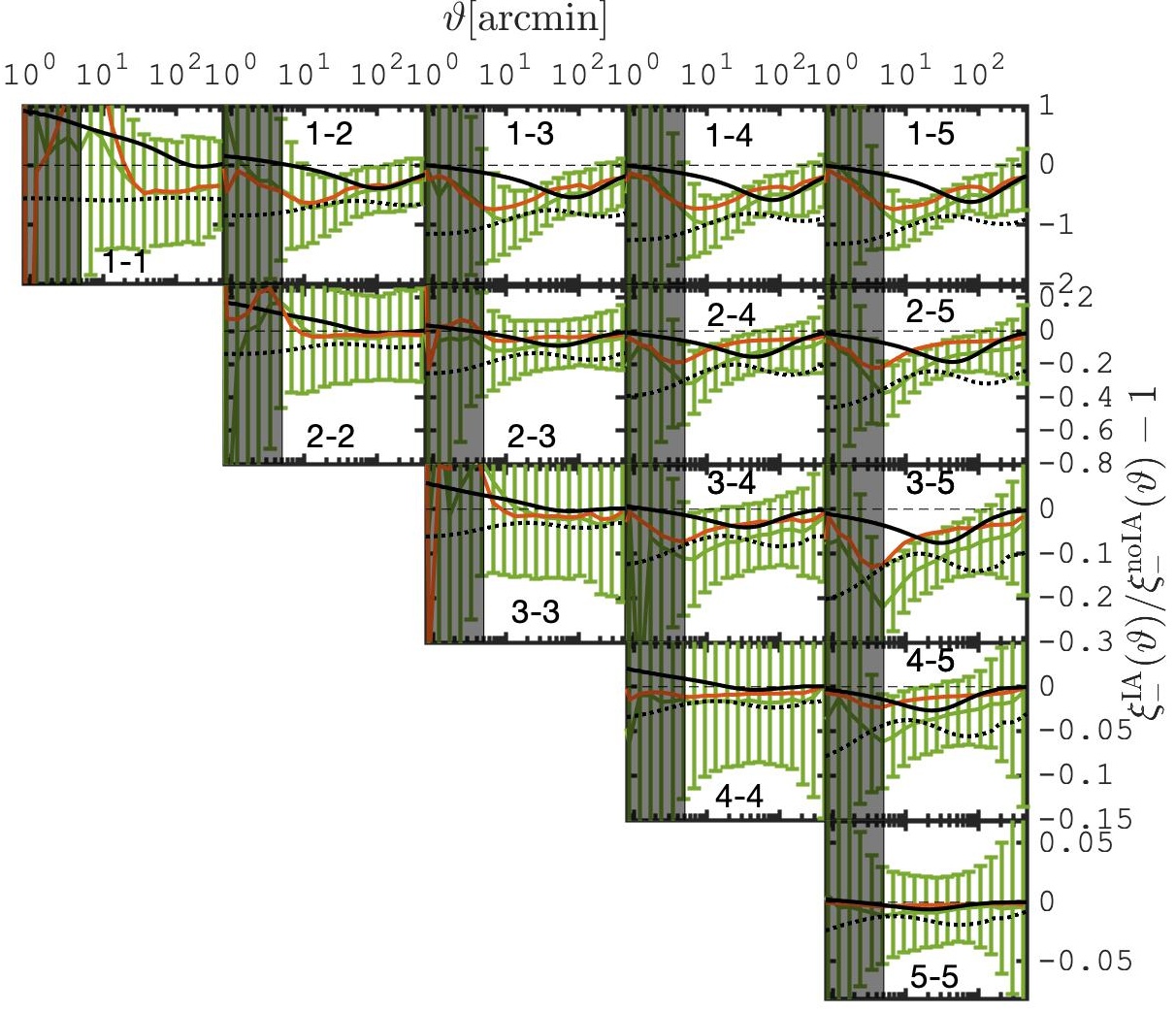}
\caption{Same as Fig. \ref{fig:xi_deltaNLA}, but comparing the TT (green) and the $\delta$-TT  (orange) infusion models, with $C_2=1.0$  and only for smoothing of 0.5$h^{-1}$Mpc. Note that the theoretical  model for the $\delta$-TT case does not exist yet.}
\label{fig:xi_deltaTT}
\end{figure*}

Results for the tidal torque model are presented in Fig. \ref{fig:xi_deltaTT} (in green). In this case, we see that the larger smoothing scales agree better with the theoretical predictions, likely due to a mismatch caused by projection effects. Indeed, the TT model, being quadratic in the tidal field, is more sensitive to smaller scales \citep{TATT}, and the missing third dimension inherent in our method is more prone to inaccuracies. We observe discrepancies in the small angular separations for $\xi_-$ at low redshifts, but achieve a better match for $\vartheta>$ 40 arcmin.

Importantly, we find that our implementation of the TT model yields alignments that are too strong at low redshifts, and hence require an empirical redshift-dependent calibration to improve the match with theory  for all choices of $C_2$.  This mismatch is likely due in part to only having the projected tidal field -- non-linear operations on the 3D tidal field, including the TT model, do not commute with projection. We achieve this calibration by rescaling $\epsilon^{\rm IA, TT}(z<0.5) \rightarrow \epsilon^{\rm IA, TT} /2.5$, acknowledging that this exhibits limits in our ability to model the tidal torquing without accessing the third dimension of the tidal field.

\subsubsection*{Extended-TT model}
\label{subsec:deltaTT}

The extended-TT model, constructed using the TT coupling on galaxies linearly tracing the underlying matter density, also needs to be calibrated. This has some degree of arbitrariness due to the absence of a theoretical model to match against it. We choose to match the TT theory on large angular separations at all redshifts, physically motivated by the fact that this modified galaxy bias should not strongly impact scales that are much larger than galaxy clusters.
This therefore requires us to lower the original ellipticities, especially at low $z$: $\epsilon^{{\rm IA}, \delta{\rm TT}}(z<0.5) \rightarrow \epsilon^{{\rm IA}, \delta{\rm TT}} /2.5$ as for the normal TT model, further followed by a global $\epsilon^{{\rm IA}, \delta{\rm TT}} \rightarrow \epsilon^{{\rm IA}, \delta{\rm TT}}/20.0$ rescaling. This is a large calibration condition, which compensates for the overly strong coupling computed from our simulations. The results are shown in Fig. \ref{fig:xi_deltaTT} (in brown), where we observe that the deviations with respect to the TT model occur at small angular scales ($\vartheta < $ 20 arcmin) in $\xi_+$, and for the lowest redshifts. Although this model is harder to match with the current theoretical IA model, it provides a good example with which we can study the impact of IA mis-modelling, e.g. analysing this model with the TATT predictions, similar to the investigations of \citet{Paopiamsap2024}.

\subsubsection*{HOD-TATT model}
\label{subsec:HOD}

\begin{figure*}

\includegraphics[width=\columnwidth]{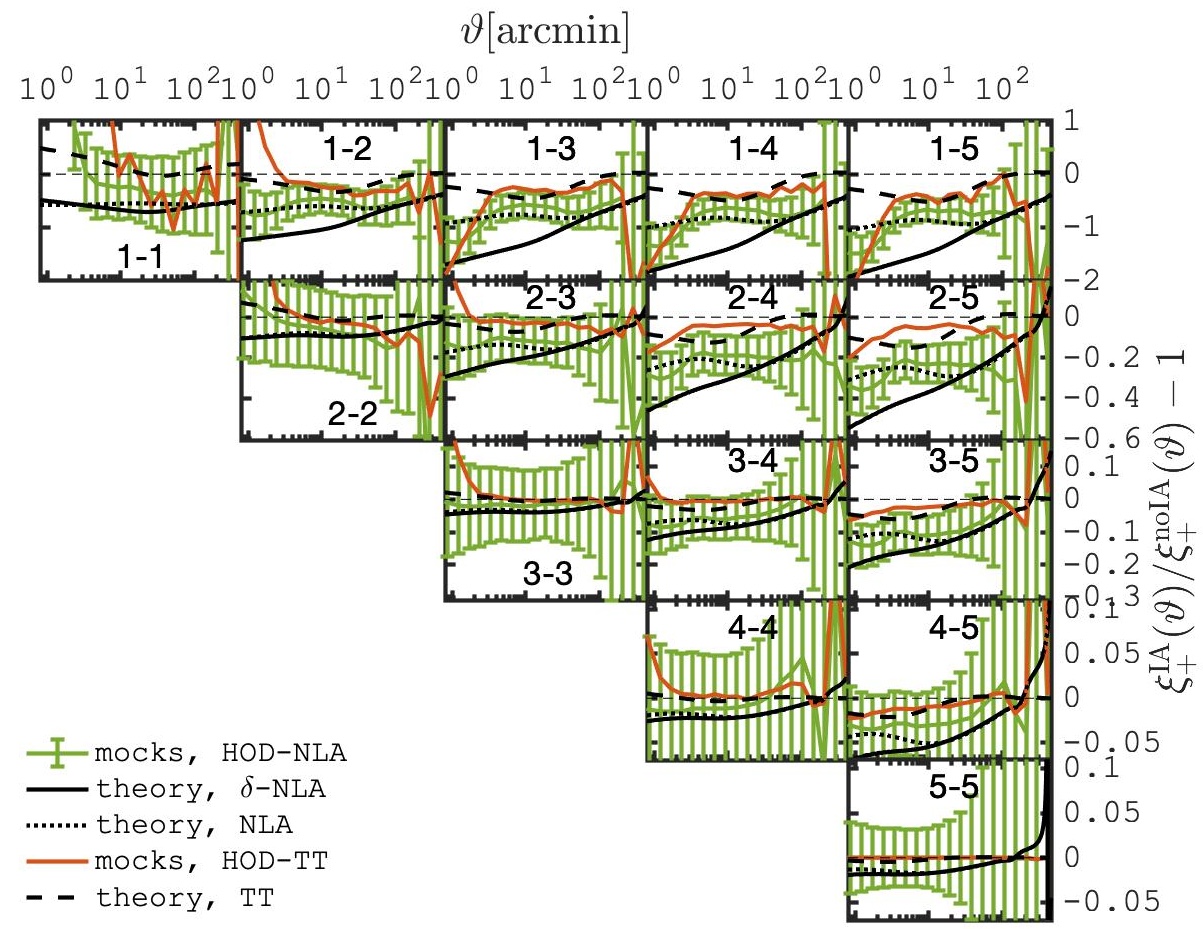}
\includegraphics[width=\columnwidth]{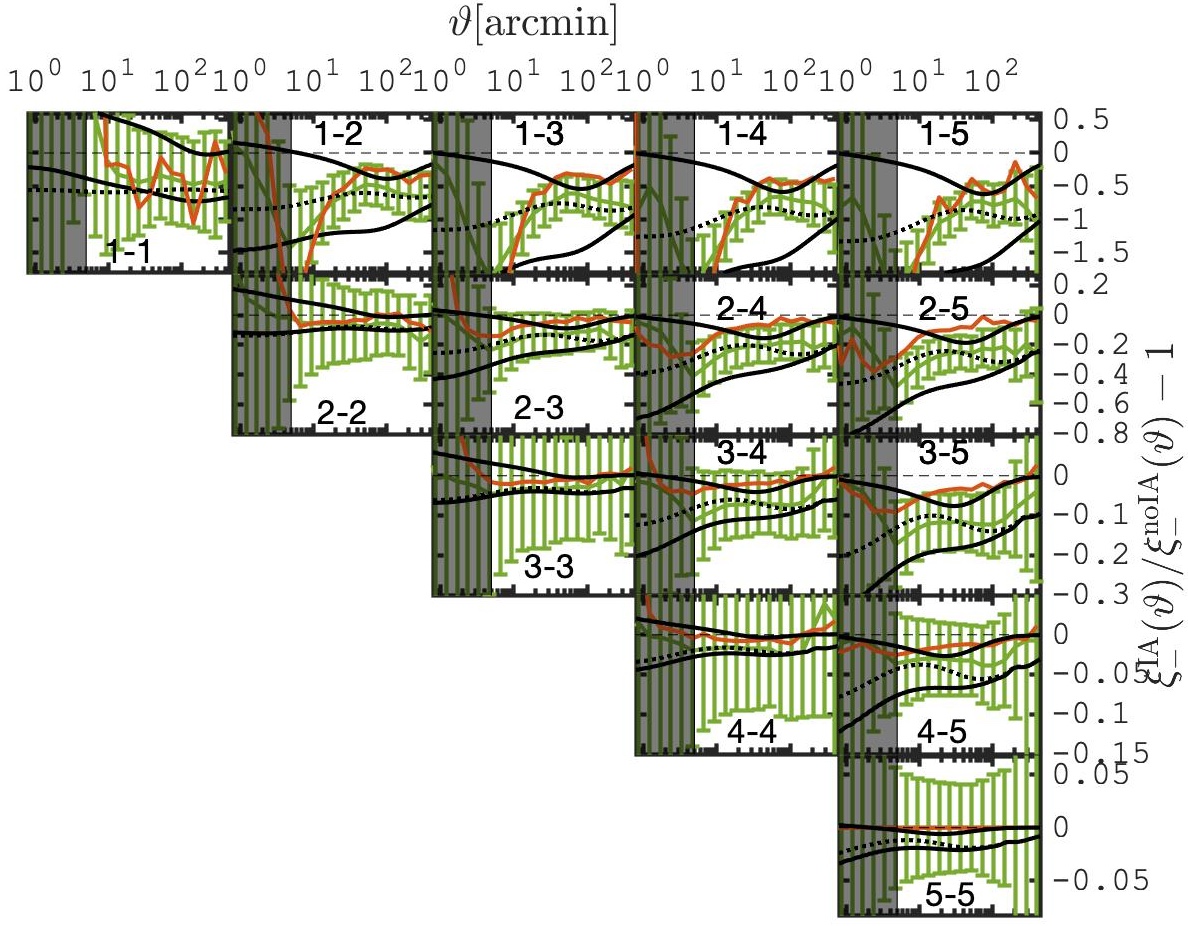}
\caption{Same as Fig. \ref{fig:xi_deltaNLA}, but for the HOD-NLA (green) and HOD-TT (orange) models. The theory lines show various TATT models that could be used to fit these results, as indicated in the legend.
}
\label{fig:xi_hod_TATT}
\end{figure*}

The largest difference between the previous models and those based on HOD galaxies is that the latter are non-linear biased tracers of the matter distribution, which means that a larger number can populate regions of large over-densities, where the tidal fields are generally stronger.
We therefore expect the impact of IA to be stronger in this case.
We show in Fig. \ref{fig:xi_hod_TATT} the results from the HOD-NLA and HOD-TT models, which together make up our HOD-TATT. One of the most interesting features is that the HOD-NLA closely follows the normal NLA theory predictions over a large range of scales, suggesting that the impact of the non-linear galaxy bias is sub-dominant down to a few arcmin.
We also observe a small-angle upturn in bins 1-1 and 2-2, which can be attributed to mismodelling the $II$ term (as is the case for the extended-NLA, see the discussion above). 
We finally note that the impact of IA measured in bin 5 is negligible.

\subsection{Validation with full inference}
 \label{sec:inference}

 \begin{table}
   \centering
   \caption{Priors used when sampling the likelihood in Sec. \ref{sec:validation}. }
   \tabcolsep=0.11cm
      \begin{tabular}{@{} cccccccc @{}} 
      \hline
      \hline
       Parameter       &  range & prior \\
       \hline
       Cosmology\\ 
       $\Omega_{\rm m}$ &[0.1, 0.5] & Flat\\
       $S_8$ & [0.6, 0.9] & Flat\\
       \hline
       IA\\
       $A_{\rm IA}$ & [-5, 5]&  Flat\\                    
        $b_{\rm TA}$ & [0, 5] & Flat\\
        $C_2$ & [-5, 5] & Flat\\
     \hline
    \hline 
    \end{tabular}
    \label{table:priors}
\end{table}

\begin{figure}
\includegraphics[width=\columnwidth]{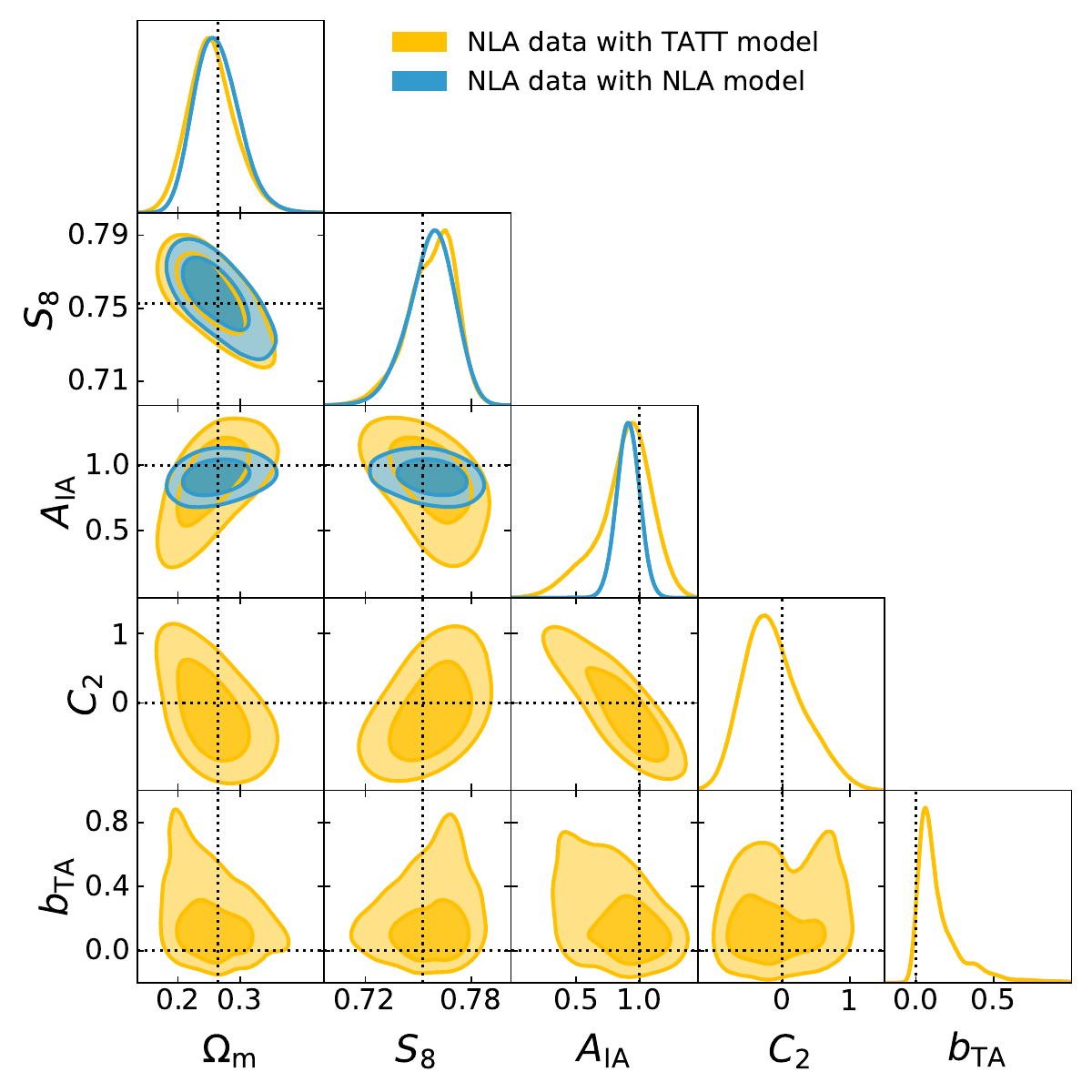}
\caption{Cosmological inference from $\gamma$-2PCF measured in the NLA-infused simulations, where either all three TATT parameters are varied (yellow) or only $A_{\rm IA}$ is varied (blue) in the model.
The cross-hairs represent the truth as fixed in the {\it Outer Rim} $N$-body simulations \citep{OuterRim} and in the infusion method. Offsets are expected due to sampling variance in our simulations.}
\label{fig:corner_nla_combo}
\end{figure}
\begin{figure}
\includegraphics[width=\columnwidth]{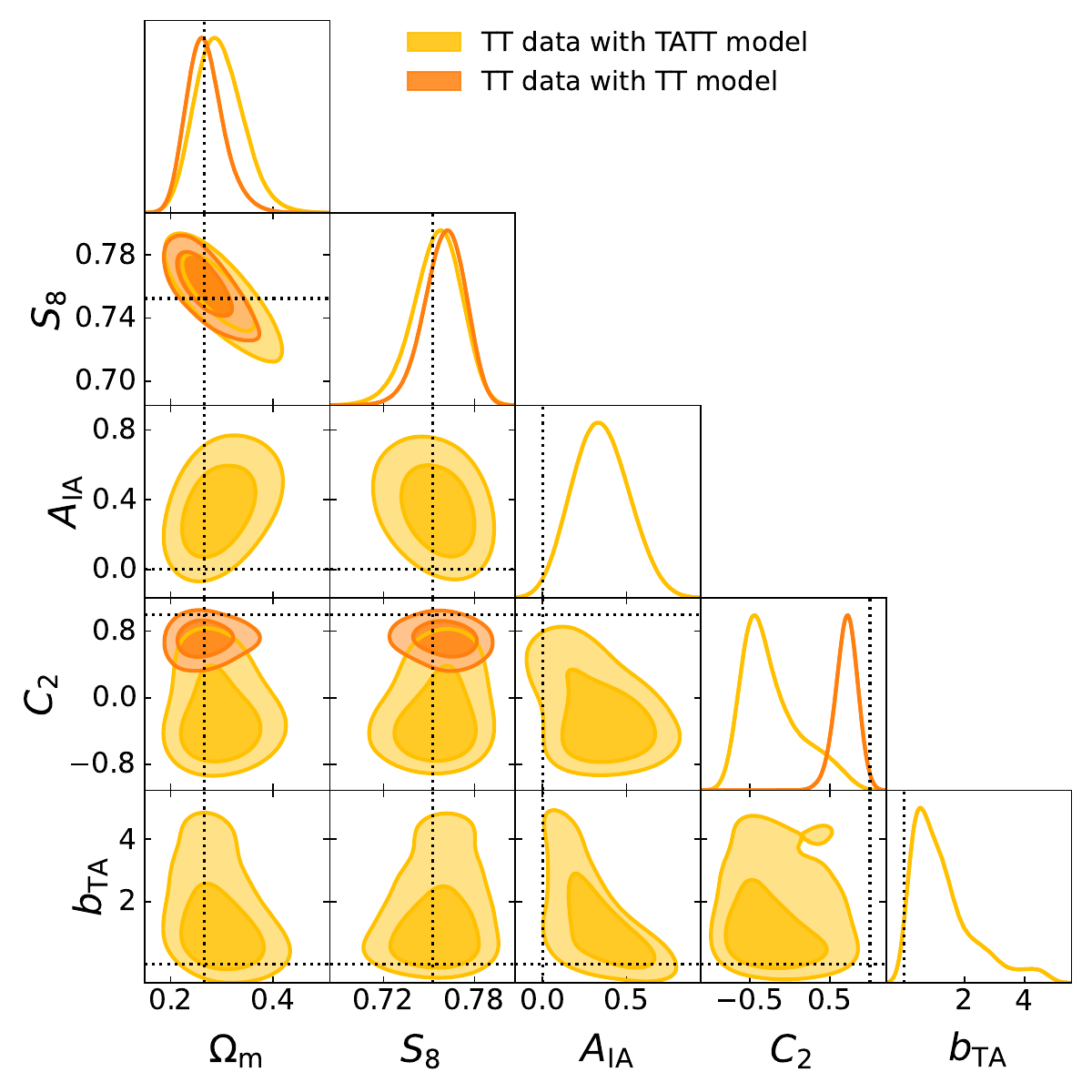}
\caption{Cosmological inference from $\gamma$-2PCF measured in the TT-infused simulations, where either all three TATT parameters are varied (yellow), or varying $C_2$ only (orange), setting $A_{\rm IA}$ and $b_{\rm TA}$ to zero.}
\label{fig:triangle_tt_combo}
\end{figure}
\begin{figure}
\includegraphics[width=\columnwidth]{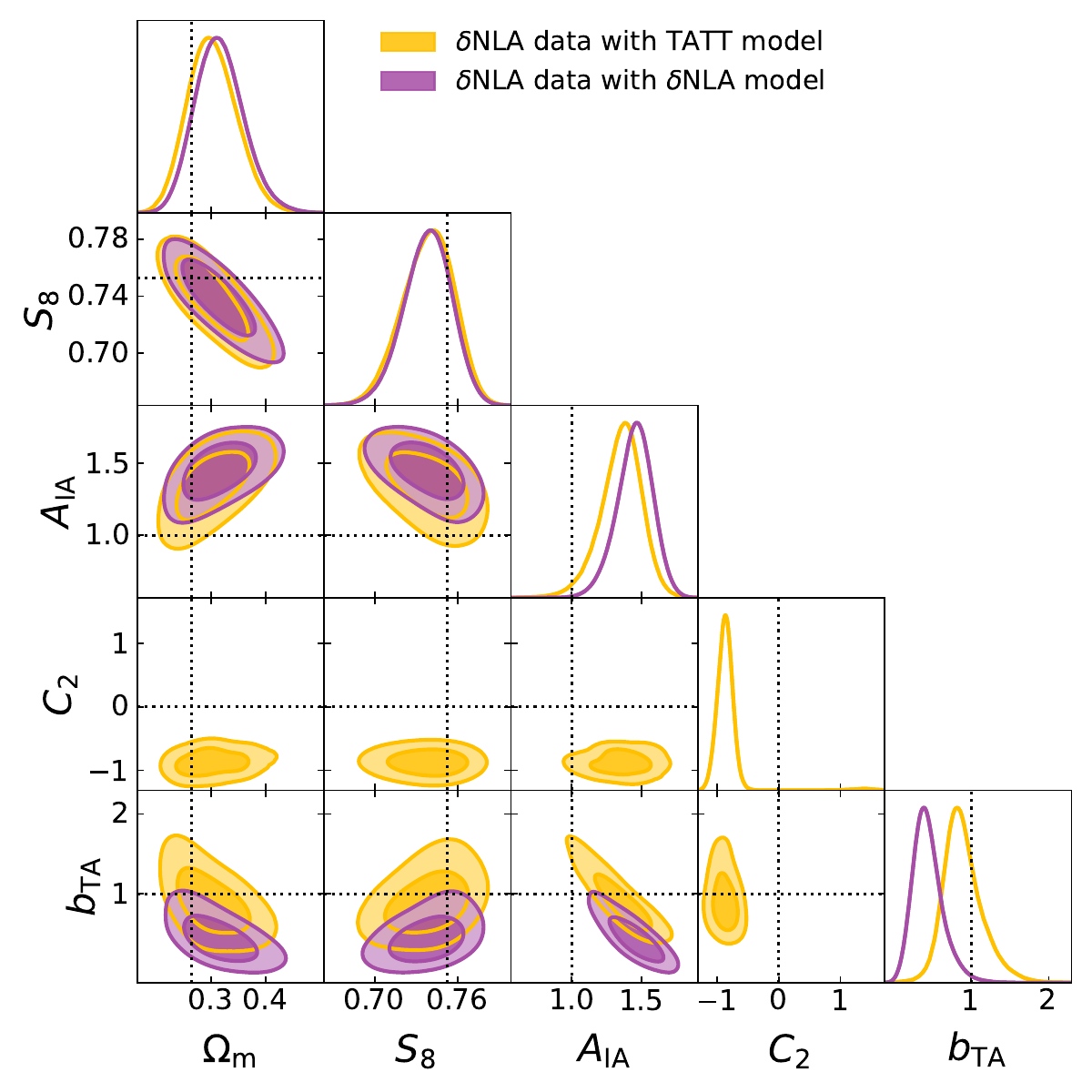}
\caption{Cosmological inference from the extended NLA-infused simulations, where either all three TATT parameters are varied (yellow) or only $A_{\rm IA}$ and $b_{\rm TA}$ (purple). }
\label{fig:triangle_deltaNLA_combo}
\end{figure}
\begin{figure}
\includegraphics[width=\columnwidth]{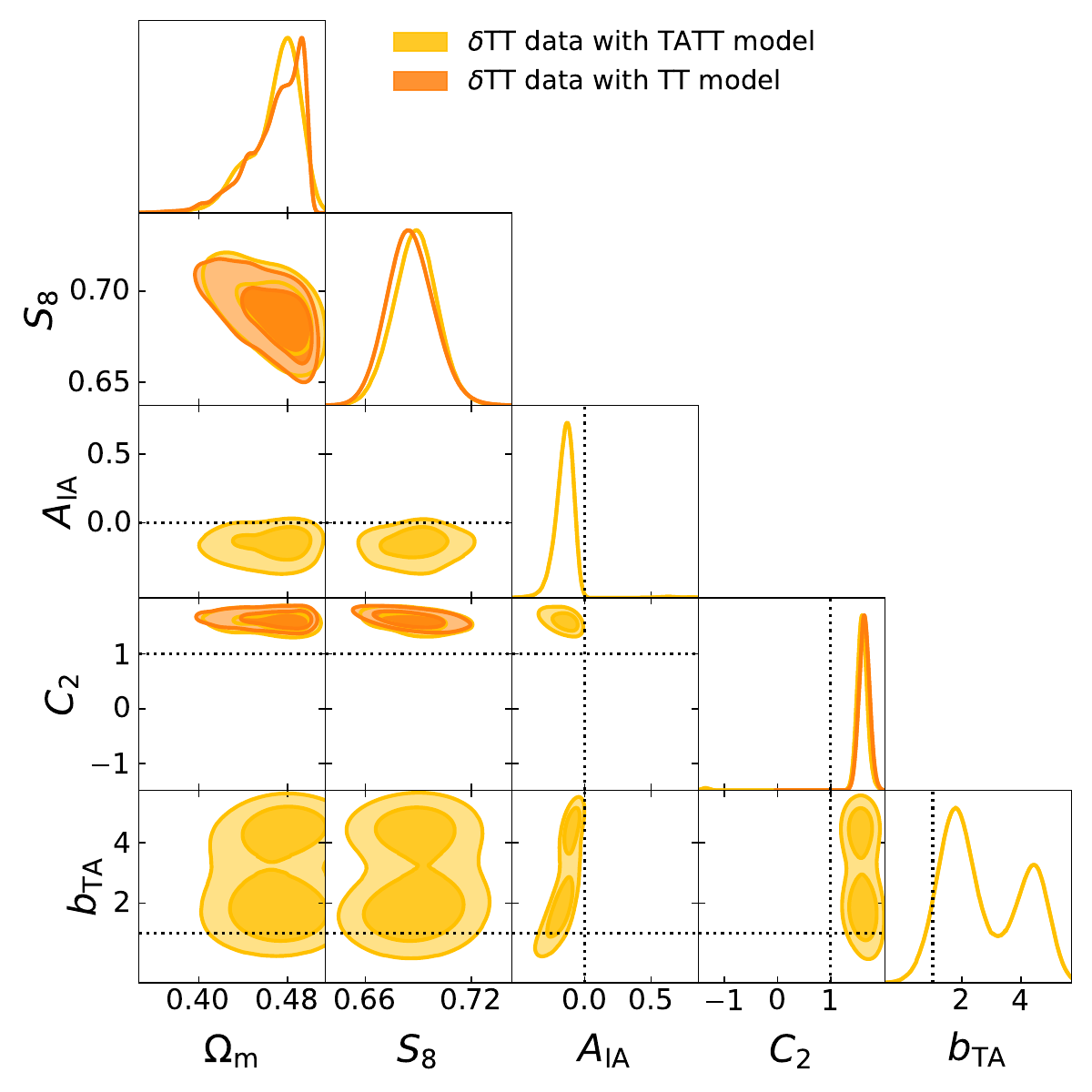}
\caption{Cosmological inference from the $\delta$-TT-infused simulations, where  all three TATT parameters are varied (yellow) or recovering the case where $A_\mathrm{IA}$ is fixed (orange), which corresponds to the TT model.}
\label{fig:corner_deltatt}
\end{figure}
\begin{figure}
\includegraphics[width=\columnwidth]{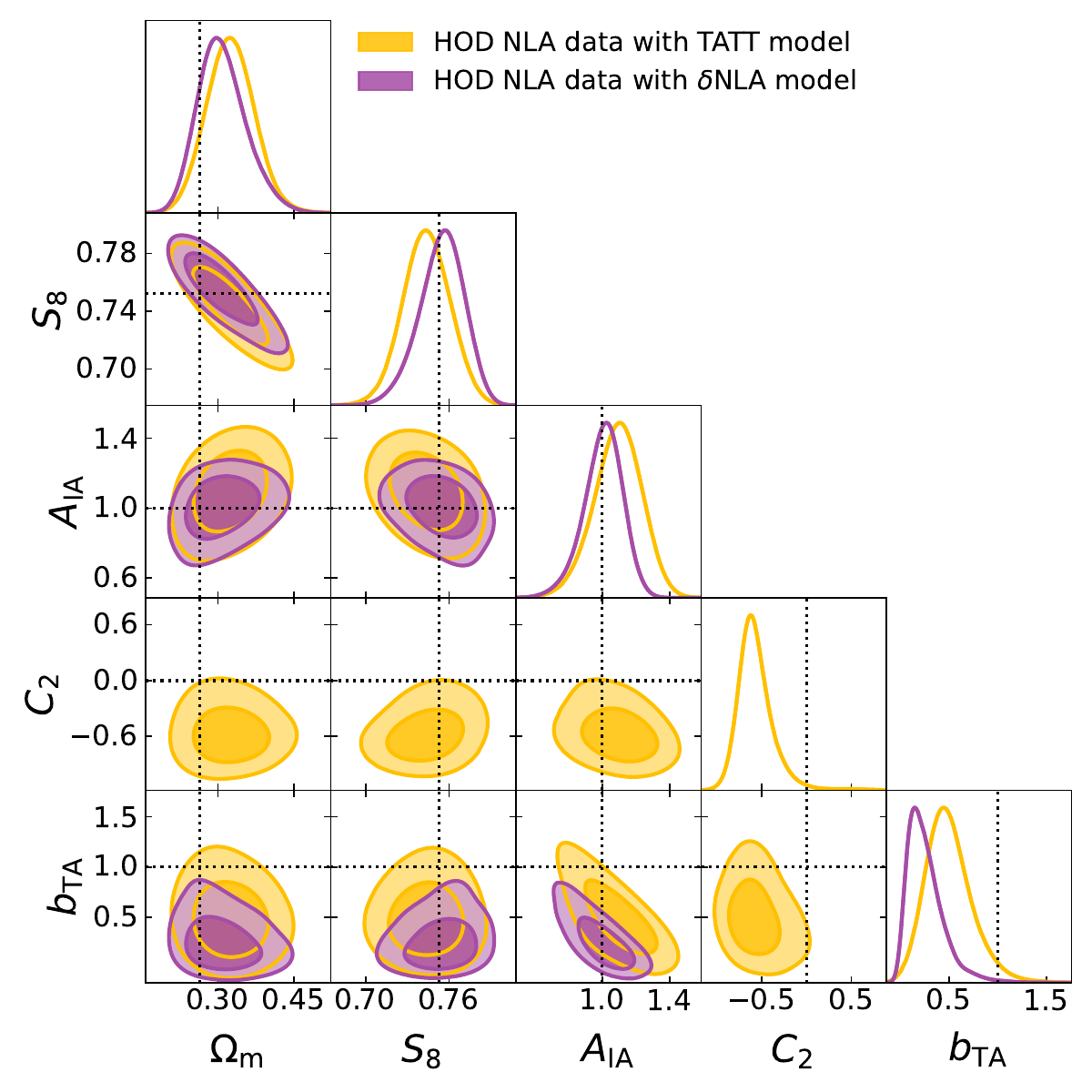}
\caption{Cosmological inference from the HOD-NLA-infused simulations, where either all three TATT parameters are varied (yellow), or fixing $C_2$ to $0.0$ (purple) .}
\label{fig:triangle_HOD_NLA_combo}
\end{figure}
\begin{figure}
\includegraphics[width=\columnwidth]{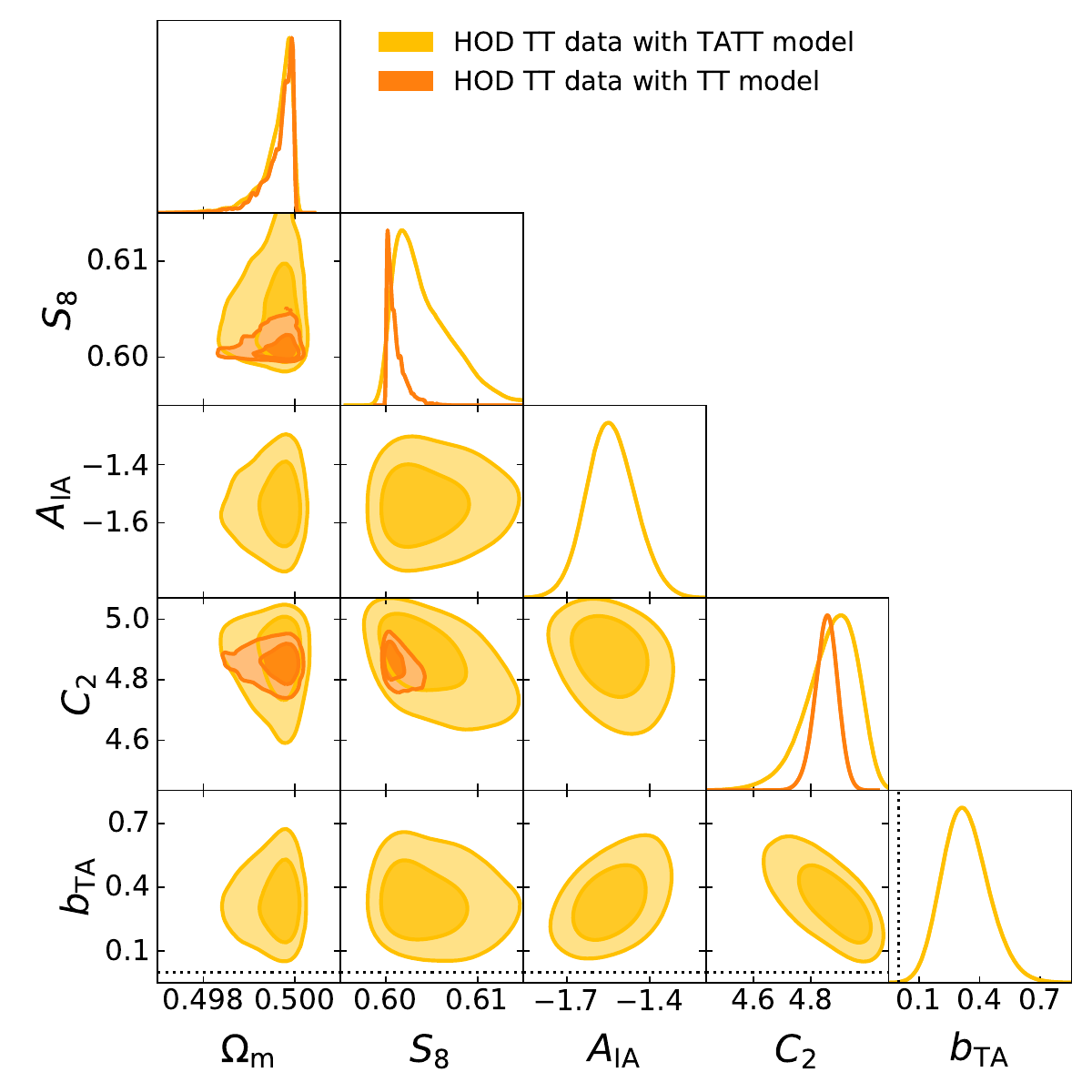}
\caption{Cosmological inference from the HOD-TT-infused simulations, where either all three TATT parameters are varied (yellow), or fixing $A_\mathrm{IA}$ and $b_{\rm TA}$ to zero (orange).}
\label{fig:triangle_HOD_TT_combo}
\end{figure}

The previous section presented comparisons between measured and theoretically modelled signals, reaching generally an excellent agreement but highlighting some small differences. In order to quantify the importance of such differences in a cosmic shear analysis, our evaluation metric must measure the bias between the true and the inferred cosmological parameters. We achieve this by running nested likelihood sampling chains with {\sc multinest} \citep{Multinest}, which returns the posterior distributions $\mathcal{P}$ on the cosmological parameters $\boldsymbol \pi$ given a data vector $\boldsymbol d$, a model vector $\boldsymbol m$, a covariance matrix C, and likelihood function $\mathcal L$ and a set of priors on the inferred parameters. We use a multivariate Gaussian likelihood, and ignore the standard cosmic shear systematic effects caused by photometric redshift errors, shape calibrations, or baryonic feedback \citep[see][for a recent example]{DESY3-KiDS1000}. We vary a subset of the vanilla $\Lambda$CDM cosmological parameters to establish the general accuracy; increasing the number of varied parameters can lead to a number of projection effects that complicate the interpretation. We therefore focus on varying the parameters best measured by lensing, namely $S_8$, $\Omega_{\rm m}$ and the IA parameters ($A_{\rm IA}, b_{\rm TA}, C_2$), and update at every step the theoretical predictions computed from Eq. (\ref{eq:xipm_th}). We adopt wide flat priors for all of these, as detailed in Table \ref{table:priors}. 

With the precision of the Rubin data, small fluctuations in the data vectors can lead to significant shifts in the inferred cosmology, which is statistically expected but makes the IA validation exercise more difficult to interpret. We therefore run our likelihood analyses on the lowest three redshift bins, deliberately excluding the highest two as they contribute the most to the cosmological information, while we want here to learn more about the recovery of the IA sector. For the same reason we ignore shape noise here, which would only create additional scatter in the posteriors inferred from $\gamma$-2PCFs. We additionally exclude angular bins where the models probe highly non-linear scales, and select $\vartheta \in [2 ; 200]$ arcmin for $\xi_+$, and $\vartheta \in [20 ; 400]$ arcmin for $\xi_-$. These rejected scales are highly affected by other systematic effects such as baryonic feedback \citep[see, e.g.][]{Semboloni11, HWVH15, 2020arXiv200715026H} and hence are often removed from the data vectors \citep[as in][]{DESY3_Amon}.

The marginalised 2D posteriors are presented in Figs. \ref{fig:corner_nla_combo}-\ref{fig:triangle_HOD_TT_combo}, analysing simulated data from the six models introduced in Sec. \ref{sec:IA_th}. In the absence of IA contamination, the inference from the simulated data is expected to yield posteriors on the cosmological parameters that are consistent with the input truth but not necessarily centered on it due to sample variance. When analysing IA-infused simulations, degeneracies are expected between IA and cosmological parameters. Moreover, we have shown in Section \ref{subsec:data_vector} that some of our IA models do not match perfectly with their analytical implementations (i.e. $\delta$-NLA, HOD-NLA and HOD-TT), hence for these we expect the inference to be biased as a direct consequence of this.

We dissect the analyses by varying either one, two  or all three TATT parameters, highlighting some of the interesting degeneracies. When analysing the NLA-infused data (Fig. \ref{fig:corner_nla_combo}), all cosmological and IA parameters are accurately recovered, both for the NLA and TATT modelling. In the latter case, the inferred values for $b_{\rm TA}$ and $C_2$ are consistent with zero, but we observe a strong degeneracy between  $A_{\rm IA}$ and $C_2$ as also found in e.g. \citet{DESY3_Secco} and \citet{Paopiamsap2024}. The joint posterior of these two parameters is slightly shifted towards lower values, a result that persists when analysing different noise realisations of the data, and is therefore a projection effect. The inferred $\Omega_{\rm m}$ is slightly shifted to higher values, however, this is not surprising given that cosmic shear alone is generally not able to constrain this parameter very well, leading to large fluctuations in the inferred value. The parameter $S_8$ however is well recovered, slightly on the high side. As expected, opening up the full TATT parameter space slightly degrades the constraints on cosmology, but no large shifts are observed. The $w$CDM chains also show an excellent convergence, as shown in Appendix \ref{app:figs}.

We next analyse the TT data with the TATT model in Fig. \ref{fig:triangle_tt_combo}, and observe an accurate recovery of $\Omega_{\rm m}$ and $S_8$, however the IA sector is strongly affected by the $A_{\rm IA}- C_2$ degeneracy, which pulls the posterior towards the NLA model, with values of $C_2$ consistent with 0 and preferring $A_{\rm IA}\sim 0.5$. When fixing $A_{\rm IA}$ and $b_{\rm TA}$ to the input truth however, the inferred $C_2$ just undershoots the input, pointing to residual differences between our implementation of the TT model and that calculated by TATT. Nevertheless, the fact that the cosmology is well recovered is promising, pointing to the fact that inaccuracies in the IA modelling are below the acceptance threshold identified in \citet{Paopiamsap2024}.

As explained above, the $\delta$-NLA model infused in our simulations is not entirely well captured by the predictions, especially the $II$ term for the lowest redshift bins. This causes problems in the cosmology inference, since the predictions attempt to compensate the difference in the IA model, leading to biases in the other parameters. Fig. \ref{fig:triangle_deltaNLA_combo} shows exactly this, failing to correctly infer most parameters. In this particular case, $S_8$ is off by $2\sigma$, $A_{\rm IA}$ and $C_2$ are biased by 4-5$\sigma$, while $\Omega_{\rm m}$ and $b_{\rm TA}$ are within 1$\sigma$. The $\delta$-TT model is even more affected by mis-modelling, with a 3$\sigma$ shift in both $\Omega_{\rm m}$ and $S_8$, as seen in Fig. \ref{fig:corner_deltatt}.
 
Finally, while the input parameters from the infused HOD-NLA model  are well recovered by both our $\delta$-NLA and our TATT inference analysis (see Fig. \ref{fig:triangle_HOD_NLA_combo}), the HOD-TT is catastrophic (see Fig. \ref{fig:triangle_HOD_TT_combo}): most of our posteriors are pushed against the prior edges, as a result of the incapacity for our theoretical IA models to describe the HOD-TT data. 
  
To summarise this section, we are able to correctly infer the cosmology from many models (NLA, $\delta$-NLA, TT and HOD-TT) but not all ($\delta$-TT and HOD-TT), which in this case lead to significant biases both in the cosmology and IA parameters.  This is caused by a strong interaction between the TT ellipticities and the galaxy bias, which is currently not implemented in the theoretical model. In other words, if the true IA model in our Universe resembled our implementation of the $\delta$-TT or the HOD-TT model, and that we would analyse the cosmic shear $\xi_\pm$ data with the NLA or TATT, our inferred $S_8$ value would be biased low by more than 0.05, significantly contributing towards an $S_8$ tension \citep[for various recent results on this tension see, e.g.][]{QUAIA_Alonso2023, KiDSLegacy}. More precisely, it would shift the parameter in the same direction that is often observed (i.e. many late-time probes observe a lower $S_8$ value compared to CMB probes). We are careful here not to claim that this is the single cause, but clearly mis-modelling in the IA sector could be a key factor. Some of this could be captured during a data analysis by the examination of the goodness-of-fit: the reduced $\chi^2$ in the catastrophic cases reach several tens to several thousands, whereas our NLA and TT models have a reduced $\chi^2$ close to unity,  indicating here that the models provide a good fit to the data.

These tests complete our validation of the IA-infusion pipeline, and we now look in the next section at the impact of IA on the weak lensing higher-order statistics mentioned in Sec. \ref{subsec:beyond-2pt}.

\section{Impact of Intrinsic alignment and source clustering on non-Gaussian statistics}
\label{sec:HOWLS}

\begin{figure*}
\includegraphics[width=5.5in]{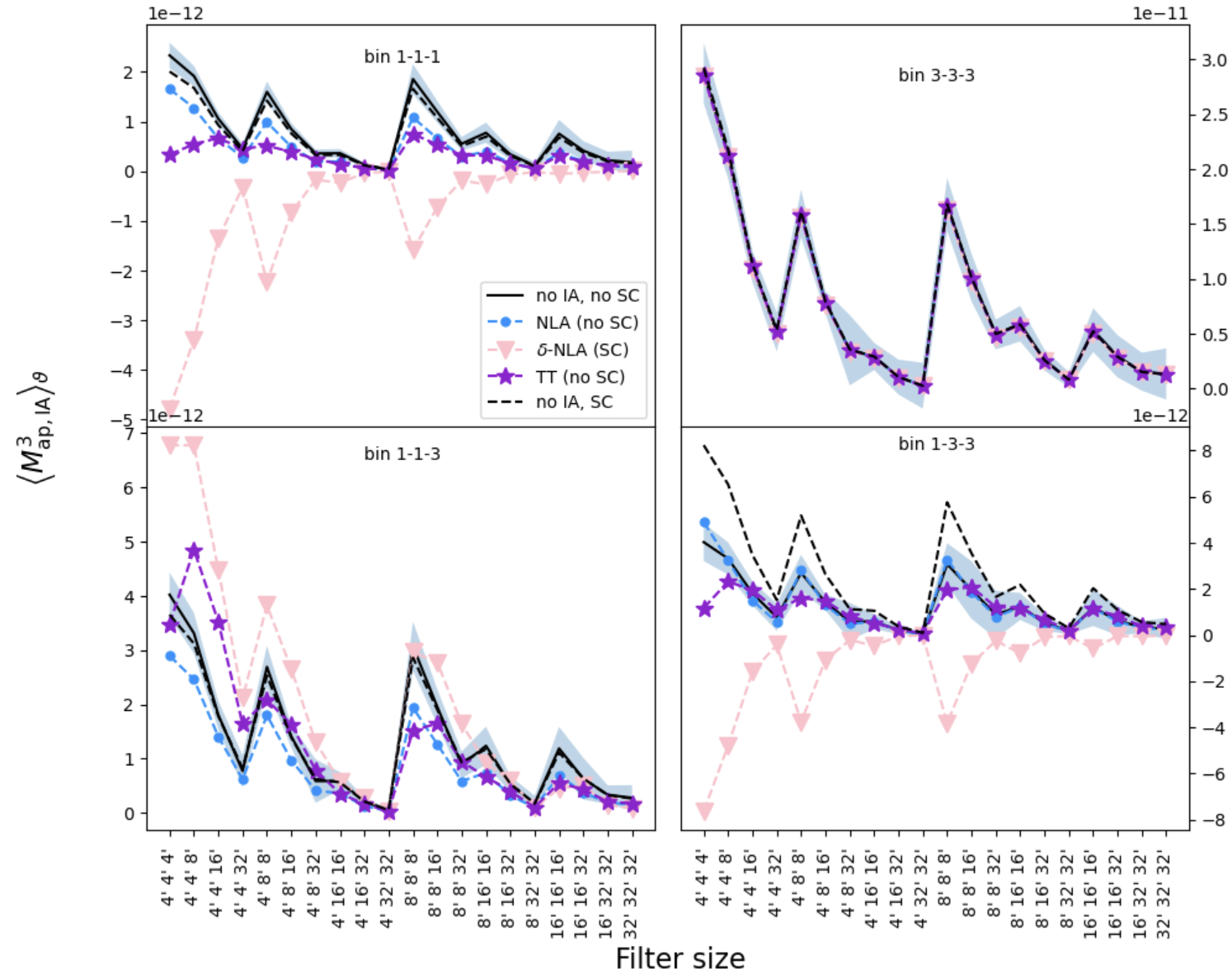}
\caption{Impact of IA and SC on the $M_{\rm ap}^3$  statistics (see Sec. \ref{subsec:beyond-2pt}) for different  IA models. The individual panels show the results from four combinations of tomographic bins, as labeled near the upper edge. These measurements are from noise-free kappa maps.   The error bars are computed from the analytical calculations described in  \citet{Linke_Map3Cov}. As seen in the other probes presented later, the impact of IA is much larger that the statistical error bar when the data includes the first redshift bin.}  
\label{fig:Map3_IA}
\end{figure*}

\begin{figure*}
\includegraphics[width=3.1in]{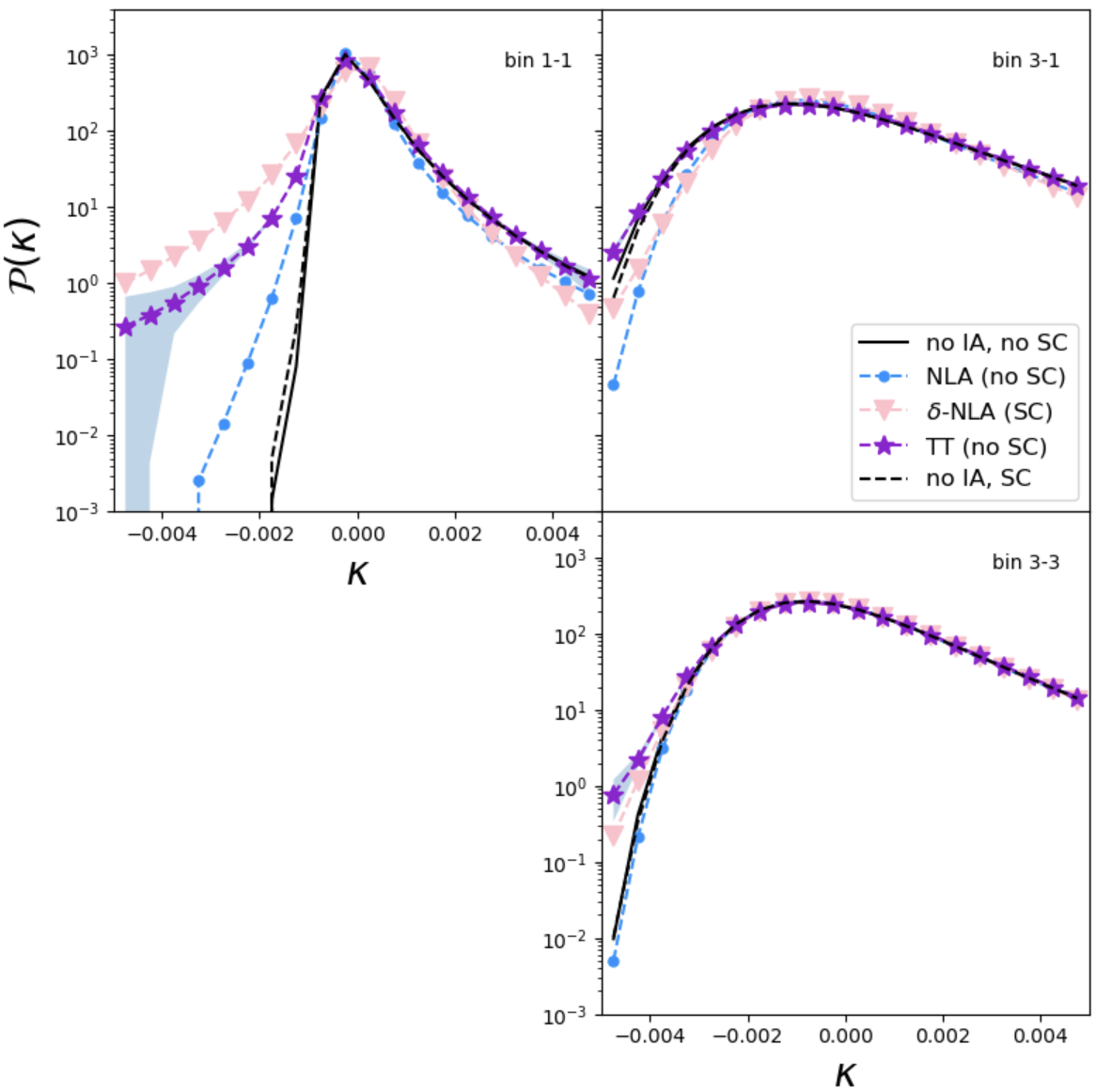}
\includegraphics[width=3.1in]{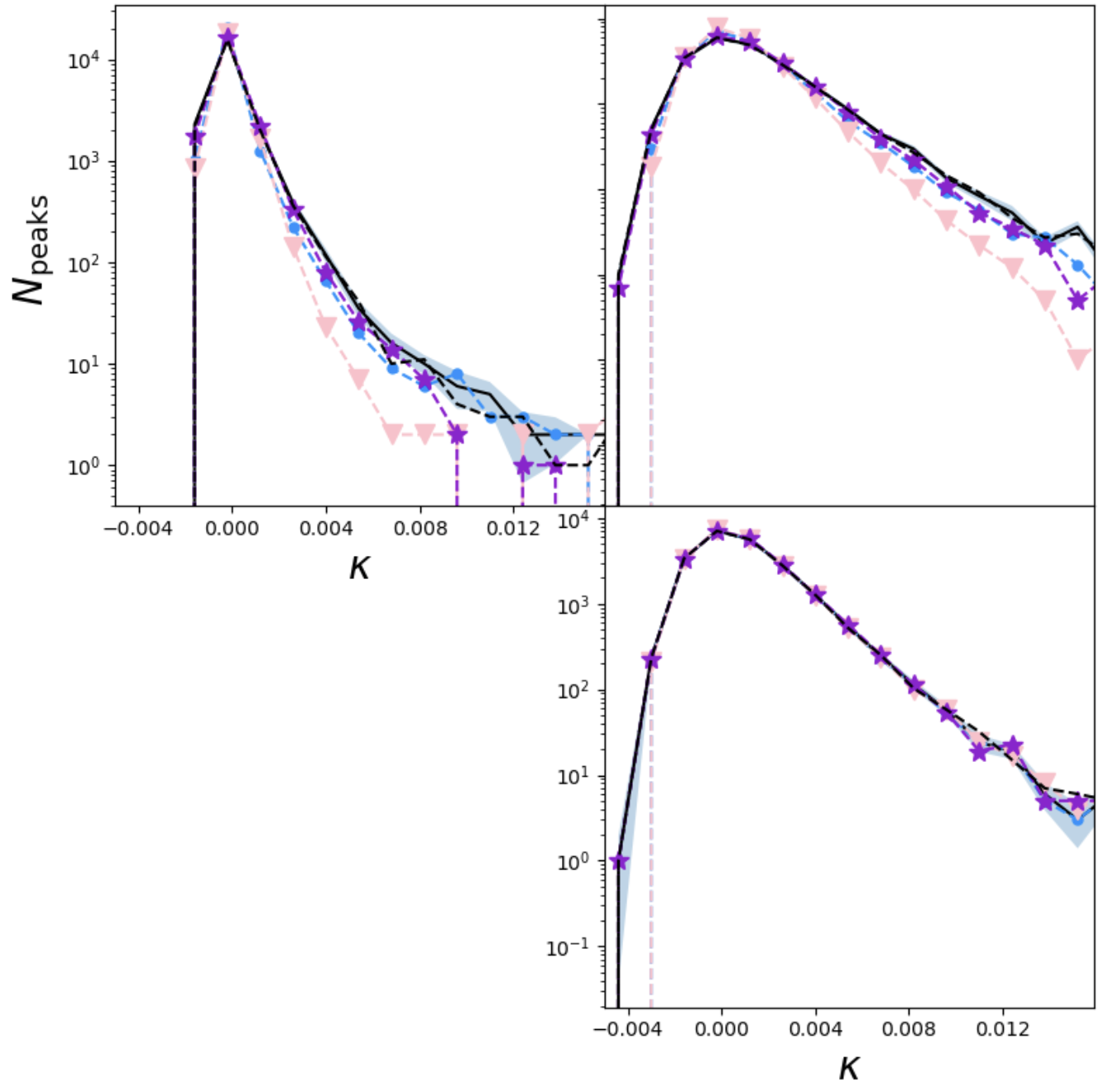}
\includegraphics[width=3.1in]{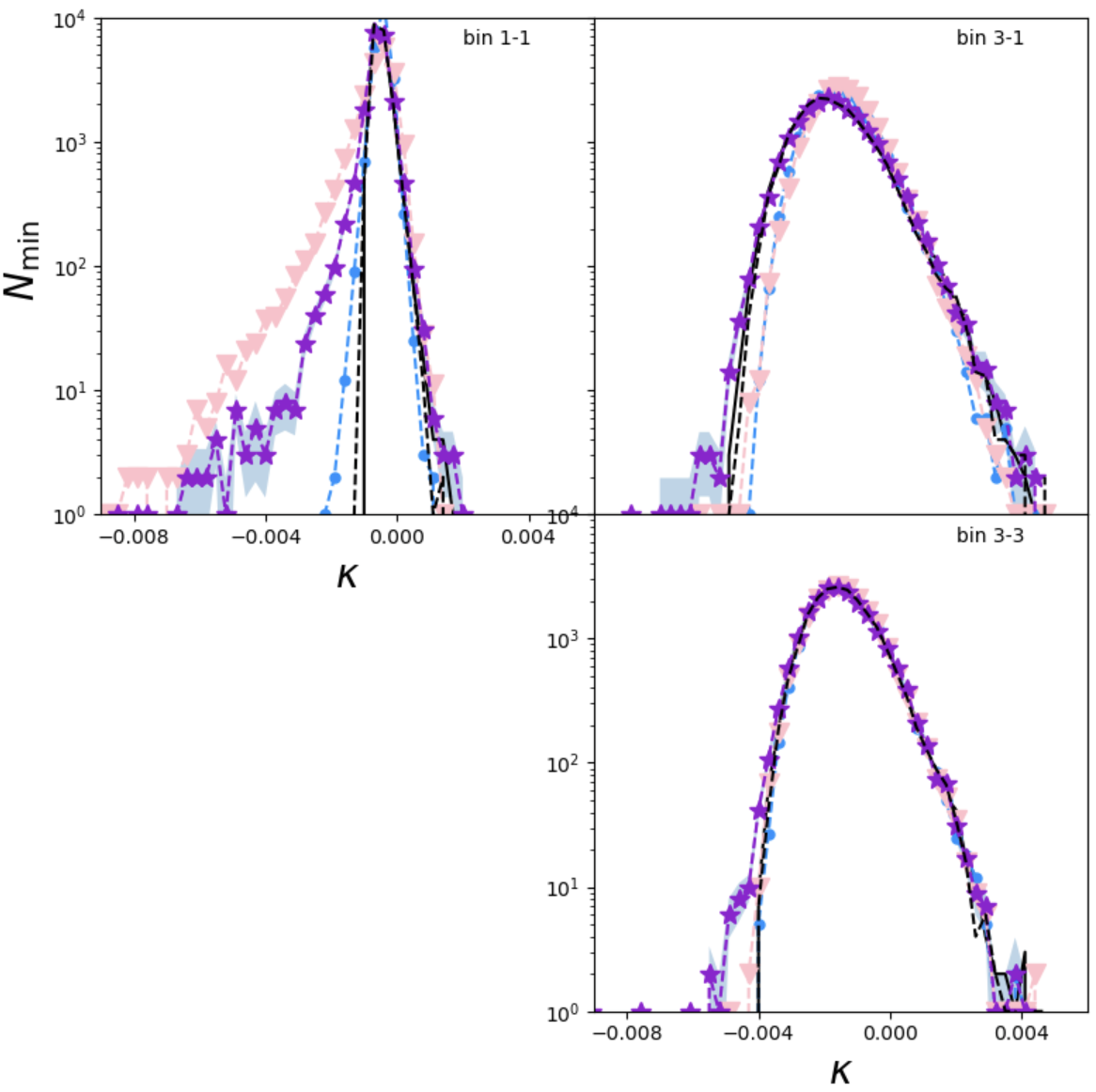}
\includegraphics[width=3.1in]{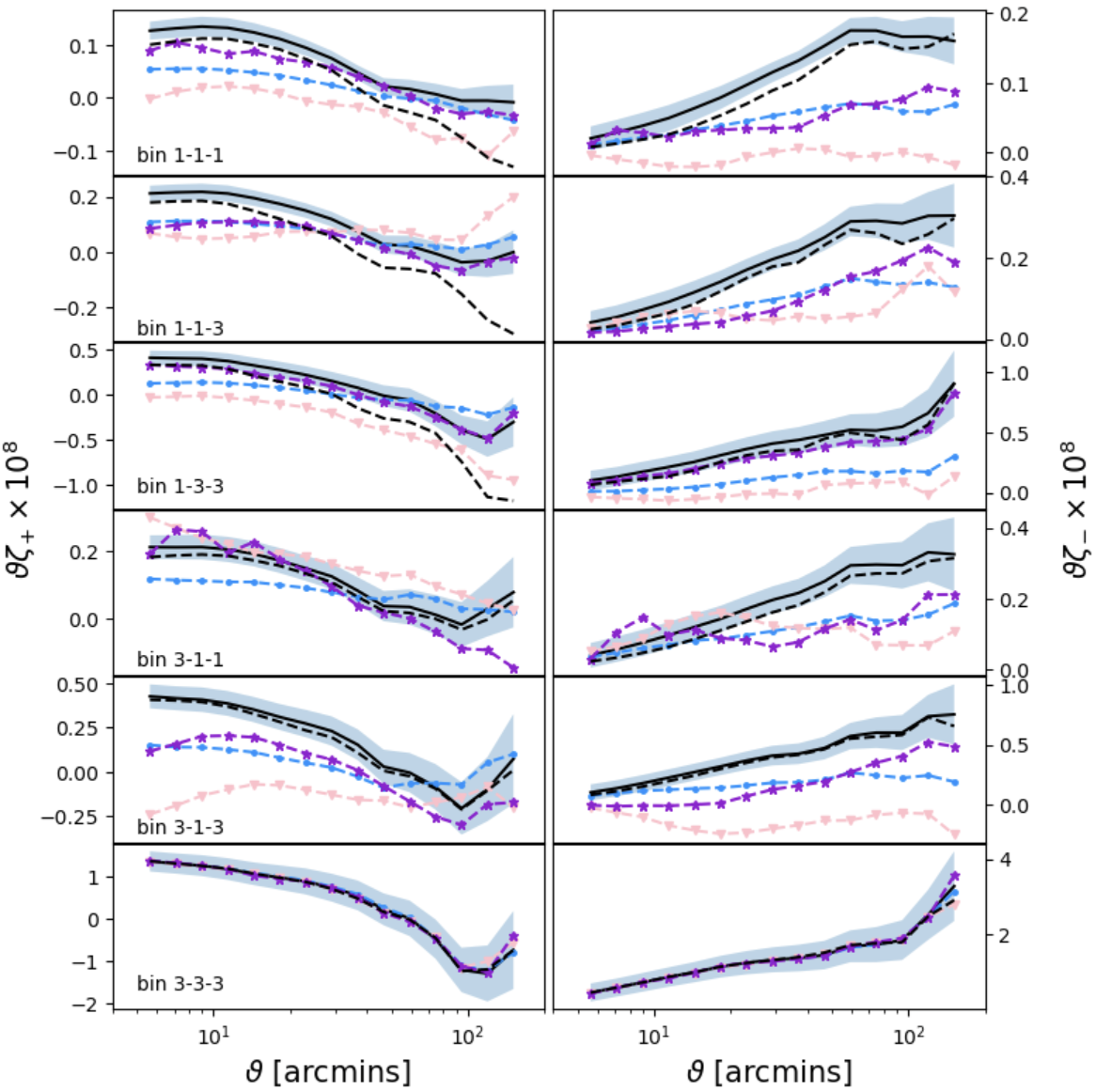}
\caption{ 
Same as Fig. \ref{fig:Map3_IA}, but for the other probes described in Sec. \ref{subsec:beyond-2pt}, namely the lensing PDF statistics ({\it upper left}), the peak count statistics ({\it upper right}), the minima count statistics ({\it lower left}) and the integrated $\gamma$-3PCFs ({\it lower right}), the latter being further split into $\zeta_+$ ({\it left sub-panels}) and $\zeta_-$ ({\it right sub-panels}). 
Shaded regions show the 1$\sigma$ error estimated either from jack-knife resampling the lensing maps (for the first three probes, see main text for details) or from the scatter between the different aperture regions when computing $\langle \zeta_\pm \rangle$, for the integrated $\gamma$-3PCF.   
For the lensing PDF and $N_{\rm min}$, the error bars are computed from the TT model to better highlight the distinguishing power in the negative $\kappa$ tail. 
The combinations of redshift bins used in the calculations are indicated in the panels; for the integrated $\gamma$-3PCF, the first index represents the bin of the aperture mass, and the last two indices are used to compute the local $\gamma$-2PCF within the patches  \citep[see Sec.\ref{subsec:beyond-2pt} and ][for more details]{Halder2021, Gong2023}.  This is for noise-free maps and catalogues; shape noise generally dilute the relative importance of IA, as discussed in Appendix \ref{app:figs}.} 
\label{fig:PDF_IA}
\end{figure*}
%
%

\begin{figure}
\includegraphics[width=3.3in]{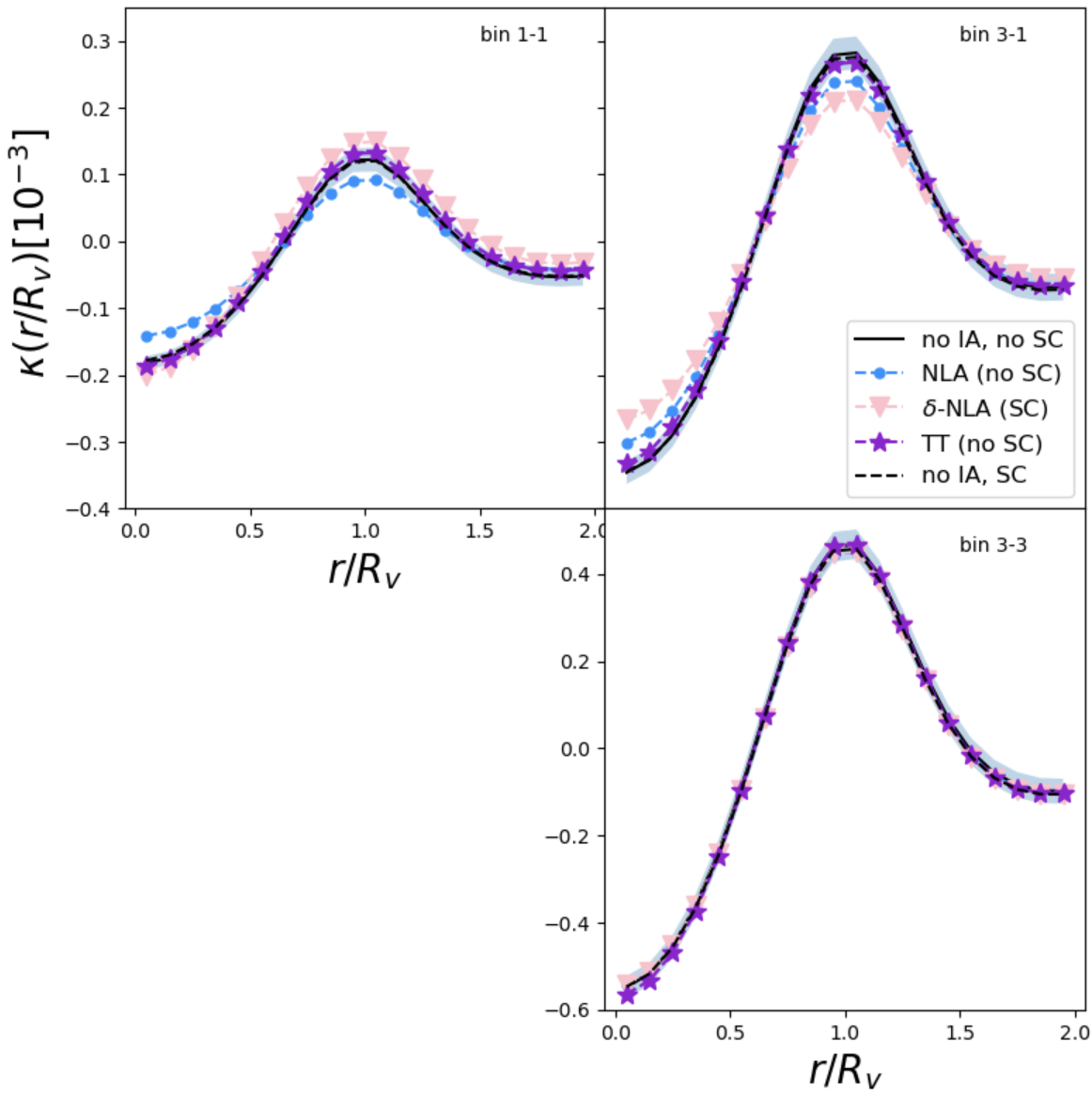}
\caption{Same as Fig. \ref{fig:Map3_IA}, but for the void profile statistics. The error bars are from jackknife resampling the void catalogues. }
\label{fig:voids}
\end{figure}


This section presents the impact of different IA models listed in Table \ref{table:IAmodels} on the measurements from higher-order weak lensing statistics introduced in Sec. \ref{subsec:beyond-2pt}. We compare as well the results between the mock galaxies that linearly trace the underlying dak matter and those populated with uniform random positions to study the impact of source clustering (SC) on these statistics.  The latter affect lensing statistics whenever clustered source galaxies at the low-redshift end of a broad $n(z)$ distribution act as lenses for the source galaxies at the high-redshift end of the same distribution, creating an uneven sampling that preferentially populates regions of high density. SC is especially severe when the $n^i(z)$ overlaps with the lensing kernel $q^i(z)$, which in our case is maximal for the first redshift bin. Moreover, as shown in \citet{DESY3_Gatti_source_clust}, these high-density regions also tend to have lower shape noise level per area, further impacting non-Gaussian statistics such as the lensing PDF. Decoupling IA and SC is not always possible, since for example they both contribute to the $\delta$-NLA and the two HOD-based models. 

We first establish in this section the relative importance of these effects, which greatly varies across redshift, and then offer a physical interpretation for these observations, when possible. We choose a Gaussian smoothing of 5 arcmin for the map-based estimators, and use the noise-free mocks to better highlight the impact on the cosmological signal.

\subsubsection*{$M_{\rm ap}^3$}

Fig. \ref{fig:Map3_IA} shows the relative bias caused by IA and SC on the $M_{\rm ap}^3$ statistics for the redshift bins combinations (1-1-1, 1-1-3, 1-3-3 and 3-3-3). The other combinations show similar features, where we can see that the impact of both IA and SC is the largest when including  the lowest redshift bin: SC alters $M_{\rm ap}^3$ by about 10\%, while it reduces to a few percent effect at higher redshifts. The NLA and TT affect the statistics at the level of 50\% in measurements that include the lowest redshift bin, and below 5\% elsewhere. 
Note that the cross-bins are heavily affected by IA, containing terms like $GGI$ and $GII$. The $\delta$-NLA model has the largest impact at low redshift as well, especially for the lowest smoothing scales, where the cosmological signal is completely suppressed by IA. This is consistent with what we observed in two-point statistics, where the same model had a very large $II$ term at low redshift. The difference is that for two-points the negative IA signal becomes squared in the $II$ term and therefore adds to the $GG$ term in auto-tomographic configurations, whereas in this case, the $III$ term has an odd power of $I$, which is negative and here removes power from the $GG$ signal. The cross-redshift bins in this figure contains different combinations of terms such as $G^1G^1G^1$, $G^1I^1I^3$, $G^1I^1I^3$ and $G^3I^1I^3$, which can be either positive or negative, depending on their relative amplitudes. In this figure, the error bars are obtained from the analytical calculation from \citet{Linke_Map3Cov}, which includes the Gaussian and non-Gaussian terms, up to sixth order in the lensing field.

\subsubsection*{$\kappa$-PDF}

Fig. \ref{fig:PDF_IA}  shows similar calculations, this time for the lensing PDF. In this case, the error bars are estimated from jackknife resampling the lensing {\sc healpix} maps in 28 regions, defined by splitting the map in large pixels of NSIDE=4. 
The SC has the largest impact on the negative tail of the PDF, reaching a 10-50\% suppression in the $\sigma \lesssim -1.5$  region (where $\sigma$ here approximates the PDF as a Gaussian). This is explained by the fact that there are fewer galaxies in under-dense regions when they linearly trace the dark matter distribution than when galaxy positions are sampled at random. As a result, the negative $\kappa$ pixels are under-sampled in the former case, leading to the observed suppression.
Worth highlighting is the fact that the NLA and TT signals are more different than for the $M_{\rm ap}^3$ statistics, suggesting that the lensing PDF is better suited for differentiating these and constraining the IA sector. Once again, the largest impact comes from the $\delta$-NLA model, at low redshift, approaching 100\% suppression of the cosmic shear signal over some of the $\kappa$-bins considered. Otherwise the global shape of the residuals closely match those found in \citet{Tidalator2D}: the IA tends to suppress the wings of the PDF, increasing instead the area of the maps with $\kappa\approx0$. 

Note that these measurements are carried out on noise-free maps; including shape noise effectively widen the variance of these distributions \citep{LensingPDF}, significantly diluting the impact. For a galaxy density of a few per arcmin$^2$, resembling current Stage-III surveys, we observe that the fractional impact is reduced  approximately tenfold, leading to up biases $\lesssim
10\%$, which are still larger than the statistical error bars.  This is further discussed in Appendix \ref{app:figs}.


\subsubsection*{Peak counts}

The top right part of Fig. \ref{fig:PDF_IA} presents results from the peak statistics, where the SC term plays a negligible role, consistent with what was found in \citet{DESY3_Gatti_source_clust}. The IA models tend to lower the number of strong negative and positive peaks in favour of those with $\mathcal{S/N}\sim 0$ in cross-redshift bins, with little impact on the auto-tomographic bins except at the lowest redshift, consistent with the previous measurements by \citet{Tidalator}. As for the other statistics, the $\delta$-NLA model is the strongest, due to the $\delta$-weighting. The TT model is hardly distinguishable from the NLA, except perhaps for a slightly shallower slope in bin 3-1. This likely yields strong degeneracies between these two IA models, and potentially with cosmological parameters. Generally, the value of $S_8$ inferred from peak statistics is not degenerate with $A_{\rm IA}$ \citep{KiDS1000_JHD}, in contrast with $\xi_{\pm}$ where this degeneracy is severe. We leave a full investigation of model differentiation for future work. We also show in Appendix \ref{app:figs} the results from noisy measurements, where the impact of SC and IA is almost completely lost. This is an advantage of two-point function, since the IA contamination has almost no effect on the data vector.

\subsubsection*{Minima counts}

We show in the bottom left part of Fig. \ref{fig:PDF_IA} the results on the minima statistics, which exhibits one of the largest distinguishing power of all probes considered here, in its negative $\kappa$-tails. The most under-dense regions are an excellent laboratory to test for different IA models, notably for their sensitivity to a special regime of the tidal fields, away from the densest region.  In fact different models could behave wildly differently there since the perturbations caused by IA are acting on very low densities;   any deviations will be strongly felt, which is exactly what we see. All terms here seem distinguishable (source clustering, NLA, $\delta$-NLA and TT).  This makes the minima one of the most useful probes to learn about the IA.  In the source plane, source clustering tends to move galaxies away from voids, and towards clustered regions.  Having a relatively low galaxy density to start with, turning on SC therefore has little impact on the lowest redshift bin. Including bin 3 brings enough galaxies to adequately sample the field, where we can now observe a reduction in the number of minima found in under-dense and over-dense regions when SC is turned on. This apparent Gaussianisation is also seen in the presence of shape noise, presented Appendix \ref{app:figs}. Turning on IA reduces the minima count in large-$\kappa$ regions, especially for bin 1-3, and boost the count in under-dense regions ($\kappa < 0.003$) by orders of magnitude. From this, it becomes clear that the tidal forces, while undoing some of the true lensing correlations, are at the same time producing a large number of smaller fluctuating structures in the void regions, which the minima count statistics naturally pick up. IA also affects our other under-density statistics, the lensing void profile, as we shall soon see.

\subsubsection*{Integrated $\gamma$-3PCF}

The bottom right part of Fig. \ref{fig:PDF_IA} shows the results for the integrated $\gamma$-3PCFs $\zeta_{\pm}$. The cross-redshift bins are defined as $\zeta_\pm^{ijk}(\theta) = \langle M^i_{\rm ap} \xi_\pm^{jk}(\theta) \rangle$, and presented for the six combinations involving bins 1 and 3: 1-1-1, 1-1-3, 1-3-3, 3-1-1, 3-1-3 and 3-3-3. The error bars here are estimated from the standard deviation measured from the scatter of the $\zeta_{\pm}$ estimates across the different apertures.

From the plot it can be seen that for IA the $\delta$-NLA model has the strongest impact, followed by NLA and then TT. The impact of these 3 IA models exceeds the sampling variance by more than $1\sigma$ for $\zeta_{\pm}$ in all panels that include the first redshift bin (and more severely for $\zeta_{-}$ which is more sensitive to small-scale modes relative to $\zeta_{+}$. These models have at times very similar signatures, hence we expect some degeneracies between the inferred values of $A_{\rm IA}$, $b_{\rm TA}$ and $C_2$ in an inference based on the TATT model. Higher redshifts are less affected by IA, with a relative impact of a few percent only,  due to the fact that the relative contribution to IA or SC from the non-linear density field is stronger for lower redshift bins, where at the same time the lensing kernel is weak. The impact of IA at low redshift is stronger than for many other probes investigated in this work and comparable to $M_{\rm ap}^3$, erasing most of the $\zeta_-$ signal. Physically this can be understood as the local measurements of $\xi_\pm$ statistics within apertures being affected by IA (see Fig. \ref{fig:xi_deltaNLA} for the impact of $\delta$-NLA to the $\gamma$-2PCFs), hence re-weighting these measurements by the local $M_{\rm ap}$ preserves the IA contribution coming from the $II$ and $GI$ terms. Moreover, the $M_{\rm ap}$ itself is contaminated both by IA and SC which together add a further impact. These observations are in line with those seen in the $M_{\rm ap}^3$ statistics, i.e. which contain the bispectrum terms $III$, $IIG$, and $IGG$ that contribute positively and negatively to the net signal depending on whether the power of $I$ is even or odd. 

Much like the IA, the SC term alone (indicated with the black dashed line) is discernible, although still within the 1$\sigma$ uncertainty, in the lowest redshift bin only, where the $n^1(z)$ that sources the $\delta$ field maximally overlaps with the corresponding lensing kernel $q^1(z)$, thus leading to a non-negligible SC contribution.

\subsubsection*{Lensing void profiles}

As described earlier, the positive and negative tails of the $\kappa$ distribution are suppressed in IA-contaminated maps, compared to the no-IA case (see Fig. \ref{fig:PDF_IA}, top left).  This was also seen in the context of the density split statistics in \citet[][see their figure A5]{KiDS1000_Burger}, where the amplitude of the stacked lensing profile around clusters and troughs were generally reduced by IA. In other words, lensing peaks appear lower, and lensing troughs shallower. The latter results are well recovered for the void profile statistics, when contaminated in the NLA, which leads to an overall flattening of the void lensing profile, as shown in Fig. \ref{fig:voids}. Clearly visible both in bins 1-1 and 3-1, the inner parts of the voids ($r/R_V<1.0$) are less negative than in the no-IA case, consistent with this picture. The ridges at $r/R_V=1.0$ are also lowered by IA, due to the lowered amplitude of peaks that form the void boundaries. As for the other non-Gaussian probes, this is not affecting the bin 3-3 measurements, where IA are significantly weaker. The $\delta$-NLA model shown in pink behaves similarly for bin 3-1, albeit with an even stronger impact, however the results in bin 1-1 differs significantly, exhibiting a higher ridge. This behavior occurs because voids are mostly enclosed by small peaks, as they are the most abundant. In the NLA model, the number of peaks with $\kappa=0$ is higher compared to the other models, so slightly negative peaks in the no-IA case gain amplitude when the $\delta$-NLA model is added, yielding the observed increase in ridge height. 
Finally, source clustering and IA from the TT model play a negligible role in all three panels. This is due to the fact that these models all yield roughly the same number of total peaks, which, as shown in \cite{Davies2018}, is a leading factor that sets the amplitude of the void lensing profile. 

\subsubsection*{Summary}

In each of these cases, neglecting IA and SC produces large biases, with the most serious effect seen on measurements including the lower redshift bin and for three-point functions ($M_{\rm ap}^3$ and integrated 3PCFs). Such effects will inevitably lead to potentially dangerous biases in cosmological inference analyses if unaccounted for, which we will investigate in the future. Our simulations can serve as a first step in this direction, providing a cosmology-independent forward modelling option. Finally, note that we focus here on the relative impact of IA on the data vectors, but our methods could easily be extended to compute the derivatives with respect to the different IA parameters $[A_{\rm IA}, b_{\rm TA}, C_2]$, which can then be used in Fisher forecasts or for forward-modelling the effect directly inside a likelihood analysis as in e.g. \citet{DESY1_Heydenreich, KiDS1000_Burger, KiDS1000_JHD}. 
%

In a typical beyond-2pt analysis, one disposes of a suite of simulated light-cones from which lensing statistics are measured, then emulated. This already involves computing convergence and shear maps, from which lensing catalogues are constructed. Including our flexible intrinsic alignment models here would mean a few additional steps:  (i) compute the tidal maps, for each mass shell, (ii) compute the quantities ($\delta$, $\epsilon^{\rm NLA}$ and $\epsilon^{\rm TT}$ for each simulated galaxy, and (iii) create mock data that sample $(A_{\rm IA}, b_{\rm TA},C_2)$ and infuse the corresponding  IA model with Eqs. (\ref{eq:tidal_th},\ref{eq:tidal_th_deltaNLA},  \ref{eq:tidal_th_TT}, and \ref{eq:eps_int}). This inevitably adds to dimensionality of the problem, however the most expensive parts -- running the $N$-body simulations, the light-cones, computing the shear maps and the tidal field maps -- need only to be done once. Variations in IA parameters could also be done outside the MCMC and emulated, as was done in i.e. \citet{DESY3_Zuercher}.
\section{Conclusions}
\label{sec:conclusion}

This paper describes a flexible method to infuse multiple intrinsic alignments models into simulated weak lensing catalogues. These are based on a linear or quadratic coupling between the galaxy intrinsic shapes and the projected tidal field, computed directly from projected mass maps, and thus are straight forward to implement in existing cosmic shear simulations where light-cone mass shells are saved to disk. The IA models are further refined by considering various tracer types, connecting mock galaxies and the underlying dark matter field,  either assuming no connection, a linear bias or an HOD-based non-linear bias. All six models (2 coupling $\times$ 3 tracer types) are studied with two-point statistics and compared to the commonly used NLA and TATT models, highlighting regimes where the agreements are at times excellent, and at times inconsistent. Discrepancies are expected as the TATT and our IA models are based on slightly different physical assumptions, notably concerning the non-linear galaxy bias model, the smoothing scale applied to the tidal fields, or the exclusion of the radial component of the tidal fields. In particular, the low-redshift $II$ term shows the largest disagreement for all models involving non-linear biased tracers,  overshooting the theory by tens to hundreds of percent at small-angle.

We quantify the impact of such differences with a full MCMC analysis, where we find that the cosmology is well recovered only for the NLA, $\delta$-NLA and HOD-NLA models, whereas all others, which involve quadratic tidal coupling, tend to push the inferred value of $S_8$ towards lower values, and $\Omega_{\rm m}$ towards higher values. The IA parameters are often degenerate with one another, and well recovered only for the NLA and HOD-NLA models. For the TT and $\delta$-NLA models, the IA parameters are 2$\sigma$ away from the input truth,  highlighting the subtle differences between the theoretical calculations and the numerical implementation of these models. Even with these limitations, all these models serve as good training sets to further our understanding of cosmic shear data analyses in which the true IA signal is not guaranteed to follow either the NLA or the TATT model. The type of cosmological biases shown here could be present in existing results reported in the literature, demanding additional tests like goodness-of-fit to determine the model(s) of choice.

Finally, we study the impact of these different IA models (as well as source coupling, SC) on non-Gaussian weak lensing statistics, including the lensing PDF, peak count, minima, integrated 3PCFs,  $M_{\rm ap}^3$  and void profiles. We find that both the coupling type (linear or quadratic) and the bias model (random, linear, or non-linear) have large effects, especially for tomographic combinations that include low-redshift galaxies. The $\delta$-NLA model has been observed to have the largest impact on most probes, by up to a factor of 2. 
Overall, the different IA models leave distinct signatures that should be differentiable with the upcoming data, in particular with probes sensitive to negative $\kappa$ values like the lensing PDF and the minima; a full demonstration of this will require an MCMC analysis based on these non-Gaussian probes, which we leave for future work.

Some caveats are worth mentioning here.  First, although validated with a full MCMC, the IA infusion models in this work were obtained at a fixed fiducial cosmology, and we expect the amplitude of the impact of both SC and IA to be cosmology-dependent. Applying our methods to mocks with varying cosmology is straightforward and will allow us to investigate the cosmology dependence of our findings. Second, our infusion method is limited by the fact that we are working with projected tidal fields, and some IA models including the tidal torque (TT) are sensitive to structures in the tidal field that we projected out along the radial line out sight direction, degrading the accuracy of these models at small scales. A natural extension of our technique would compute the $s_{ij}$ maps on thinner shells, or utilize the full 3D fields, to incorporate the components that we left out. Our results mostly focus on noise-free simulations, which exacerbate the impact of intrinsic alignment; adding shape noise is straightforward, and has been done in Appendix \ref{app:figs} for a relatively low galaxy density. Repeating this at scale will allow in the future to model exactly the statistical power of real data.

Future work will also include a split by color/morphology to better reflect the fact that different galaxy populations are affected differently by IA, redshift dependence of the IA parameters (e.g. $A_{\rm IA} \rightarrow A_{\rm IA}\left(\frac{1+z}{1+z_0}\right)^\eta$), as in e.g.  \citet{DESY3_Amon,DESY3_Secco}. Ideally, upgrades on direct IA measurements from spectroscopic surveys such as \citet{Johnston_IA} and \citet{IA_direct} could allow us to set priors on some of our model parameters, which should reflect in tighter cosmological constraints, but this is complicated by the complex selection of source galaxies, hence a full calibration is yet to be found.

The current work will be key for upcoming non-Gaussian cosmic shear analyses, where the choice of IA model used will impact the inferred cosmology. Other sources of uncertainty including baryons and photo-$z$ errors also need to be considered, and these are typically degenerate with some of the IA parameters \citep{Leonard2024}, which therefore must be jointly modelled. This can be achieved by combining the techniques presented in this paper with simulations that include baryon feedback, either from fully hydro-dynamical calculations \citep{kappaMTNG, MTNG-IA} or from baryonified $N$-body simulations \citep{Baryonification2}. In all cases, it is highly likely that our baryonic and IA models will remain approximations to the real physical models. 
As such, statistical tools such as empirical model selection \citep{Campos_Samuroff_IA} or Bayesian model averaging \citep{Grandon2024} should be employed to protect the inference from IA model mis-specifications.
\section*{Acknowledgements}

We would like to thank Avijit Bera, Andrew Hearin and Simon Samuroff for their thoughtful comments on the manuscript. This paper has undergone internal review in the LSST Dark Energy Science Collaboration. 
The internal reviewers were Daniela Grand\'on and Christos Georgiou, who challenged our methods and assumptions. 
All authors contributed to the development and writing of this paper. JHD led the analysis; NS ran the MCMC chains; LMV led testing of the IA infusion; JA measured most HOS from the different simulations; CTD produced the jacknife covariance matrices and many map-based measurement tools; PL infused the tidal field quantities on the HOD galaxy catalogue, while NvA helped with the interface; CMG developed the algorithm to select HOD galaxies that best match the analytical $N(z)$;  LC, AH, LL, JL, LP and CU developed many of the HOS measurement tools;  JB assisted in comparisons with analytic IA models; KH and SR produced the {\it OuterRim} products used in this paper.

JHD acknowledges support from an STFC Ernest Rutherford Fellowship (project reference ST/S004858/1). 
NS is supported in part by the OpenUniverse effort, which is funded by NASA under JPL Contract Task 70-711320, “Maximizing Science Exploitation of Simulated Cosmological Survey Data Across Surveys. 
JB and NvA are supported by NSF award AST-2206563, DOE grant DE-SC0024787, and the Roman Research and Support Participation program under NASA grant 80NSSC24K0088.
The work of KH and PL at Argonne National Laboratory was supported under the U.S. DOE contract DE-AC02-06CH11357.
This research used resources of the Argonne Leadership Computing Facility, which is a DOE Office of Science User Facility supported under Contract DE-AC02-06CH11357. 
LC and CU were supported by the STFC Astronomy Theory Consolidated Grant ST/W001020/1 from UK Research \& Innovation,  CU was also supported by the European Union (ERC StG, LSS\_BeyondAverage, 101075919). 
LL is supported by the Austrian Science Fund (FWF) [ESP 357-N]. 
CMG is funded by a Lady Bertha Jeffreys Studentship at Newcastle University. 
LP acknowledges support from the DLR grant 50QE2002.

We thank Mike Jarvis and Joe Zuntz for their excellent maintenance of the {\sc TreeCorr} \citep{treecorr_jarvis} correlation function measurement tool and {\sc CosmoSIS} \citep{Zuntz_2015_cosmosis} inference package, respectively. We are also grateful to Pierre Burger for sharing his code to sample linearly-biased tracers from mass maps.
We finally thank Elisa Chisari, Benjamin Joachimi and other members of the echo-IA task force for useful discussions on IA modelling in general.

The DESC acknowledges ongoing support from the Institut National de 
Physique Nucl\'eaire et de Physique des Particules in France; the 
Science \& Technology Facilities Council in the United Kingdom; and the
Department of Energy and the LSST Discovery Alliance
in the United States.  DESC uses resources of the IN2P3 
Computing Center (CC-IN2P3--Lyon/Villeurbanne - France) funded by the Centre National de la Recherche Scientifique; the National Energy Research Scientific Computing Center, a DOE Office of Science User 
Facility supported by the Office of Science of the U.S.\ Department of Energy under Contract No.\ DE-AC02-05CH11231; STFC DiRAC HPC Facilities, funded by UK BEIS National E-infrastructure capital grants; and the UK 
particle physics grid, supported by the GridPP Collaboration.  This work was performed in part under DOE Contract DE-AC02-76SF00515.

Some of the results in this paper have been derived using the {\sc healpy} and {\sc HEALPix} packages.
The authors acknowledge the developers and contributors of the open-source software packages that made this work possible, including but not limited to {\sc NumPy} \citep{numpy_harris2020array}, {\sc SciPy} \citep{2020SciPy-NMeth}, {\sc Astropy} \citep{The_Astropy_Collaboration_2022}, {\sc Matplotlib} \citep{Hunter:2007_matplotlib}, {\sc pyCCL} \citep{CCL}, and {\sc GetDist} \citep{GetDist}.
These tools were invaluable in performing the analysis and generating the results presented in this work.

\section*{Data Availability}

IA infusion codes will be made publicly available after acceptance for publication. Simulation products are DESC proprietary and can be made available upon request under certain conditions.


\bibliographystyle{mnras}
\bibliography{paper_IA} 

\begin{thebibliography}{}
\makeatletter
\relax
\def\mn@urlcharsother{\let\do\@makeother \do\$\do\&\do\#\do\^\do\_\do\%\do\~}
\def\mn@doi{\begingroup\mn@urlcharsother \@ifnextchar [ {\mn@doi@} {\mn@doi@[]}}
\def\mn@doi@[#1]#2{\def\@tempa{#1}\ifx\@tempa\@empty \href {http://dx.doi.org/#2} {doi:#2}\else \href {http://dx.doi.org/#2} {#1}\fi \endgroup}
\def\mn@eprint#1#2{\mn@eprint@#1:#2::\@nil}
\def\mn@eprint@arXiv#1{\href {http://arxiv.org/abs/#1} {{\tt arXiv:#1}}}
\def\mn@eprint@dblp#1{\href {http://dblp.uni-trier.de/rec/bibtex/#1.xml} {dblp:#1}}
\def\mn@eprint@#1:#2:#3:#4\@nil{\def\@tempa {#1}\def\@tempb {#2}\def\@tempc {#3}\ifx \@tempc \@empty \let \@tempc \@tempb \let \@tempb \@tempa \fi \ifx \@tempb \@empty \def\@tempb {arXiv}\fi \@ifundefined {mn@eprint@\@tempb}{\@tempb:\@tempc}{\expandafter \expandafter \csname mn@eprint@\@tempb\endcsname \expandafter{\@tempc}}}

\bibitem[\protect\citeauthoryear{{Alonso}, {Fabbian}, {Storey-Fisher}, {Eilers}, {Garc{\'\i}a-Garc{\'\i}a}, {Hogg}  \& {Rix}}{{Alonso} et~al.}{2023}]{QUAIA_Alonso2023}
{Alonso} D.,  {Fabbian} G.,  {Storey-Fisher} K.,  {Eilers} A.-C.,  {Garc{\'\i}a-Garc{\'\i}a} C.,  {Hogg} D.~W.,   {Rix} H.-W.,  2023, \mn@doi [arXiv e-prints] {10.48550/arXiv.2306.17748}, \href {https://ui.adsabs.harvard.edu/abs/2023arXiv230617748A} {p. arXiv:2306.17748}

\bibitem[\protect\citeauthoryear{{Amon} et~al.,}{{Amon} et~al.}{2022}]{DESY3_Amon}
{Amon} A.,  et~al., 2022, \mn@doi [\prd] {10.1103/PhysRevD.105.023514}, \href {https://ui.adsabs.harvard.edu/abs/2022PhRvD.105b3514A} {105, 023514}

\bibitem[\protect\citeauthoryear{{Angulo}, {Zennaro}, {Contreras}, {Aric{\`o}}, {Pellejero-Iba{\~n}ez}  \& {St{\"u}cker}}{{Angulo} et~al.}{2021}]{BACCOEmulator}
{Angulo} R.~E.,  {Zennaro} M.,  {Contreras} S.,  {Aric{\`o}} G.,  {Pellejero-Iba{\~n}ez} M.,   {St{\"u}cker} J.,  2021, \mn@doi [\mnras] {10.1093/mnras/stab2018}, \href {https://ui.adsabs.harvard.edu/abs/2021MNRAS.507.5869A} {507, 5869}

\bibitem[\protect\citeauthoryear{{Asgari} et~al.,}{{Asgari} et~al.}{2021}]{KiDS1000_Asgari}
{Asgari} M.,  et~al., 2021, \mn@doi [\aap] {10.1051/0004-6361/202039070}, \href {https://ui.adsabs.harvard.edu/abs/2021A&A...645A.104A} {645, A104}

\bibitem[\protect\citeauthoryear{{Barthelemy}, {Codis}, {Uhlemann}, {Bernardeau}  \& {Gavazzi}}{{Barthelemy} et~al.}{2020}]{LensingPDF_Nulling}
{Barthelemy} A.,  {Codis} S.,  {Uhlemann} C.,  {Bernardeau} F.,   {Gavazzi} R.,  2020, \mn@doi [\mnras] {10.1093/mnras/staa053}, \href {https://ui.adsabs.harvard.edu/abs/2020MNRAS.492.3420B} {492, 3420}

\bibitem[\protect\citeauthoryear{{Barthelemy}, {Halder}, {Gong}  \& {Uhlemann}}{{Barthelemy} et~al.}{2024}]{Barthelemy2024}
{Barthelemy} A.,  {Halder} A.,  {Gong} Z.,   {Uhlemann} C.,  2024, \mn@doi [\jcap] {10.1088/1475-7516/2024/03/060}, \href {https://ui.adsabs.harvard.edu/abs/2024JCAP...03..060B} {2024, 060}

\bibitem[\protect\citeauthoryear{{Bera}, {Medina Varela}, {Sooriyaarachchi}, {Ishak}, {Williams}  \& {The LSST Dark Energy Science Collaboration}}{{Bera} et~al.}{2025}]{Bera2025}
{Bera} A.,  {Medina Varela} L.,  {Sooriyaarachchi} V.,  {Ishak} M.,  {Williams} C.,   {The LSST Dark Energy Science Collaboration} 2025, \mn@doi [arXiv e-prints] {10.48550/arXiv.2503.24269}, \href {https://ui.adsabs.harvard.edu/abs/2025arXiv250324269B} {p. arXiv:2503.24269}

\bibitem[\protect\citeauthoryear{{Bernardeau}}{{Bernardeau}}{1998}]{SC_Bernardeau}
{Bernardeau} F.,  1998, \mn@doi [\aap] {10.48550/arXiv.astro-ph/9712115}, \href {https://ui.adsabs.harvard.edu/abs/1998A&A...338..375B} {338, 375}

\bibitem[\protect\citeauthoryear{{Blazek}, {Vlah}  \& {Seljak}}{{Blazek} et~al.}{2015}]{Blazek2015}
{Blazek} J.,  {Vlah} Z.,   {Seljak} U.,  2015, \mn@doi [\jcap] {10.1088/1475-7516/2015/08/015}, \href {https://ui.adsabs.harvard.edu/abs/2015JCAP...08..015B} {2015, 015}

\bibitem[\protect\citeauthoryear{{Blazek}, {MacCrann}, {Troxel}  \& {Fang}}{{Blazek} et~al.}{2019}]{TATT}
{Blazek} J.~A.,  {MacCrann} N.,  {Troxel} M.~A.,   {Fang} X.,  2019, \mn@doi [\prd] {10.1103/PhysRevD.100.103506}, \href {https://ui.adsabs.harvard.edu/abs/2019PhRvD.100j3506B} {100, 103506}

\bibitem[\protect\citeauthoryear{Boyle, Uhlemann, Friedrich, Barthelemy, Codis, Bernardeau, Giocoli  \& Baldi}{Boyle et~al.}{2021}]{LensingPDF_Cosmo}
Boyle A.,  Uhlemann C.,  Friedrich O.,  Barthelemy A.,  Codis S.,  Bernardeau F.,  Giocoli C.,   Baldi M.,  2021, \mn@doi [Mon. Not. Roy. Astron. Soc.] {10.1093/mnras/stab1381}, 505, 2886

\bibitem[\protect\citeauthoryear{{Bridle} \& {King}}{{Bridle} \& {King}}{2007}]{NLA}
{Bridle} S.,  {King} L.,  2007, \mn@doi [New Journal of Physics] {10.1088/1367-2630/9/12/444}, \href {http://adsabs.harvard.edu/abs/2007NJPh....9..444B} {9, 444}

\bibitem[\protect\citeauthoryear{{Brown}, {Taylor}, {Hambly}  \& {Dye}}{{Brown} et~al.}{2002}]{Brown2002}
{Brown} M.~L.,  {Taylor} A.~N.,  {Hambly} N.~C.,   {Dye} S.,  2002, \mn@doi [\mnras] {10.1046/j.1365-8711.2002.05354.x}, \href {https://ui.adsabs.harvard.edu/abs/2002MNRAS.333..501B} {333, 501}

\bibitem[\protect\citeauthoryear{{Burger} et~al.,}{{Burger} et~al.}{2022}]{KiDS1000_Burger}
{Burger} P.~A.,  et~al., 2022, arXiv e-prints, \href {https://ui.adsabs.harvard.edu/abs/2022arXiv220802171B} {p. arXiv:2208.02171}

\bibitem[\protect\citeauthoryear{{Burger} et~al.,}{{Burger} et~al.}{2023}]{KiDS1000_Map3}
{Burger} P.~A.,  et~al., 2023, \mn@doi [arXiv e-prints] {10.48550/arXiv.2309.08602}, \href {https://ui.adsabs.harvard.edu/abs/2023arXiv230908602B} {p. arXiv:2309.08602}

\bibitem[\protect\citeauthoryear{{Cacciato}, {van den Bosch}, {More}, {Mo}  \& {Yang}}{{Cacciato} et~al.}{2013}]{2013MNRAS.430..767C}
{Cacciato} M.,  {van den Bosch} F.~C.,  {More} S.,  {Mo} H.,   {Yang} X.,  2013, \mn@doi [\mnras] {10.1093/mnras/sts525}, \href {http://adsabs.harvard.edu/abs/2013MNRAS.430..767C} {430, 767}

\bibitem[\protect\citeauthoryear{{Campos}, {Samuroff}  \& {Mandelbaum}}{{Campos} et~al.}{2023}]{Campos_Samuroff_IA}
{Campos} A.,  {Samuroff} S.,   {Mandelbaum} R.,  2023, \mn@doi [\mnras] {10.1093/mnras/stad2213}, \href {https://ui.adsabs.harvard.edu/abs/2023MNRAS.525.1885C} {525, 1885}

\bibitem[\protect\citeauthoryear{{Castiblanco}, {Uhlemann}, {Harnois-D{\'e}raps}  \& {Barthelemy}}{{Castiblanco} et~al.}{2024}]{LensingPDF}
{Castiblanco} L.,  {Uhlemann} C.,  {Harnois-D{\'e}raps} J.,   {Barthelemy} A.,  2024, \mn@doi [arXiv e-prints] {10.48550/arXiv.2405.09651}, \href {https://ui.adsabs.harvard.edu/abs/2024arXiv240509651C} {p. arXiv:2405.09651}

\bibitem[\protect\citeauthoryear{{Catelan}, {Kamionkowski}  \& {Blandford}}{{Catelan} et~al.}{2001}]{Catelan_IA_Tidal}
{Catelan} P.,  {Kamionkowski} M.,   {Blandford} R.~D.,  2001, \mn@doi [\mnras] {10.1046/j.1365-8711.2001.04105.x}, \href {https://ui.adsabs.harvard.edu/abs/2001MNRAS.320L...7C} {320, L7}

\bibitem[\protect\citeauthoryear{{Chiang}, {Coles}  \& {Naselsky}}{{Chiang} et~al.}{2002}]{2002MNRAS.337..488C}
{Chiang} L.-Y.,  {Coles} P.,   {Naselsky} P.,  2002, \mn@doi [\mnras] {10.1046/j.1365-8711.2002.05931.x}, \href {http://adsabs.harvard.edu/abs/2002MNRAS.337..488C} {337, 488}

\bibitem[\protect\citeauthoryear{{Chisari} et~al.,}{{Chisari} et~al.}{2018}]{HorizonAGN}
{Chisari} N.~E.,  et~al., 2018, \mn@doi [\mnras] {10.1093/mnras/sty2093}, \href {https://ui.adsabs.harvard.edu/abs/2018MNRAS.480.3962C} {480, 3962}

\bibitem[\protect\citeauthoryear{{Chisari} et~al.,}{{Chisari} et~al.}{2019}]{CCL}
{Chisari} N.~E.,  et~al., 2019, \mn@doi [\apjs] {10.3847/1538-4365/ab1658}, \href {https://ui.adsabs.harvard.edu/abs/2019ApJS..242....2C} {242, 2}

\bibitem[\protect\citeauthoryear{Collaboration et~al.,}{Collaboration et~al.}{2022a}]{DESY3_3x2}
Collaboration D.,  et~al., 2022a, \mn@doi [\prd] {10.1103/PhysRevD.105.023520}, \href {https://ui.adsabs.harvard.edu/abs/2022PhRvD.105b3520A} {105, 023520}

\bibitem[\protect\citeauthoryear{Collaboration et~al.,}{Collaboration et~al.}{2022b}]{The_Astropy_Collaboration_2022}
Collaboration T.~A.,  et~al., 2022b, \mn@doi [The Astrophysical Journal] {10.3847/1538-4357/ac7c74}, 935, 167

\bibitem[\protect\citeauthoryear{Crittenden, Natarajan, Pen  \& Theuns}{Crittenden et~al.}{2002}]{Crittenden2002}
Crittenden R.~G.,  Natarajan P.,  Pen U.-L.,   Theuns T.,  2002, \mn@doi [The Astrophysical Journal] {10.1086/338838}, 568, 20

\bibitem[\protect\citeauthoryear{{Dalal} et~al.,}{{Dalal} et~al.}{2023}]{HSCY3_Cl}
{Dalal} R.,  et~al., 2023, \mn@doi [arXiv e-prints] {10.48550/arXiv.2304.00701}, \href {https://ui.adsabs.harvard.edu/abs/2023arXiv230400701D} {p. arXiv:2304.00701}

\bibitem[\protect\citeauthoryear{{Dark Energy Survey Collaboration} et~al.,}{{Dark Energy Survey Collaboration} et~al.}{2023}]{DESY3-KiDS1000}
{Dark Energy Survey Collaboration} et~al., 2023, \mn@doi [arXiv e-prints] {10.48550/arXiv.2305.17173}, \href {https://ui.adsabs.harvard.edu/abs/2023arXiv230517173E} {p. arXiv:2305.17173}

\bibitem[\protect\citeauthoryear{{Davies}, {Cautun}  \& {Li}}{{Davies} et~al.}{2018}]{Davies2018}
{Davies} C.~T.,  {Cautun} M.,   {Li} B.,  2018, \mn@doi [\mnras] {10.1093/mnrasl/sly135}, \href {https://ui.adsabs.harvard.edu/abs/2018MNRAS.480L.101D} {480, L101}

\bibitem[\protect\citeauthoryear{{Davies}, {Paillas}, {Cautun}  \& {Li}}{{Davies} et~al.}{2021}]{Davies2021}
{Davies} C.~T.,  {Paillas} E.,  {Cautun} M.,   {Li} B.,  2021, \mn@doi [\mnras] {10.1093/mnras/staa3262}, \href {https://ui.adsabs.harvard.edu/abs/2021MNRAS.500.2417D} {500, 2417}

\bibitem[\protect\citeauthoryear{{Davies}, {Cautun}, {Giblin}, {Li}, {Harnois-D{\'e}raps}  \& {Cai}}{{Davies} et~al.}{2022}]{Davies2022}
{Davies} C.~T.,  {Cautun} M.,  {Giblin} B.,  {Li} B.,  {Harnois-D{\'e}raps} J.,   {Cai} Y.-C.,  2022, \mn@doi [\mnras] {10.1093/mnras/stac1204}, \href {https://ui.adsabs.harvard.edu/abs/2022MNRAS.513.4729D} {513, 4729}

\bibitem[\protect\citeauthoryear{{Delgado} et~al.,}{{Delgado} et~al.}{2023}]{MTNG-IA}
{Delgado} A.~M.,  et~al., 2023, \mn@doi [\mnras] {10.1093/mnras/stad1781}, \href {https://ui.adsabs.harvard.edu/abs/2023MNRAS.523.5899D} {523, 5899}

\bibitem[\protect\citeauthoryear{{Euclid Collaboration: Ajani} et~al.,}{{Euclid Collaboration: Ajani} et~al.}{2023}]{HOWLS_paper1}
{Euclid Collaboration: Ajani} V.,  et~al., 2023, \mn@doi [\aap] {10.1051/0004-6361/202346017}, \href {https://ui.adsabs.harvard.edu/abs/2023A&A...675A.120E} {675, A120}

\bibitem[\protect\citeauthoryear{{Euclid Collaboration: Knabenhans} et~al.,}{{Euclid Collaboration: Knabenhans} et~al.}{2019}]{EuclidEmulator}
{Euclid Collaboration: Knabenhans} M.,  et~al., 2019, \mn@doi [\mnras] {10.1093/mnras/stz197}, \href {https://ui.adsabs.harvard.edu/abs/2019MNRAS.484.5509E} {484, 5509}

\bibitem[\protect\citeauthoryear{{Ferlito} et~al.,}{{Ferlito} et~al.}{2023}]{kappaMTNG}
{Ferlito} F.,  et~al., 2023, \mn@doi [\mnras] {10.1093/mnras/stad2205}, \href {https://ui.adsabs.harvard.edu/abs/2023MNRAS.524.5591F} {524, 5591}

\bibitem[\protect\citeauthoryear{{Feroz}, {Hobson}  \& {Bridges}}{{Feroz} et~al.}{2009}]{Multinest}
{Feroz} F.,  {Hobson} M.~P.,   {Bridges} M.,  2009, \mn@doi [\mnras] {10.1111/j.1365-2966.2009.14548.x}, \href {https://ui.adsabs.harvard.edu/abs/2009MNRAS.398.1601F} {398, 1601}

\bibitem[\protect\citeauthoryear{{Fluri}, {Kacprzak}, {Lucchi}, {Refregier}, {Amara}, {Hofmann}  \& {Schneider}}{{Fluri} et~al.}{2019}]{Fluri2019}
{Fluri} J.,  {Kacprzak} T.,  {Lucchi} A.,  {Refregier} A.,  {Amara} A.,  {Hofmann} T.,   {Schneider} A.,  2019, \mn@doi [\prd] {10.1103/PhysRevD.100.063514}, \href {https://ui.adsabs.harvard.edu/abs/2019PhRvD.100f3514F} {100, 063514}

\bibitem[\protect\citeauthoryear{{Fortuna}, {Hoekstra}, {Joachimi}, {Johnston}, {Chisari}, {Georgiou}  \& {Mahony}}{{Fortuna} et~al.}{2020}]{Fortuna2020}
{Fortuna} M.~C.,  {Hoekstra} H.,  {Joachimi} B.,  {Johnston} H.,  {Chisari} N.~E.,  {Georgiou} C.,   {Mahony} C.,  2020, arXiv e-prints, \href {https://ui.adsabs.harvard.edu/abs/2020arXiv200302700F} {p. arXiv:2003.02700}

\bibitem[\protect\citeauthoryear{{Fu} et~al.,}{{Fu} et~al.}{2014}]{Fu2014}
{Fu} L.,  et~al., 2014, \mn@doi [\mnras] {10.1093/mnras/stu754}, \href {http://adsabs.harvard.edu/abs/2014MNRAS.441.2725F} {441, 2725}

\bibitem[\protect\citeauthoryear{{Gatti} et~al.,}{{Gatti} et~al.}{2020}]{Gatti20}
{Gatti} M.,  et~al., 2020, \mn@doi [\mnras] {10.1093/mnras/staa2680}, \href {https://ui.adsabs.harvard.edu/abs/2020MNRAS.498.4060G} {498, 4060}

\bibitem[\protect\citeauthoryear{{Gatti} et~al.,}{{Gatti} et~al.}{2023}]{DESY3_Gatti_source_clust}
{Gatti} M.,  et~al., 2023, \mn@doi [arXiv e-prints] {10.48550/arXiv.2307.13860}, \href {https://ui.adsabs.harvard.edu/abs/2023arXiv230713860G} {p. arXiv:2307.13860}

\bibitem[\protect\citeauthoryear{{Georgiou}, {Chisari}, {Bilicki}, {La Barbera}, {Napolitano}, {Roy}  \& {Tortora}}{{Georgiou} et~al.}{2025}]{Georgiou2025}
{Georgiou} C.,  {Chisari} N.~E.,  {Bilicki} M.,  {La Barbera} F.,  {Napolitano} N.~R.,  {Roy} N.,   {Tortora} C.,  2025, \mn@doi [arXiv e-prints] {10.48550/arXiv.2502.09452}, \href {https://ui.adsabs.harvard.edu/abs/2025arXiv250209452G} {p. arXiv:2502.09452}

\bibitem[\protect\citeauthoryear{{Giblin}, {Cai}  \& {Harnois-D{\'e}raps}}{{Giblin} et~al.}{2023}]{Giblin_PDF}
{Giblin} B.,  {Cai} Y.-C.,   {Harnois-D{\'e}raps} J.,  2023, \mn@doi [\mnras] {10.1093/mnras/stad230}, \href {https://ui.adsabs.harvard.edu/abs/2023MNRAS.520.1721G} {520, 1721}

\bibitem[\protect\citeauthoryear{{Gong}, {Halder}, {Barreira}, {Seitz}  \& {Friedrich}}{{Gong} et~al.}{2023}]{Gong2023}
{Gong} Z.,  {Halder} A.,  {Barreira} A.,  {Seitz} S.,   {Friedrich} O.,  2023, \mn@doi [\jcap] {10.1088/1475-7516/2023/07/040}, \href {https://ui.adsabs.harvard.edu/abs/2023JCAP...07..040G} {2023, 040}

\bibitem[\protect\citeauthoryear{{G{\'o}rski}, {Hivon}, {Banday}, {Wandelt}, {Hansen}, {Reinecke}  \& {Bartelmann}}{{G{\'o}rski} et~al.}{2005}]{healpix}
{G{\'o}rski} K.~M.,  {Hivon} E.,  {Banday} A.~J.,  {Wandelt} B.~D.,  {Hansen} F.~K.,  {Reinecke} M.,   {Bartelmann} M.,  2005, \mn@doi [\apj] {10.1086/427976}, \href {http://adsabs.harvard.edu/abs/2005ApJ...622..759G} {622, 759}

\bibitem[\protect\citeauthoryear{{Grand{\'o}n} \& {Sellentin}}{{Grand{\'o}n} \& {Sellentin}}{2024}]{Grandon2024}
{Grand{\'o}n} D.,  {Sellentin} E.,  2024, \mn@doi [arXiv e-prints] {10.48550/arXiv.2407.20448}, \href {https://ui.adsabs.harvard.edu/abs/2024arXiv240720448G} {p. arXiv:2407.20448}

\bibitem[\protect\citeauthoryear{{Gruen} et~al.,}{{Gruen} et~al.}{2018}]{Gruen2017}
{Gruen} D.,  et~al., 2018, \mn@doi [\prd] {10.1103/PhysRevD.98.023507}, \href {https://ui.adsabs.harvard.edu/abs/2018PhRvD..98b3507G} {98, 023507}

\bibitem[\protect\citeauthoryear{{Halder} \& {Barreira}}{{Halder} \& {Barreira}}{2022}]{Halder2022}
{Halder} A.,  {Barreira} A.,  2022, \mn@doi [\mnras] {10.1093/mnras/stac2046}, \href {https://ui.adsabs.harvard.edu/abs/2022MNRAS.515.4639H} {515, 4639}

\bibitem[\protect\citeauthoryear{{Halder}, {Friedrich}, {Seitz}  \& {Varga}}{{Halder} et~al.}{2021}]{Halder2021}
{Halder} A.,  {Friedrich} O.,  {Seitz} S.,   {Varga} T.~N.,  2021, \mn@doi [\mnras] {10.1093/mnras/stab1801}, \href {https://ui.adsabs.harvard.edu/abs/2021MNRAS.506.2780H} {506, 2780}

\bibitem[\protect\citeauthoryear{{Halder}, {Gong}, {Barreira}, {Friedrich}, {Seitz}  \& {Gruen}}{{Halder} et~al.}{2023}]{Halder2023}
{Halder} A.,  {Gong} Z.,  {Barreira} A.,  {Friedrich} O.,  {Seitz} S.,   {Gruen} D.,  2023, \mn@doi [\jcap] {10.1088/1475-7516/2023/10/028}, \href {https://ui.adsabs.harvard.edu/abs/2023JCAP...10..028H} {2023, 028}

\bibitem[\protect\citeauthoryear{{Harnois-D\'eraps}, {van Waerbeke}, {Viola}  \& {Heymans}}{{Harnois-D\'eraps} et~al.}{2015}]{HWVH15}
{Harnois-D\'eraps} J.,  {van Waerbeke} L.,  {Viola} M.,   {Heymans} C.,  2015, \mnras, 450, 1212

\bibitem[\protect\citeauthoryear{{Harnois-D{\'e}raps} et~al.,}{{Harnois-D{\'e}raps} et~al.}{2018}]{SLICS}
{Harnois-D{\'e}raps} J.,  et~al., 2018, \mn@doi [\mnras] {10.1093/mnras/sty2319}, \href {http://adsabs.harvard.edu/abs/2018MNRAS.481.1337H} {481, 1337}

\bibitem[\protect\citeauthoryear{{Harnois-D{\'e}raps}, {Martinet}  \& {Reischke}}{{Harnois-D{\'e}raps} et~al.}{2021a}]{Tidalator}
{Harnois-D{\'e}raps} J.,  {Martinet} N.,   {Reischke} R.,  2021a, \mn@doi [\mnras] {10.1093/mnras/stab3222}, \href {https://ui.adsabs.harvard.edu/abs/2021MNRAS.tmp.2970H} {}

\bibitem[\protect\citeauthoryear{{Harnois-D{\'e}raps}, {Martinet}, {Castro}, {Dolag}, {Giblin}, {Heymans}, {Hildebrandt}  \& {Xia}}{{Harnois-D{\'e}raps} et~al.}{2021b}]{HD21}
{Harnois-D{\'e}raps} J.,  {Martinet} N.,  {Castro} T.,  {Dolag} K.,  {Giblin} B.,  {Heymans} C.,  {Hildebrandt} H.,   {Xia} Q.,  2021b, \mn@doi [\mnras] {10.1093/mnras/stab1623}, \href {https://ui.adsabs.harvard.edu/abs/2021MNRAS.506.1623H} {506, 1623}

\bibitem[\protect\citeauthoryear{{Harnois-D{\'e}raps}, {Martinet}  \& {Reischke}}{{Harnois-D{\'e}raps} et~al.}{2022}]{Tidalator2D}
{Harnois-D{\'e}raps} J.,  {Martinet} N.,   {Reischke} R.,  2022, \mn@doi [\mnras] {10.1093/mnras/stab3222}, \href {https://ui.adsabs.harvard.edu/abs/2022MNRAS.509.3868H} {509, 3868}

\bibitem[\protect\citeauthoryear{{Harnois-D\'eraps} et~al.,}{{Harnois-D\'eraps} et~al.}{2024}]{KiDS1000_JHD}
{Harnois-D\'eraps} J.,  et~al., 2024, \mn@doi [arXiv e-prints] {10.48550/arXiv.2405.10312}, \href {https://ui.adsabs.harvard.edu/abs/2024arXiv240510312H} {p. arXiv:2405.10312}

\bibitem[\protect\citeauthoryear{Harris et~al.,}{Harris et~al.}{2020}]{numpy_harris2020array}
Harris C.~R.,  et~al., 2020, \mn@doi [Nature] {10.1038/s41586-020-2649-2}, 585, 357

\bibitem[\protect\citeauthoryear{{Heitmann} et~al.,}{{Heitmann} et~al.}{2019}]{OuterRim}
{Heitmann} K.,  et~al., 2019, \mn@doi [\apjs] {10.3847/1538-4365/ab4da1}, \href {https://ui.adsabs.harvard.edu/abs/2019ApJS..245...16H} {245, 16}

\bibitem[\protect\citeauthoryear{{Heydenreich}, {Br{\"u}ck}, {Burger}, {Harnois-D{\'e}raps}, {Unruh}, {Castro}, {Dolag}  \& {Martinet}}{{Heydenreich} et~al.}{2022}]{DESY1_Heydenreich}
{Heydenreich} S.,  {Br{\"u}ck} B.,  {Burger} P.,  {Harnois-D{\'e}raps} J.,  {Unruh} S.,  {Castro} T.,  {Dolag} K.,   {Martinet} N.,  2022, arXiv e-prints, \href {https://ui.adsabs.harvard.edu/abs/2022arXiv220411831H} {p. arXiv:2204.11831}

\bibitem[\protect\citeauthoryear{{Heydenreich}, {Linke}, {Burger}  \& {Schneider}}{{Heydenreich} et~al.}{2023}]{Heydenreichetal2023}
{Heydenreich} S.,  {Linke} L.,  {Burger} P.,   {Schneider} P.,  2023, \mn@doi [\aap] {10.1051/0004-6361/202244820}, \href {https://ui.adsabs.harvard.edu/abs/2023A&A...672A..44H} {672, A44}

\bibitem[\protect\citeauthoryear{{Heymans} et~al.,}{{Heymans} et~al.}{2020}]{KiDS1000_Heymans}
{Heymans} C.,  et~al., 2020, arXiv e-prints, \href {https://ui.adsabs.harvard.edu/abs/2020arXiv200715632H} {p. arXiv:2007.15632}

\bibitem[\protect\citeauthoryear{{Hilbert}, {Xu}, {Schneider}, {Springel}, {Vogelsberger}  \& {Hernquist}}{{Hilbert} et~al.}{2017}]{Hilbert_IA2017}
{Hilbert} S.,  {Xu} D.,  {Schneider} P.,  {Springel} V.,  {Vogelsberger} M.,   {Hernquist} L.,  2017, \mn@doi [\mnras] {10.1093/mnras/stx482}, \href {https://ui.adsabs.harvard.edu/abs/2017MNRAS.468..790H} {468, 790}

\bibitem[\protect\citeauthoryear{{Hirata} \& {Seljak}}{{Hirata} \& {Seljak}}{2004}]{Hirata2004}
{Hirata} C.~M.,  {Seljak} U.,  2004, \mn@doi [\prd] {10.1103/PhysRevD.70.063526}, \href {https://ui.adsabs.harvard.edu/abs/2004PhRvD..70f3526H} {70, 063526}

\bibitem[\protect\citeauthoryear{{Hoffmann} et~al.,}{{Hoffmann} et~al.}{2022}]{MICE_IA}
{Hoffmann} K.,  et~al., 2022, \mn@doi [\prd] {10.1103/PhysRevD.106.123510}, \href {https://ui.adsabs.harvard.edu/abs/2022PhRvD.106l3510H} {106, 123510}

\bibitem[\protect\citeauthoryear{{Huang} et~al.,}{{Huang} et~al.}{2020}]{2020arXiv200715026H}
{Huang} H.-J.,  et~al., 2020, arXiv e-prints, \href {https://ui.adsabs.harvard.edu/abs/2020arXiv200715026H} {p. arXiv:2007.15026}

\bibitem[\protect\citeauthoryear{Hunter}{Hunter}{2007}]{Hunter:2007_matplotlib}
Hunter J.~D.,  2007, \mn@doi [Computing in Science \& Engineering] {10.1109/MCSE.2007.55}, 9, 90

\bibitem[\protect\citeauthoryear{{Ivezi{\'c}} et~al.,}{{Ivezi{\'c}} et~al.}{2019}]{LSST-Design}
{Ivezi{\'c}} {\v{Z}}.,  et~al., 2019, \mn@doi [\apj] {10.3847/1538-4357/ab042c}, \href {https://ui.adsabs.harvard.edu/abs/2019ApJ...873..111I} {873, 111}

\bibitem[\protect\citeauthoryear{{Jagvaral}, {Lanusse}  \& {Mandelbaum}}{{Jagvaral} et~al.}{2024}]{IA_ML}
{Jagvaral} Y.,  {Lanusse} F.,   {Mandelbaum} R.,  2024, \mn@doi [arXiv e-prints] {10.48550/arXiv.2409.18761}, \href {https://ui.adsabs.harvard.edu/abs/2024arXiv240918761J} {p. arXiv:2409.18761}

\bibitem[\protect\citeauthoryear{{Jarvis}}{{Jarvis}}{2015}]{treecorr_jarvis}
{Jarvis} M.,  2015, {TreeCorr: Two-point correlation functions}, Astrophysics Source Code Library, record ascl:1508.007

\bibitem[\protect\citeauthoryear{{Jarvis}, {Bernstein}  \& {Jain}}{{Jarvis} et~al.}{2004}]{TreeCorr}
{Jarvis} M.,  {Bernstein} G.,   {Jain} B.,  2004, \mn@doi [\mnras] {10.1111/j.1365-2966.2004.07926.x}, \href {http://adsabs.harvard.edu/abs/2004MNRAS.352..338J} {352, 338}

\bibitem[\protect\citeauthoryear{{Joachimi}, {Semboloni}, {Hilbert}, {Bett}, {Hartlap}, {Hoekstra}  \& {Schneider}}{{Joachimi} et~al.}{2013}]{Joachimi_IA2013}
{Joachimi} B.,  {Semboloni} E.,  {Hilbert} S.,  {Bett} P.~E.,  {Hartlap} J.,  {Hoekstra} H.,   {Schneider} P.,  2013, \mn@doi [\mnras] {10.1093/mnras/stt1618}, \href {https://ui.adsabs.harvard.edu/abs/2013MNRAS.436..819J} {436, 819}

\bibitem[\protect\citeauthoryear{Joachimi et~al.,}{Joachimi et~al.}{2015}]{Joachimi_IA_review_2015}
Joachimi B.,  et~al., 2015, \mn@doi [Space Science Reviews] {10.1007/s11214-015-0177-4}, 193, 1–65

\bibitem[\protect\citeauthoryear{{Joachimi} et~al.,}{{Joachimi} et~al.}{2020}]{KiDS1000_Joachimi}
{Joachimi} B.,  et~al., 2020, arXiv e-prints, \href {https://ui.adsabs.harvard.edu/abs/2020arXiv200701844J} {p. arXiv:2007.01844}

\bibitem[\protect\citeauthoryear{{Johnston} et~al.,}{{Johnston} et~al.}{2019}]{Johnston_IA}
{Johnston} H.,  et~al., 2019, \mn@doi [\aap] {10.1051/0004-6361/201834714}, \href {https://ui.adsabs.harvard.edu/abs/2019A&A...624A..30J} {624, A30}

\bibitem[\protect\citeauthoryear{{Kaiser} \& {Squires}}{{Kaiser} \& {Squires}}{1993}]{KaiserSquires}
{Kaiser} N.,  {Squires} G.,  1993, \mn@doi [\apj] {10.1086/172297}, \href {http://adsabs.harvard.edu/abs/1993ApJ...404..441K} {404, 441}

\bibitem[\protect\citeauthoryear{Kiessling et~al.,}{Kiessling et~al.}{2015}]{Kiessling_IA_review_2015}
Kiessling A.,  et~al., 2015, \mn@doi [Space Science Reviews] {10.1007/s11214-015-0203-6}, 193, 67–136

\bibitem[\protect\citeauthoryear{{Kilbinger} et~al.,}{{Kilbinger} et~al.}{2017}]{Kilbinger17}
{Kilbinger} M.,  et~al., 2017, \mn@doi [\mnras] {10.1093/mnras/stx2082}, \href {https://ui.adsabs.harvard.edu/abs/2017MNRAS.472.2126K} {472, 2126}

\bibitem[\protect\citeauthoryear{{Kirk}, {Rassat}, {Host}  \& {Bridle}}{{Kirk} et~al.}{2012}]{Kirk2012}
{Kirk} D.,  {Rassat} A.,  {Host} O.,   {Bridle} S.,  2012, \mn@doi [\mnras] {10.1111/j.1365-2966.2012.21099.x}, \href {https://ui.adsabs.harvard.edu/abs/2012MNRAS.424.1647K} {424, 1647}

\bibitem[\protect\citeauthoryear{Kirk et~al.,}{Kirk et~al.}{2015}]{Kirk_IA_review_2015}
Kirk D.,  et~al., 2015, \mn@doi [Space Science Reviews] {10.1007/s11214-015-0213-4}, 193, 139–211

\bibitem[\protect\citeauthoryear{{Korytov} et~al.,}{{Korytov} et~al.}{2019}]{cosmoDC2}
{Korytov} D.,  et~al., 2019, \mn@doi [\apjs] {10.3847/1538-4365/ab510c}, \href {https://ui.adsabs.harvard.edu/abs/2019ApJS..245...26K} {245, 26}

\bibitem[\protect\citeauthoryear{{Krause}, {Eifler}  \& {Blazek}}{{Krause} et~al.}{2016}]{Krause2016}
{Krause} E.,  {Eifler} T.,   {Blazek} J.,  2016, \mn@doi [\mnras] {10.1093/mnras/stv2615}, \href {https://ui.adsabs.harvard.edu/abs/2016MNRAS.456..207K} {456, 207}

\bibitem[\protect\citeauthoryear{Lamman, Tsaprazi, Shi, \v{S}ar\v{c}evi\'c, Pyne, Legnani  \& Ferreira}{Lamman et~al.}{2023}]{Lamman_IA_guide}
Lamman C.,  Tsaprazi E.,  Shi J.,  \v{S}ar\v{c}evi\'c N.~N.,  Pyne S.,  Legnani E.,   Ferreira T.,  2023, \mn@doi [arXiv e-prints] {10.21105/astro.2309.08605}

\bibitem[\protect\citeauthoryear{{Lanzieri}, {Lanusse}, {Modi}, {Horowitz}, {Harnois-D{\'e}raps}, {Starck}  \& {LSST Dark Energy Science Collaboration (LSST DESC)}}{{Lanzieri} et~al.}{2023}]{Lanzieri2023}
{Lanzieri} D.,  {Lanusse} F.,  {Modi} C.,  {Horowitz} B.,  {Harnois-D{\'e}raps} J.,  {Starck} J.-L.,   {LSST Dark Energy Science Collaboration (LSST DESC)} 2023, \mn@doi [\aap] {10.1051/0004-6361/202346888}, \href {https://ui.adsabs.harvard.edu/abs/2023A&A...679A..61L} {679, A61}

\bibitem[\protect\citeauthoryear{{Laureijs} et~al.,}{{Laureijs} et~al.}{2011}]{RedBook}
{Laureijs} R.,  et~al., 2011, \mn@doi [arXiv e-prints] {10.48550/arXiv.1110.3193}, \href {https://ui.adsabs.harvard.edu/abs/2011arXiv1110.3193L} {p. arXiv:1110.3193}

\bibitem[\protect\citeauthoryear{{Lee}, {Haiman}, {Pandey}  \& {Genel}}{{Lee} et~al.}{2025}]{Lee2025_TNG-IA}
{Lee} M.~E.,  {Haiman} Z.,  {Pandey} S.,   {Genel} S.,  2025, arXiv e-prints, \href {https://ui.adsabs.harvard.edu/abs/2025arXiv250412460L} {p. arXiv:2504.12460}

\bibitem[\protect\citeauthoryear{{Leonard}, {Rau}  \& {Mandelbaum}}{{Leonard} et~al.}{2024}]{Leonard2024}
{Leonard} C.~D.,  {Rau} M.~M.,   {Mandelbaum} R.,  2024, \mn@doi [\prd] {10.1103/PhysRevD.109.083528}, \href {https://ui.adsabs.harvard.edu/abs/2024PhRvD.109h3528L} {109, 083528}

\bibitem[\protect\citeauthoryear{{Lewis}}{{Lewis}}{2019}]{GetDist}
{Lewis} A.,  2019, \mn@doi [arXiv e-prints] {10.48550/arXiv.1910.13970}, \href {https://ui.adsabs.harvard.edu/abs/2019arXiv191013970L} {p. arXiv:1910.13970}

\bibitem[\protect\citeauthoryear{Li et~al.,}{Li et~al.}{2023a}]{HSCY3_2pcf}
Li X.,  et~al., 2023a, \mn@doi [Phys. Rev. D] {10.1103/PhysRevD.108.123518}, 108, 123518

\bibitem[\protect\citeauthoryear{Li et~al.,}{Li et~al.}{2023b}]{KiDS1000_Li}
Li S.-S.,  et~al., 2023b, \mn@doi [Astronomy &amp; Astrophysics] {10.1051/0004-6361/202347236}, 679, A133

\bibitem[\protect\citeauthoryear{{Linke}, {Heydenreich}, {Burger}  \& {Schneider}}{{Linke} et~al.}{2023}]{Linke_Map3Cov}
{Linke} L.,  {Heydenreich} S.,  {Burger} P.~A.,   {Schneider} P.,  2023, \mn@doi [\aap] {10.1051/0004-6361/202245652}, \href {https://ui.adsabs.harvard.edu/abs/2023A&A...672A.185L} {672, A185}

\bibitem[\protect\citeauthoryear{{Linke}, {Pyne}, {Joachimi}, {Georgiou}, {Hoffmann}, {Mandelbaum}  \& {Singh}}{{Linke} et~al.}{2024}]{Linke_IA_3pt}
{Linke} L.,  {Pyne} S.,  {Joachimi} B.,  {Georgiou} C.,  {Hoffmann} K.,  {Mandelbaum} R.,   {Singh} S.,  2024, \mn@doi [\aap] {10.1051/0004-6361/202451032}, \href {https://ui.adsabs.harvard.edu/abs/2024A&A...691A.312L} {691, A312}

\bibitem[\protect\citeauthoryear{{Linke} et~al.,}{{Linke} et~al.}{2025}]{Linke_SC}
{Linke} L.,  et~al., 2025, \mn@doi [\aap] {10.1051/0004-6361/202451494}, \href {https://ui.adsabs.harvard.edu/abs/2025A&A...693A.210L} {693, A210}

\bibitem[\protect\citeauthoryear{{Liu}, {Yuan}, {Pan}, {Zhang}, {Wang}  \& {Fan}}{{Liu} et~al.}{2023}]{HSCY1_Peaks_th}
{Liu} X.,  {Yuan} S.,  {Pan} C.,  {Zhang} T.,  {Wang} Q.,   {Fan} Z.,  2023, \mn@doi [\mnras] {10.1093/mnras/stac2971}, \href {https://ui.adsabs.harvard.edu/abs/2023MNRAS.519..594L} {519, 594}

\bibitem[\protect\citeauthoryear{{Mackey}, {White}  \& {Kamionkowski}}{{Mackey} et~al.}{2002}]{TT_V0}
{Mackey} J.,  {White} M.,   {Kamionkowski} M.,  2002, \mn@doi [\mnras] {10.1046/j.1365-8711.2002.05337.x}, \href {https://ui.adsabs.harvard.edu/abs/2002MNRAS.332..788M} {332, 788}

\bibitem[\protect\citeauthoryear{Mandelbaum et~al.,}{Mandelbaum et~al.}{2011}]{BlueIA}
Mandelbaum R.,  et~al., 2011, MNRAS, 410, 844

\bibitem[\protect\citeauthoryear{{Marques} et~al.,}{{Marques} et~al.}{2023}]{HSCY1_peaks_sims}
{Marques} G.~A.,  et~al., 2023, arXiv e-prints, \href {https://ui.adsabs.harvard.edu/abs/2023arXiv230810866M} {p. arXiv:2308.10866}

\bibitem[\protect\citeauthoryear{{Martinet}, {Harnois-D{\'e}raps}, {Jullo}  \& {Schneider}}{{Martinet} et~al.}{2020}]{Martinet20}
{Martinet} N.,  {Harnois-D{\'e}raps} J.,  {Jullo} E.,   {Schneider} P.,  2020, arXiv e-prints, \href {https://ui.adsabs.harvard.edu/abs/2020arXiv201007376M} {p. arXiv:2010.07376}

\bibitem[\protect\citeauthoryear{{Mead}, {Brieden}, {Tr{\"o}ster}  \& {Heymans}}{{Mead} et~al.}{2020}]{HMCode2020}
{Mead} A.,  {Brieden} S.,  {Tr{\"o}ster} T.,   {Heymans} C.,  2020, arXiv e-prints, \href {https://ui.adsabs.harvard.edu/abs/2020arXiv200901858M} {p. arXiv:2009.01858}

\bibitem[\protect\citeauthoryear{{Miyatake} et~al.,}{{Miyatake} et~al.}{2023}]{HSCY3_3x2}
{Miyatake} H.,  et~al., 2023, \mn@doi [\prd] {10.1103/PhysRevD.108.123517}, \href {https://ui.adsabs.harvard.edu/abs/2023PhRvD.108l3517M} {108, 123517}

\bibitem[\protect\citeauthoryear{{Moran} et~al.,}{{Moran} et~al.}{2023}]{miratitan_final}
{Moran} K.~R.,  et~al., 2023, \mn@doi [\mnras] {10.1093/mnras/stac3452}, \href {https://ui.adsabs.harvard.edu/abs/2023MNRAS.520.3443M} {520, 3443}

\bibitem[\protect\citeauthoryear{{Nicola} et~al.,}{{Nicola} et~al.}{2024}]{NonlinearBiasNicola}
{Nicola} A.,  et~al., 2024, \mn@doi [\jcap] {10.1088/1475-7516/2024/02/015}, \href {https://ui.adsabs.harvard.edu/abs/2024JCAP...02..015N} {2024, 015}

\bibitem[\protect\citeauthoryear{{Pandya}, {Yang}, {Van Alfen}, {Blazek}  \& {Walters}}{{Pandya} et~al.}{2025}]{Pandya2025}
{Pandya} S.,  {Yang} Y.,  {Van Alfen} N.,  {Blazek} J.,   {Walters} R.,  2025, \mn@doi [arXiv e-prints] {10.48550/arXiv.2504.05235}, \href {https://ui.adsabs.harvard.edu/abs/2025arXiv250405235P} {p. arXiv:2504.05235}

\bibitem[\protect\citeauthoryear{{Paopiamsap}, {Porqueres}, {Alonso}, {Harnois-D\'eraps}  \& {Leonard}}{{Paopiamsap} et~al.}{2023}]{Paopiamsap2024}
{Paopiamsap} A.,  {Porqueres} N.,  {Alonso} D.,  {Harnois-D\'eraps} J.,   {Leonard} C.~D.,  2023, \mn@doi [arXiv e-prints] {10.48550/arXiv.2311.16812}, \href {https://ui.adsabs.harvard.edu/abs/2023arXiv231116812P} {p. arXiv:2311.16812}

\bibitem[\protect\citeauthoryear{{Pedersen}, {Yao}, {Ishak}  \& {Zhang}}{{Pedersen} et~al.}{2020}]{SelCalibrationPedersen}
{Pedersen} E.~M.,  {Yao} J.,  {Ishak} M.,   {Zhang} P.,  2020, \mn@doi [\apjl] {10.3847/2041-8213/aba51b}, \href {https://ui.adsabs.harvard.edu/abs/2020ApJ...899L...5P} {899, L5}

\bibitem[\protect\citeauthoryear{{Prat} et~al.,}{{Prat} et~al.}{2023}]{TXPipe}
{Prat} J.,  et~al., 2023, \mn@doi [The Open Journal of Astrophysics] {10.21105/astro.2212.09345}, \href {https://ui.adsabs.harvard.edu/abs/2023OJAp....6E..13P} {6, 13}

\bibitem[\protect\citeauthoryear{{Samuroff} et~al.,}{{Samuroff} et~al.}{2019}]{DESY1_IA_Samuroff}
{Samuroff} S.,  et~al., 2019, \mn@doi [\mnras] {10.1093/mnras/stz2197}, \href {https://ui.adsabs.harvard.edu/abs/2019MNRAS.489.5453S} {489, 5453}

\bibitem[\protect\citeauthoryear{{Samuroff} et~al.,}{{Samuroff} et~al.}{2023}]{IA_direct}
{Samuroff} S.,  et~al., 2023, \mn@doi [\mnras] {10.1093/mnras/stad2013}, \href {https://ui.adsabs.harvard.edu/abs/2023MNRAS.524.2195S} {524, 2195}

\bibitem[\protect\citeauthoryear{{Schmidt}, {Chisari}  \& {Dvorkin}}{{Schmidt} et~al.}{2015}]{Schmidt2015}
{Schmidt} F.,  {Chisari} N.~E.,   {Dvorkin} C.,  2015, \mn@doi [\jcap] {10.1088/1475-7516/2015/10/032}, \href {https://ui.adsabs.harvard.edu/abs/2015JCAP...10..032S} {2015, 032}

\bibitem[\protect\citeauthoryear{{Schmitz}, {Hirata}, {Blazek}  \& {Krause}}{{Schmitz} et~al.}{2018}]{Schmitz18}
{Schmitz} D.~M.,  {Hirata} C.~M.,  {Blazek} J.,   {Krause} E.,  2018, \mn@doi [\jcap] {10.1088/1475-7516/2018/07/030}, \href {https://ui.adsabs.harvard.edu/abs/2018JCAP...07..030S} {2018, 030}

\bibitem[\protect\citeauthoryear{{Schneider}, {van Waerbeke}, {Jain}  \& {Kruse}}{{Schneider} et~al.}{1998}]{Schneider1998}
{Schneider} P.,  {van Waerbeke} L.,  {Jain} B.,   {Kruse} G.,  1998, \mnras, 296, 873

\bibitem[\protect\citeauthoryear{{Schneider}, {Teyssier}, {Stadel}, {Chisari}, {Le Brun}, {Amara}  \& {Refregier}}{{Schneider} et~al.}{2019}]{Baryonification2}
{Schneider} A.,  {Teyssier} R.,  {Stadel} J.,  {Chisari} N.~E.,  {Le Brun} A. M.~C.,  {Amara} A.,   {Refregier} A.,  2019, \mn@doi [\jcap] {10.1088/1475-7516/2019/03/020}, \href {https://ui.adsabs.harvard.edu/abs/2019JCAP...03..020S} {2019, 020}

\bibitem[\protect\citeauthoryear{{Secco} et~al.,}{{Secco} et~al.}{2022}]{DESY3_Secco}
{Secco} L.~F.,  et~al., 2022, \mn@doi [\prd] {10.1103/PhysRevD.105.023515}, \href {https://ui.adsabs.harvard.edu/abs/2022PhRvD.105b3515S} {105, 023515}

\bibitem[\protect\citeauthoryear{{Semboloni}, {Hoekstra}, {Schaye}, {van Daalen}  \& {McCarthy}}{{Semboloni} et~al.}{2011}]{Semboloni11}
{Semboloni} E.,  {Hoekstra} H.,  {Schaye} J.,  {van Daalen} M.~P.,   {McCarthy} I.~G.,  2011, \mn@doi [\mnras] {10.1111/j.1365-2966.2011.19385.x}, \href {http://adsabs.harvard.edu/abs/2011MNRAS.tmp.1461S} {pp 1461--+}

\bibitem[\protect\citeauthoryear{{Shan} et~al.,}{{Shan} et~al.}{2018}]{Shan18}
{Shan} H.,  et~al., 2018, \mn@doi [\mnras] {10.1093/mnras/stx2837}, \href {http://adsabs.harvard.edu/abs/2018MNRAS.474.1116S} {474, 1116}

\bibitem[\protect\citeauthoryear{{Sif{\'o}n} et~al.,}{{Sif{\'o}n} et~al.}{2015}]{2015MNRAS.454.3938S}
{Sif{\'o}n} C.,  et~al., 2015, \mn@doi [\mnras] {10.1093/mnras/stv2051}, \href {http://adsabs.harvard.edu/abs/2015MNRAS.454.3938S} {454, 3938}

\bibitem[\protect\citeauthoryear{{Singh}, {Mandelbaum}  \& {More}}{{Singh} et~al.}{2015}]{Singh_IA_LOWZ}
{Singh} S.,  {Mandelbaum} R.,   {More} S.,  2015, \mn@doi [\mnras] {10.1093/mnras/stv778}, \href {https://ui.adsabs.harvard.edu/abs/2015MNRAS.450.2195S} {450, 2195}

\bibitem[\protect\citeauthoryear{{Suchyta} et~al.,}{{Suchyta} et~al.}{2016}]{Balrog}
{Suchyta} E.,  et~al., 2016, \mn@doi [\mnras] {10.1093/mnras/stv2953}, \href {https://ui.adsabs.harvard.edu/abs/2016MNRAS.457..786S} {457, 786}

\bibitem[\protect\citeauthoryear{{Takahashi}, {Sato}, {Nishimichi}, {Taruya}  \& {Oguri}}{{Takahashi} et~al.}{2012}]{Takahashi2012}
{Takahashi} R.,  {Sato} M.,  {Nishimichi} T.,  {Taruya} A.,   {Oguri} M.,  2012, \mn@doi [\apj] {10.1088/0004-637X/761/2/152}, \href {http://adsabs.harvard.edu/abs/2012ApJ...761..152T} {761, 152}

\bibitem[\protect\citeauthoryear{{The LSST Dark Energy Science Collaboration} et~al.,}{{The LSST Dark Energy Science Collaboration} et~al.}{2018}]{LSST-SRD}
{The LSST Dark Energy Science Collaboration} et~al., 2018, \mn@doi [arXiv e-prints] {10.48550/arXiv.1809.01669}, \href {https://ui.adsabs.harvard.edu/abs/2018arXiv180901669T} {p. arXiv:1809.01669}

\bibitem[\protect\citeauthoryear{{Thiele}, {Marques}, {Liu}  \& {Shirasaki}}{{Thiele} et~al.}{2023}]{Thiele_HSC_PDF}
{Thiele} L.,  {Marques} G.~A.,  {Liu} J.,   {Shirasaki} M.,  2023, \mn@doi [arXiv e-prints] {10.48550/arXiv.2304.05928}, \href {https://ui.adsabs.harvard.edu/abs/2023arXiv230405928T} {p. arXiv:2304.05928}

\bibitem[\protect\citeauthoryear{Troxel \& Ishak}{Troxel \& Ishak}{2015}]{Troxel_IA_review_2015}
Troxel M.,  Ishak M.,  2015, \mn@doi [Physics Reports] {10.1016/j.physrep.2014.11.001}, 558, 1–59

\bibitem[\protect\citeauthoryear{{Van Alfen}, {Campbell}, {Blazek}, {Leonard}, {Lanusse}, {Hearin}, {Mandelbaum}  \& {The LSST Dark Energy Science Collaboration}}{{Van Alfen} et~al.}{2023}]{vanAlfen2023}
{Van Alfen} N.,  {Campbell} D.,  {Blazek} J.,  {Leonard} C.~D.,  {Lanusse} F.,  {Hearin} A.,  {Mandelbaum} R.,   {The LSST Dark Energy Science Collaboration} 2023, \mn@doi [arXiv e-prints] {10.48550/arXiv.2311.07374}, \href {https://ui.adsabs.harvard.edu/abs/2023arXiv231107374V} {p. arXiv:2311.07374}

\bibitem[\protect\citeauthoryear{Virtanen et~al.,}{Virtanen et~al.}{2020}]{2020SciPy-NMeth}
Virtanen P.,  et~al., 2020, \mn@doi [Nature Methods] {10.1038/s41592-019-0686-2}, \href {https://rdcu.be/b08Wh} {17, 261}

\bibitem[\protect\citeauthoryear{{Vlah}, {Chisari}  \& {Schmidt}}{{Vlah} et~al.}{2020}]{IA_EFT}
{Vlah} Z.,  {Chisari} N.~E.,   {Schmidt} F.,  2020, \mn@doi [\jcap] {10.1088/1475-7516/2020/01/025}, \href {https://ui.adsabs.harvard.edu/abs/2020JCAP...01..025V} {2020, 025}

\bibitem[\protect\citeauthoryear{{Wright} et~al.,}{{Wright} et~al.}{2025}]{KiDSLegacy}
{Wright} A.~H.,  et~al., 2025, \mn@doi [arXiv e-prints] {10.48550/arXiv.2503.19441}, \href {https://ui.adsabs.harvard.edu/abs/2025arXiv250319441W} {p. arXiv:2503.19441}

\bibitem[\protect\citeauthoryear{{Yao}, {Pedersen}, {Ishak}, {Zhang}, {Agashe}, {Xu}  \& {Shan}}{{Yao} et~al.}{2020a}]{SelfCalibrationYao1}
{Yao} J.,  {Pedersen} E.~M.,  {Ishak} M.,  {Zhang} P.,  {Agashe} A.,  {Xu} H.,   {Shan} H.,  2020a, \mn@doi [\mnras] {10.1093/mnras/staa1354}, \href {https://ui.adsabs.harvard.edu/abs/2020MNRAS.495.3900Y} {495, 3900}

\bibitem[\protect\citeauthoryear{{Yao}, {Shan}, {Zhang}, {Kneib}  \& {Jullo}}{{Yao} et~al.}{2020b}]{SelfCalibrationYao2}
{Yao} J.,  {Shan} H.,  {Zhang} P.,  {Kneib} J.-P.,   {Jullo} E.,  2020b, \mn@doi [\apj] {10.3847/1538-4357/abc175}, \href {https://ui.adsabs.harvard.edu/abs/2020ApJ...904..135Y} {904, 135}

\bibitem[\protect\citeauthoryear{{Yu}, {Zhang}, {Lin}  \& {Cui}}{{Yu} et~al.}{2015}]{source_lens_clustering}
{Yu} Y.,  {Zhang} P.,  {Lin} W.,   {Cui} W.,  2015, \mn@doi [\apj] {10.1088/0004-637X/803/1/46}, \href {https://ui.adsabs.harvard.edu/abs/2015ApJ...803...46Y} {803, 46}

\bibitem[\protect\citeauthoryear{{Yuan}, {Garrison}, {Hadzhiyska}, {Bose}  \& {Eisenstein}}{{Yuan} et~al.}{2022}]{ABACUS_HOD}
{Yuan} S.,  {Garrison} L.~H.,  {Hadzhiyska} B.,  {Bose} S.,   {Eisenstein} D.~J.,  2022, \mn@doi [\mnras] {10.1093/mnras/stab3355}, \href {https://ui.adsabs.harvard.edu/abs/2022MNRAS.510.3301Y} {510, 3301}

\bibitem[\protect\citeauthoryear{{Yuan} et~al.,}{{Yuan} et~al.}{2024}]{DESI_HOD}
{Yuan} S.,  et~al., 2024, \mn@doi [\mnras] {10.1093/mnras/stae359}, \href {https://ui.adsabs.harvard.edu/abs/2024MNRAS.530..947Y} {530, 947}

\bibitem[\protect\citeauthoryear{Zonca, Singer, Lenz, Reinecke, Rosset, Hivon  \& Gorski}{Zonca et~al.}{2019}]{Zonca2019}
Zonca A.,  Singer L.,  Lenz D.,  Reinecke M.,  Rosset C.,  Hivon E.,   Gorski K.,  2019, \mn@doi [Journal of Open Source Software] {10.21105/joss.01298}, 4, 1298

\bibitem[\protect\citeauthoryear{{Zuntz} et~al.,}{{Zuntz} et~al.}{2015a}]{cosmoSIS}
{Zuntz} J.,  et~al., 2015a, \mn@doi [Astronomy and Computing] {10.1016/j.ascom.2015.05.005}, \href {http://adsabs.harvard.edu/abs/2015A%26C....12...45Z} {12, 45}

\bibitem[\protect\citeauthoryear{Zuntz et~al.,}{Zuntz et~al.}{2015b}]{Zuntz_2015_cosmosis}
Zuntz J.,  et~al., 2015b, \mn@doi [Astronomy and Computing] {10.1016/j.ascom.2015.05.005}, 12, 45–59

\bibitem[\protect\citeauthoryear{{Z{\"u}rcher}, {Fluri}, {Sgier}, {Kacprzak}  \& {Refregier}}{{Z{\"u}rcher} et~al.}{2020}]{Zuercher2020a}
{Z{\"u}rcher} D.,  {Fluri} J.,  {Sgier} R.,  {Kacprzak} T.,   {Refregier} A.,  2020, arXiv e-prints, \href {https://ui.adsabs.harvard.edu/abs/2020arXiv200612506Z} {p. arXiv:2006.12506}

\bibitem[\protect\citeauthoryear{{Z{\"u}rcher} et~al.,}{{Z{\"u}rcher} et~al.}{2022}]{DESY3_Zuercher}
{Z{\"u}rcher} D.,  et~al., 2022, \mn@doi [\mnras] {10.1093/mnras/stac078}, \href {https://ui.adsabs.harvard.edu/abs/2022MNRAS.511.2075Z} {511, 2075}

\bibitem[\protect\citeauthoryear{{van Uitert} et~al.,}{{van Uitert} et~al.}{2016}]{2016MNRAS.459.3251V}
{van Uitert} E.,  et~al., 2016, \mn@doi [\mnras] {10.1093/mnras/stw747}, \href {http://adsabs.harvard.edu/abs/2016MNRAS.459.3251V} {459, 3251}

\bibitem[\protect\citeauthoryear{{van den Busch} et~al.,}{{van den Busch} et~al.}{2022}]{KiDS1000_vdB}
{van den Busch} J.~L.,  et~al., 2022, \mn@doi [\aap] {10.1051/0004-6361/202142083}, \href {https://ui.adsabs.harvard.edu/abs/2022A&A...664A.170V} {664, A170}

\makeatother
\end{thebibliography}


\appendix

\section{Additional figures}
\label{app:figs}

\begin{figure}
\includegraphics[width=\columnwidth]{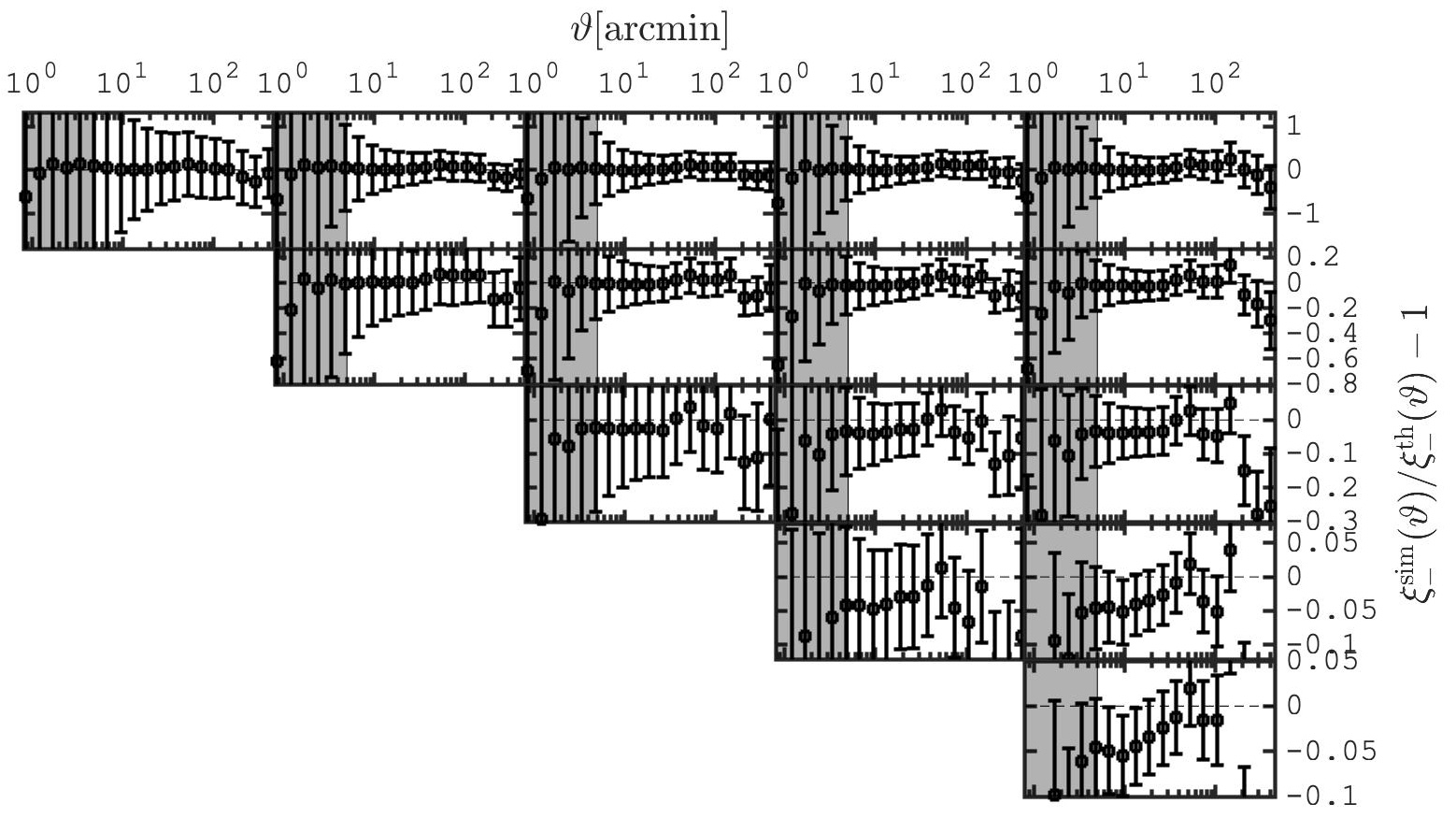}
\caption{Fractional difference on  the $\xi_-$ cosmic shear statistics between the measurements from our SkySim5000 HOD galaxies and the model predictions based on {\sc halofit}. The error bars are estimated from the analytical covariance matrix including shape noise. Fluctuations at large angles are expected from sample variance, as this is measured over a single $N$-body simulation.}
\label{fig:frac_err_sims_th}
\end{figure}

This appendix contains additional figures to further showcase the accuracy of our simulations. Fig. \ref{fig:frac_err_sims_th} presents  the fractional difference on $\xi_-$ between the measurements from our SkySim5000 HOD galaxies (noIA) and the model predictions; the agreement on these highly non-linear scales is excellent. 

\begin{figure}
\includegraphics[width=\columnwidth]{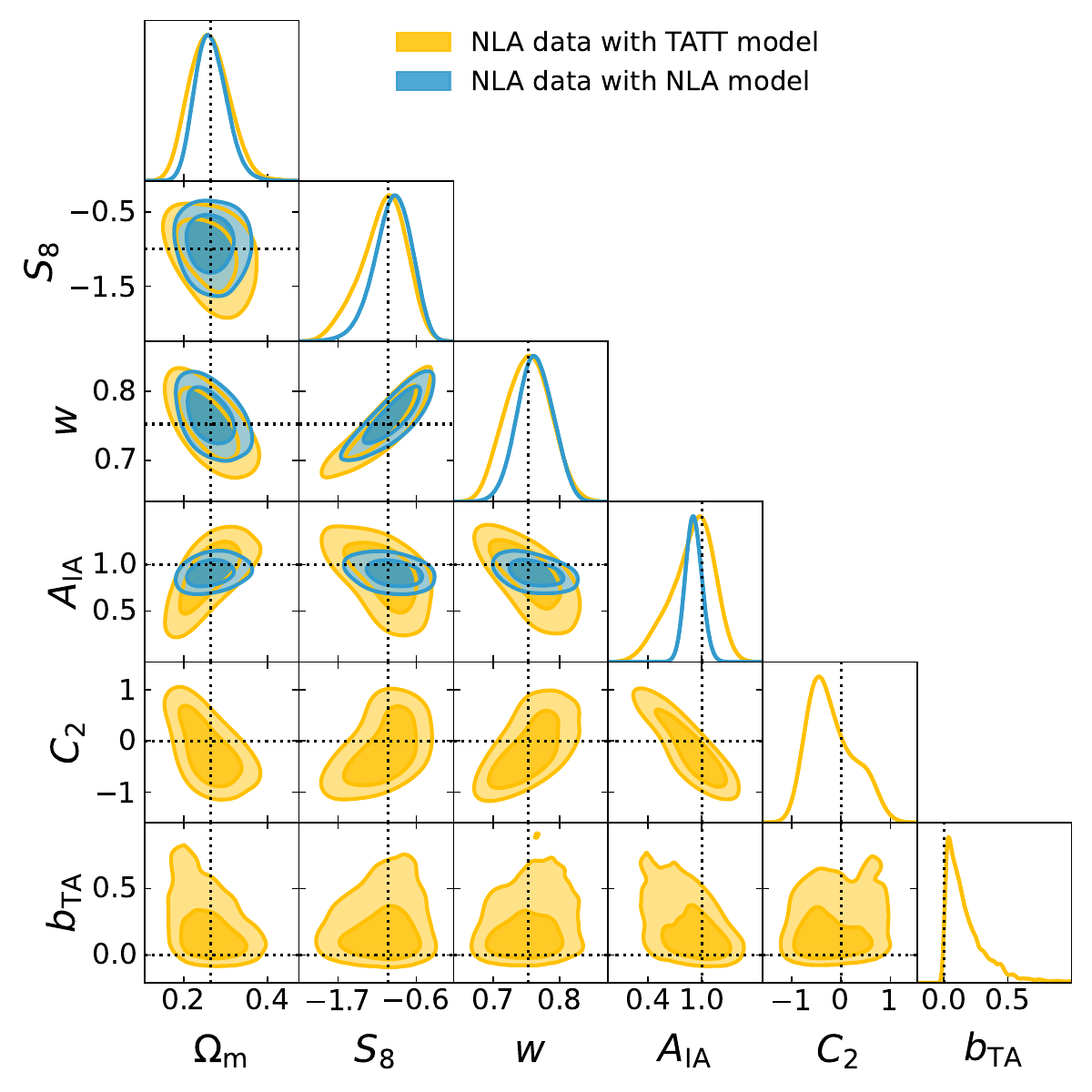}
\caption{Cosmological inference from the NLA-infused simulations, where the dark energy equation-of-state parameter $w$ is also varied.}
\label{fig:corner_nla_w0}
\end{figure}

\begin{figure*}
\includegraphics[width=3.15in]{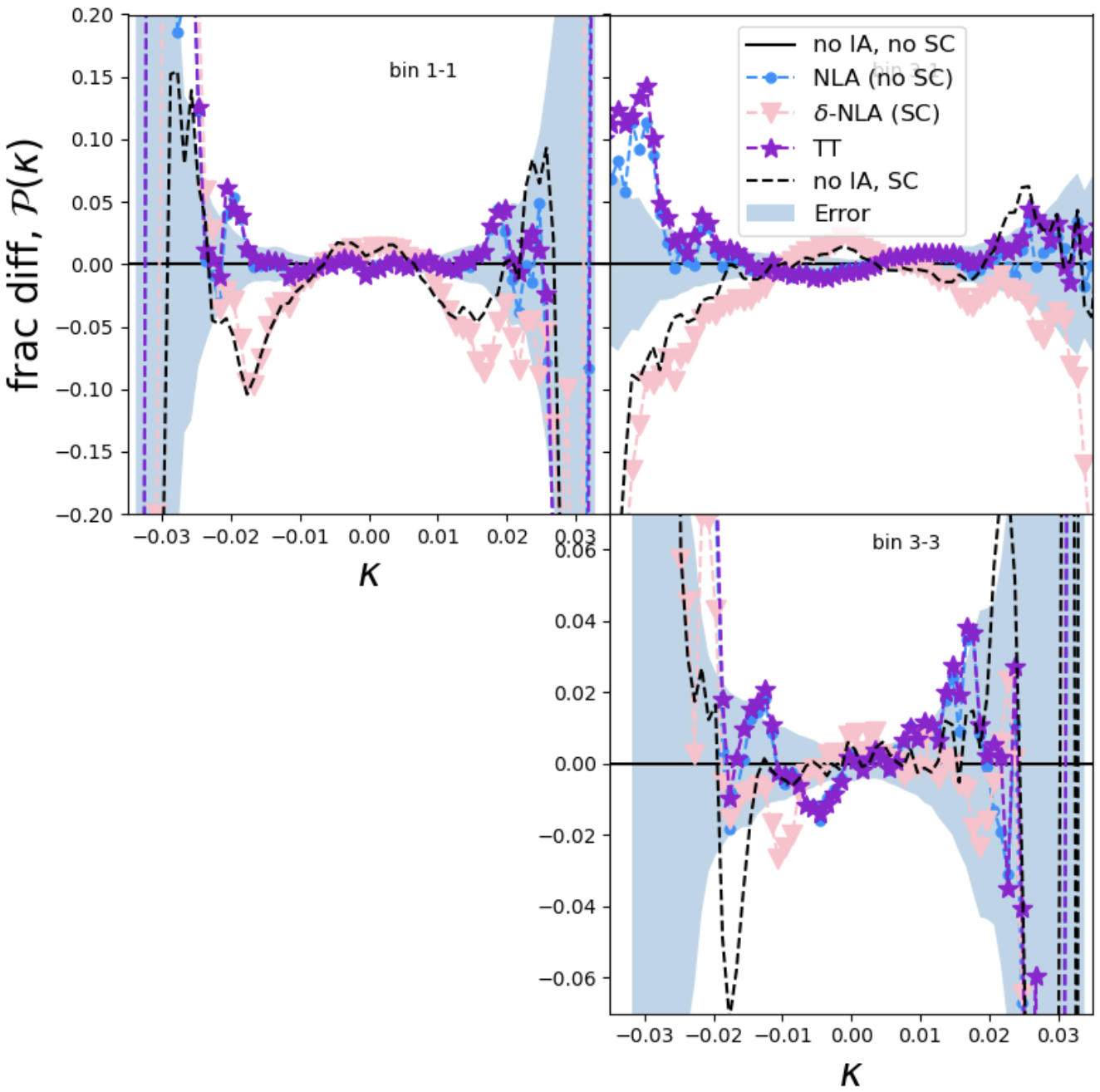}
\includegraphics[width=3.15in]{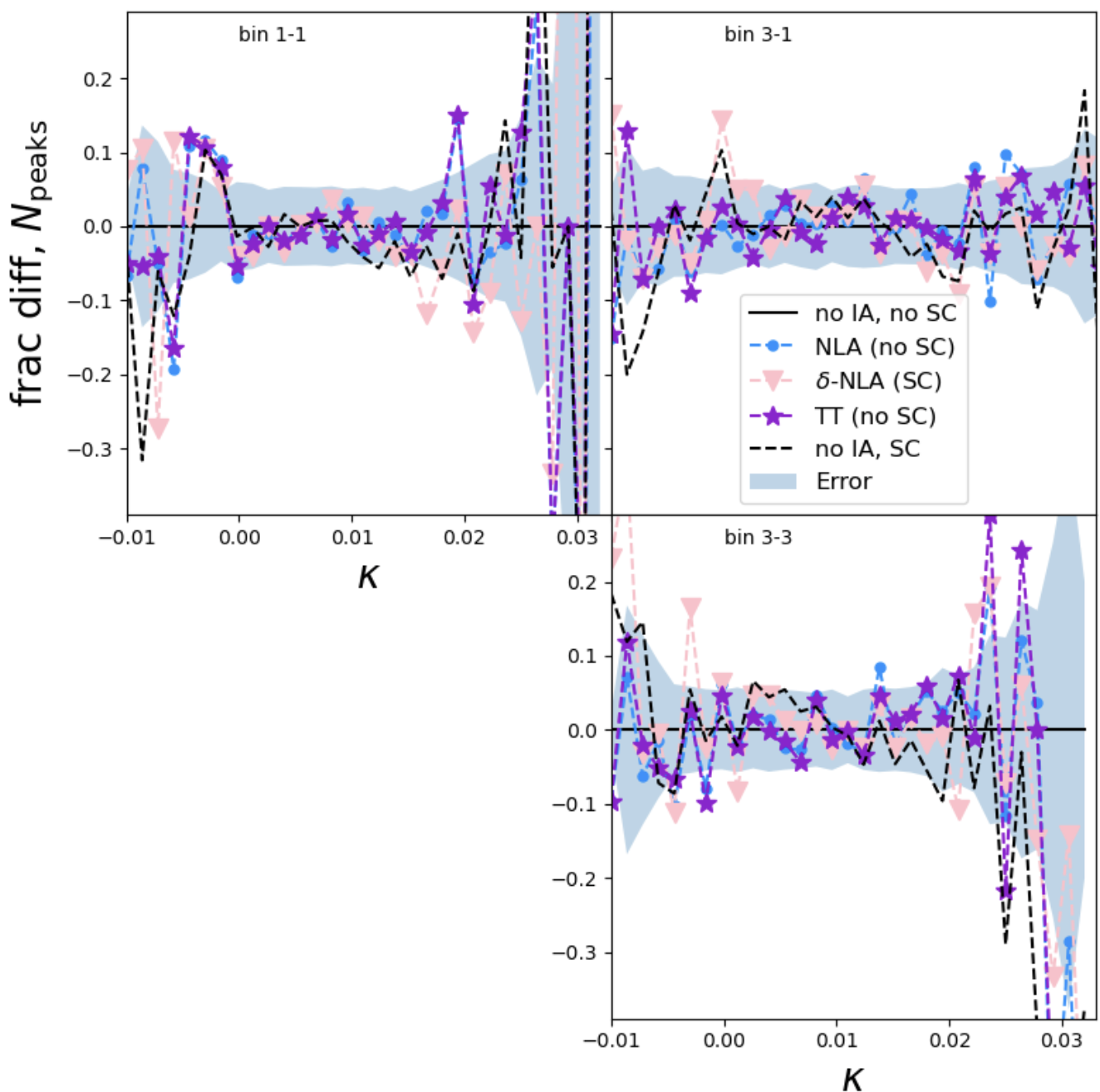}
\includegraphics[width=3.15in]{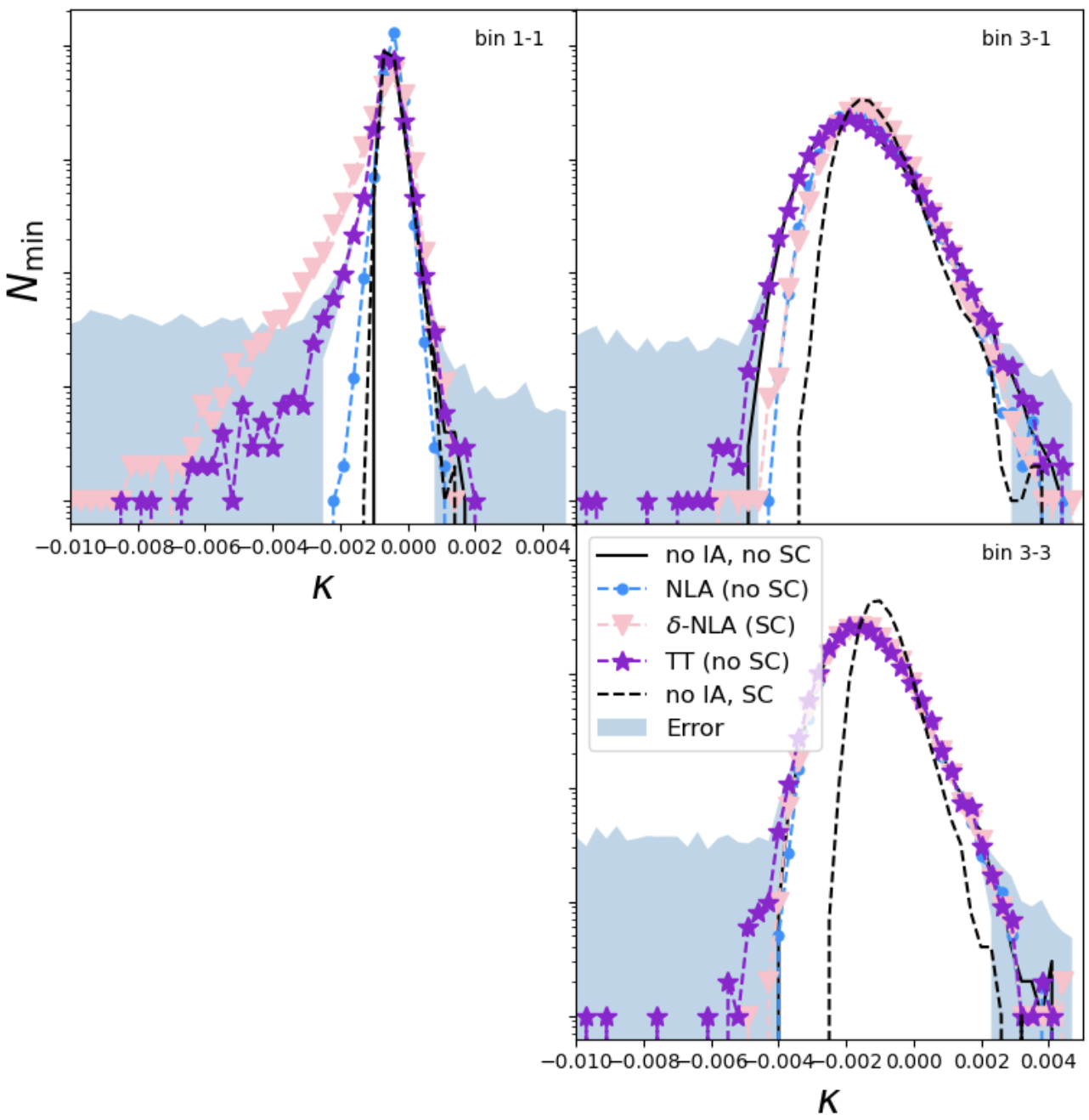}
\caption{Same as Fig. \ref{fig:PDF_IA}, but in presence of shape noise. Lensing PDF (top left) and miniuma count (bottom) are more affected by source clustering and IA than peaks (top right) is noisy data.}
\label{fig:HOS_IA_noisy}
\end{figure*}

We also show in Fig. \ref{fig:corner_nla_w0} the contours from a $w$CDM inference analysis,  where the dark energy equation-of-state parameter $w$ is varied in the range [-2;-0.3] in the MCMC, showing that the cosmological parameters are accurately recovered even in this scenario.
Finally, we show in Fig. \ref{fig:HOS_IA_noisy} the impact of 
IA on higher-order weak lensing statistics in presence of shape noise, assuming a galaxy density of $n_{\rm gal} = 3.0 \, {\rm arcmin}^{-2}$. This is about 10 times lower than the upcoming data from Rubin and {\it Euclid}, and illustrate that IA and SC contamination can be important even in this scenario for some probes (e.g. lensing PDF, minima count) but not all (peak count). This is also consistent with \citet{DESY3_Gatti_source_clust} which shows that PDF are more affected by SC than peaks. 

Note that in our calculations, the statistics are computed directly from KS-reconstructed maps, themselves generated from noisy galaxy catalogues. It is common practice to measure the statistics instead on signal-to-noise maps $\nu = \kappa/\mathcal{N}$, where pure noise maps $\mathcal{N}$ are constructed from generating pure Gaussian maps with mean set to zero and variance set to $\sigma^2 =  \sigma_\epsilon^2 / (2 \Delta \Omega_{\rm pix} n_{\rm gal})$. Here, $\Delta \Omega_{\rm pix}$ is the average pixel area, $n_{\rm gal}$ is the mean galaxy density in our survey (0.6 gal / arcmin$^2$ per tomographic bin) and $\sigma_\epsilon$ is the intrinsic dispersion in galaxy shapes, per component, which would be set to 0.27. We did not opt for this approach as it would be equivalent to relabelling our bin coordinates, which has no impact on the relative importance of the effect we are measuring in our simulations.

\section{Projected Tidal Field}
\label{app:2d_TT}

We provide in this Appendix further details on the calculation of the  projected tidal fields, and on their coupling with the intrinsic galaxy shapes.

The prescription to assign the NLA intrinsic alignment components $\widetilde{\epsilon}_1^{\rm NLA}$ and $\widetilde{\epsilon}_2^{\rm NLA}$, described in Eq. (\ref{eq:tidal_th}), involves the combinations $(s_{11} - s_{12})$ and $s_{12}$, which correspond to:
\begin{eqnarray}
 \widetilde{\epsilon}_1^{\rm NLA} (\boldsymbol \ell)  &\propto& \left(\frac{\ell_1^2 - \ell_2^2}{\ell^2} \right) \widetilde{\delta}_{\rm 2D}(\boldsymbol \ell) \, \mathcal{G_{\rm 2D}}(\sigma_{\rm G})  \, ,\nonumber \\
 \widetilde{\epsilon}_2^{\rm NLA} (\boldsymbol \ell)  &\propto& \left(\frac{\ell_1 \ell_2}{\ell^2} \right) \widetilde {\delta}_{\rm 2D}(\boldsymbol \ell)\,  \mathcal{G_{\rm 2D}}(\sigma_{\rm G}) \, .
\end{eqnarray}

These are the same filters that are used for converting convergence maps into shear maps under the KS inversion:
 \begin{eqnarray}
 \widetilde{\gamma_1} (\boldsymbol \ell)  = \left(\frac{\ell_1^2 - \ell_2^2}{\ell^2} \right) \widetilde {\kappa}(\boldsymbol \ell) \, , \hspace{1cm} \widetilde{\gamma_2} (\boldsymbol \ell)  = \left(\frac{\ell_1 \ell_2}{\ell^2} \right) \widetilde {\kappa}(\boldsymbol \ell) \, ,
\end{eqnarray}
meaning that one can linearly combine the ${\delta}_{\rm 2D}$ mass sheets with the correct coefficients and obtain NLA intrinsic ellipticities from a KS inversion code, as done in e.g. \citet{Zuercher2020a}. 
This is only true for the linear coupling (NLA,  $\delta$NLA and HOD-NLA) models, however. For the quadratic coupling (TT, $\delta$TT and HOD-TT) models, one need to revisit the above formalism.
Projecting out the $z$ components (e.i. $s_{0i}$=$s_{i0}$=0 for all $i$),  the tidal torque terms from Eq. (\ref{eq:tidal_th_TT}) can be expanded as:
 \begin{eqnarray}
\gamma^{\rm TT}_{ij}&=& C_2 \left[\sum_{k=1,2} s_{ik} s_{kj} -\frac{1}{3} \delta_{ij} s^2\right] \\                                  &=&C_2 \left[ s_{i1}s_{1j}  + s_{i2}s_{2j}  -  \frac{1}{3} \delta_{ij} \left( s_{11}^2 + s_{22}^2 +2s_{12}^2 \right)  \right] \, .
\end{eqnarray}
Specifically, this yields:
 \begin{eqnarray}
\gamma^{\rm TT}_{11}&=& C_2 \left[\frac{2}{3}s_{11}^2  -  \frac{1}{3}s_{22}^2 + \frac{1}{3}s_{12}^2 \right]\, ,  \nonumber \\
\gamma^{\rm TT}_{22}&=& C_2 \left[-\frac{1}{3}s_{11}^2  +  \frac{2}{3}s_{22}^2 + \frac{1}{3}s_{12}^2  \right]\, , \\
\gamma^{\rm TT}_{12}&=& C_2 s_{12}\left[s_{11}+s_{22}  \right]\, . \nonumber                                  
\end{eqnarray}
With the standard ellipticity definitions $\epsilon_1^{\rm TT}\equiv \gamma_{11}^{\rm TT} - \gamma_{22}^{\rm TT}$ and $\epsilon_2^{\rm TT}\equiv -\gamma_{12}$, we obtain:
\begin{eqnarray}
\epsilon_1^{\rm TT} = C_2  \left[ s_{11}^2 - s_{22}^2\right] \, , \epsilon_2^{\rm TT} = -C_2 s_{12}\left[s_{11}+s_{22}  \right]\, .
\end{eqnarray}
These are the equation we use in this paper to infuse a quadratic coupling between galaxies and tidal fields.

\bsp	
\label{lastpage}
\end{document}